\documentclass[12pt]{article}

\usepackage[T1]{fontenc}
\usepackage[utf8]{inputenc}
\usepackage[margin=1in]{geometry}
\usepackage{amsmath,amssymb}
\usepackage{graphicx}
\usepackage{placeins}
\usepackage{needspace}
\usepackage{longtable}
\usepackage[version=4]{mhchem}
\usepackage{siunitx}
\usepackage{bm}
\usepackage{microtype}
\usepackage[numbers]{natbib}
\usepackage[hidelinks]{hyperref}
\usepackage{lineno}

\let\cite\citep
\newcommand{\ppm}{ppm}
\newcommand{\sccm}{sccm}
\newcommand{\slm}{sLm}
\DeclareSIUnit\bar{bar}
\DeclareSIUnit\molecule{molecule}
\DeclareMathOperator\erf{erf}

\begin{document}

\begin{center}
{\LARGE\bfseries Simplifying Gas-Phase Kinetic Measurements with a Dual-Arm Flow Tube\par}
\vspace{1em}
{\large Olivier Durif$^{1,*}$ \quad Barbara Nozi\`ere$^{1}$\par}
\vspace{0.5em}
{\normalsize $^{1}$Department of Chemistry, KTH Royal Institute of Technology, Stockholm 10044, Sweden\par}
\vspace{0.25em}
{\normalsize $^{*}$Corresponding author: \href{mailto:olivier@durif.fr}{olivier@durif.fr}\par}
\end{center}

\begin{abstract}

We demonstrate a simple dual-arm flow tube reactor that provides well-defined reaction times ranging from one second to several minutes for gas-phase kinetic measurements near ambient temperature and pressure. Constructed entirely from standard tubing, the flow tube reactor eliminates the need for injector translation. It operates in two configurations: a variable-length configuration, in which the tube length is changed at fixed flow, and a variable-flow configuration, in which an exhaust branch allows the flow through the reaction section to be adjusted while maintaining constant reactant inlet flows.
With \SI{3.96}{\milli\meter} i.d. tubing, the measured tracer pulses had a typical full width at half maximum of about 10\% of their mean arrival time. A kinetic demonstration using the ozonolysis of 2,3-dimethyl-2-butene (TME) yields an apparent rate coefficient for TME loss $k_\mathrm{app}=\SI{2.0(5)e-15}{\centi\meter\cubed\per\molecule\per\second}$. Fitting the same kinetic data with a reaction network including secondary OH and peroxy-radical chemistry yields $k=\SI{1.18(19)e-15}{\centi\meter\cubed\per\molecule\per\second}$ at \SI{293}{\kelvin} for TME + \ce{O3}, consistent with the IUPAC recommendation.
This compact, low-cost reactor provides an alternative to conventional movable-injector flow tubes for atmospheric chemistry, fundamental kinetics, and gas--liquid or gas--solid uptake studies.

\end{abstract}

\section{Introduction}

Reliable kinetic measurements require accurate control and characterization of the time during which reactants interact. In flow-based reactors, this time is governed by the residence time distribution (RTD), whose mean defines the effective reaction time and whose width determines the temporal resolution of the measurement \cite{danckwertsContinuousFlowSystems1953,hopleyResidenceTimeDistribution2019}.

Gas-phase kinetic measurements use several types of reactors. The most common are reaction chambers \cite{dehaanHeterogeneousChemistryTroposphere1999,glowackiDesignInitialResults2007b}, flash photolysis cells \cite{kleyBlitzlichtPhotolyseanordnungFurSchumannUV1963,braunFlashPhotolysisMethane1966}, and flow tube reactors \cite{tollefsonReactionAtomicHydrogen1948,howardKineticMeasurementsUsing1979b}. For extreme temperatures, such as those encountered in combustion, jet-stirred reactors \cite{longwellHighTemperatureReaction1955,bartokMixingJetstirredReactor1960,ayassMixingstructureRelationshipJetstirred2016} or shock tubes \cite{reslerInstrumentStudyRelaxation1955,tranterDesignHighpressureSingle2001,davidsonRecentAdvancesShock2009} are often used, and at very low temperatures, room-temperature approaches can be adapted by cryogenically cooling the reactor \cite{smithRateMeasurementsReactions1973,trainorGasPhaseRecombination1973} or, more creatively, using adiabatic expansions \cite{dupeyratDesignTestingAxisymmetric1985,roweTechniquesStudyReaction1995,durifDesignLavalNozzles2022}.

\paragraph{Tank-type reactors.}

Reaction chambers \cite{glowackiDesignInitialResults2007a,osseiranDevelopmentValidationThermally2020}, continuous stirred tank reactors (CSTRs) \cite{geryContinuousStirredTank1985}, and flash or pulsed photolysis cells \cite{luoSimultaneousDeterminationTransient2020} share a common principle: reactions occur within a fixed volume, and changes in gas composition are monitored over time. To illustrate kinetic measurements in such reactors, consider the bimolecular reaction $\ce{A + B ->[k] P}$, where P denotes products and $k$ is the bimolecular rate coefficient.



In many gas-phase studies, A is supplied continuously, while a pulse generates B, thereby initiating the reaction and defining time zero for the kinetic measurement. In practice, B is commonly generated in situ by pulsed photolysis of a precursor, with flash lamps used historically \cite{horneRateHabstractionOH1967,greinerHydroxylRadicalKineticsKinetic1967} and lasers now more commonly employed \cite{shepsUVAbsorptionProbing2014,lewisNovelMultiplexAbsorption2018,debnathInvestigationKineticsMechanistic2023,fernholzKineticsReactionOH2024}. These measurements are generally performed in benchtop reaction cells, where gas residence times range from hundreds of milliseconds to a minute.

Rate coefficients are commonly determined by placing A in sufficient excess over B that $[A]$ remains effectively constant. Under these conditions, B decays exponentially. A background decay rate $k^\prime_0$, corresponding to the intercept at $[A]=0$, is first measured without A; the observed decay rate $k^\prime_1$ is then measured with A in excess. This so-called pseudo-first-order approach yields $k=(k^\prime_1-k^\prime_0)/[A]$. Because only decay rates are used, this procedure does not require the absolute concentration of B, which is often a transient radical. However, gas renewal in a flowing tank-type reactor contributes to the observed decay and is associated with a broad residence time distribution. For an ideal well-mixed tank, the mean residence time is $\tau=V/Q$, where $V$ is the reactor volume and $Q$ is the volumetric flow rate, and the normalized RTD is $E(t)=\exp(-t/\tau)/\tau$. Neglecting wall and chemical losses, gas renewal alone gives a background first-order decay rate $k^\prime_0=1/\tau$. This renewal contribution is removed in the subtraction $k^\prime_1-k^\prime_0$, provided that the paired measurements are performed under otherwise identical experimental conditions, including the same reactor pressure, temperature, volumetric flow rate, and mixing geometry, so that the RTD remains unchanged when A is added.

At small $V/Q$, rapid gas renewal makes a large contribution to $k^\prime_0$, thereby shortening the observable lifetime of B even without A. Increasing $V/Q$ reduces this background contribution and permits slower chemistry to be monitored. The trade-off is a lower maximum repetition rate: the gas must be fully renewed between successive measurements, so averaging repeated experiments becomes increasingly time-consuming as $V/Q$ increases \cite{danckwertsContinuousFlowSystems1953,danckwertsEffectIncompleteMixing1958,yablonskyNewApproachDiagnostics2009}. Further details of the pseudo-first-order method are provided in Appendix~\ref{app:pseudo_first_order}.

An alternative way to define time zero is to use continuous photolysis to build up a radical population and then monitor its decay after the lamps are switched off \cite{onelIntercomparisonHO2Measurements2017}. This method can establish a steady state under illumination, but secondary chemistry initiated during photolysis may persist after switch-off and complicate the interpretation of the decay.

Another strategy for retrieving a bimolecular rate coefficient, distinct from the pseudo-first-order method, is to employ a relative-rate technique. Rather than following a post-photolysis decay, this method maintains radical production continuously and derives the target rate coefficient relative to that of a reference compound \cite{gorsePhotochemistryGaseousHydrogen1972,simonaitisReactionO1DH2O1973,atkinsonKineticsMechanismsGasphase1986}. These experiments are generally performed in much larger volumes, often environmental chambers (or \textit{smog chambers}) widely used in atmospheric research, with durations ranging from minutes to days \cite{doyleGasPhaseKinetic1975,nilssonPhotochemicalReactorStudies2009,wangDesignCharacterizationSmog2014,romanInvestigationsGasphasePhotolysis2022,xinRateCoefficientsReactions2024}.

Despite their conceptual simplicity, tank-type reactors remain sensitive to gas inhomogeneity, wall interactions, and synchronization in pulsed experiments; data analysis also becomes more complex beyond relative-rate or pseudo-first-order methods because gas renewal and mixing must be rigorously accounted for \cite{danckwertsContinuousFlowSystems1953,danckwertsEffectIncompleteMixing1958}.

\paragraph{Tubular flow reactors and the plug-flow limit.}


By contrast, tubular flow reactors aim to approximate plug-flow behavior and have been a prolific source of kinetic data \cite{atkinsonEvaluatedKineticPhotochemical2004,atkinsonEvaluatedKineticPhotochemical2006c}. In the ideal plug-flow limit, the residence time distribution (RTD) is a Dirac impulse; experimentally, axial dispersion broadens it into a finite, narrow Gaussian profile \cite{raviChapter2Flow2017}. Their narrow RTD and point-by-point measurement principle avoid the explicit gas-renewal loss term associated with fixed-volume reactors and simplify kinetic retrieval beyond pseudo-first-order conditions. We focus here on flow tubes designed to measure fundamental rate coefficients, rather than on flow reactors used for atmospheric aging or aerosol formation where kinetic parameters are not the primary output.

Two main strategies are used to define reaction time in flow-tube kinetics. In pulsed experiments, a periodic impulse (e.g., excimer laser photolysis) initiates the reaction, and the time evolution is reconstructed at a fixed point downstream \cite{osbornMultiplexedChemicalKinetic2008a}. In continuous steady-state experiments, reaction time is varied by changing the physical distance between injection and detection, typically with a movable injector \cite{howardLaserMagneticResonance1974a,noziereUptakeMethylVinyl2006}.

Pulsed flow-tube methods resemble measurements in photolysis cells, but the gas can be renewed efficiently between pulses. The effective first-order rate constant $k^\prime$ can therefore, in principle, be determined directly from the observed decay without subtracting a gas-renewal term (the renewal contribution to $k^\prime_0$ is zero), although accurate measurements still require averaging many pulses rather than relying on a single observation.

Wall effects remain a concern, particularly in pulsed flow tubes and chamber experiments, because adsorbing species can gradually accumulate on reactor surfaces \cite{krechmerAlwaysLostNever2020}. Over time, the resulting surface modification can introduce heterogeneous-chemistry artifacts that are difficult to detect under non-equilibrium wall conditions.

Compared with pulsed methods, continuous-flow operation offers a complementary advantage: once steady state is established, the signal can be averaged as long as needed at each reaction time, while the preceding transient response can provide information on gas-wall interactions \cite{durifStrongUptakeGasPhase2024a}. Its spatial and temporal resolution is nevertheless limited by the injection mixing zone, and the movable injectors commonly used to vary reaction time can introduce leakage at sliding seals and mechanical instabilities \cite{westenbergAtomMoleculeKinetics1967,andersonGasPhaseRecombination1974,hansonNH3MassAccommodation2003}.

From a measurement perspective, the ideal reactor for gas-phase kinetics would combine (i) a narrow and directly verifiable RTD to minimize timing uncertainty, (ii) stable and independently controllable pressure and flow conditions, (iii) low wall interactions to avoid systematic biases, and (iv) mechanical simplicity to ensure reproducibility and facilitate adoption. Existing designs meet these requirements only partially: tank-type reactors suffer from broad RTDs, and conventional flow tubes rely on movable injectors that introduce mechanical complexity and potential leakage. The uncertainty propagated from reactor characterization into derived rate coefficients remains a significant concern \cite{nagyUncertaintyArrheniusParameters2011}.

\begin{figure}
    \centering
    \includegraphics[width=\linewidth]{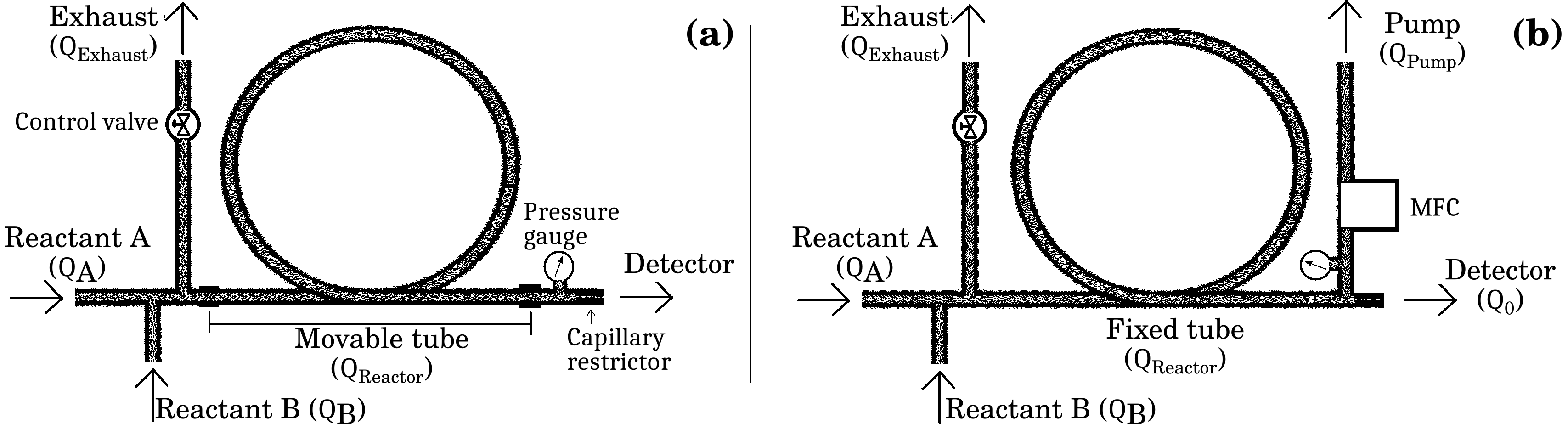}
    \caption{\textbf{Schematic of the dual-arm flow tube reactor operated in two configurations.} In both cases, reactants A and B are introduced at flow rates $Q_A$ and $Q_B$ and mixed at a tee junction defining the start of the reaction section. (a) \textbf{Variable-length configuration:} the residence time is adjusted by changing the length of the reaction tube. An optional exhaust arm facilitates operation at stable reactor pressure. (b) \textbf{Variable-flow configuration:} the residence time is adjusted by changing the flow through the reaction tube. A downstream mass flow controller (MFC) regulates $Q_\text{pump}$ and hence $Q_\text{reactor}$, while the exhaust arm maintains constant inlet reactant flows and stable reactor pressure.}
    \label{fig:flowtube}
\end{figure}

These trade-offs motivate the simple dual-arm flow tube demonstrated here (Figure~\ref{fig:flowtube}), which enables continuous-flow kinetic measurements without injector translation. Its exhaust branch decouples reactor pressure from inlet flows, allowing the reaction section to be set to and maintained at the required pressure while the internal reactor flow is varied. The reactor is operated in two configurations:
\begin{enumerate}
    \item \textbf{Variable-length configuration:} The reactor volume is adjusted by physically changing the tube length at a fixed flow rate.
    \item \textbf{Variable-flow configuration:} The flow rate within the reaction tube is varied while an exhaust line maintains constant upstream reactant concentrations.
\end{enumerate}

This design combines well-defined reaction times through a narrow RTD, short radial diffusion distances in small-diameter disposable perfluoroalkoxy alkane (PFA) tubing compared with conventional larger-diameter flow tubes, reduced wall interactions, low cost, and mechanical simplicity.


\section{Dual-arm reactor principle}

In both configurations, reactant streams A and B, supplied at flow rates $Q_A$ and $Q_B$, are combined at a tee junction that marks the entrance to the PFA reaction tube. The resulting gas mixture flows through the reaction tube at a volumetric flow rate $Q_\text{reactor}$. At the downstream end, a sampling flow $Q_0$ is directed through a capillary restrictor to the detector. The residence time is determined by the reaction-tube volume and $Q_\text{reactor}$. The roles of the exhaust and pump branches are described separately for each configuration below.

\subsection{Variable-length configuration}


In the variable-length configuration (Figure~\ref{fig:flowtube}a), the residence time is adjusted by changing the length $L$ of the PFA reaction tube while maintaining a constant flow $Q_\text{reactor}$. For a cylindrical tube of internal radius $r$, and neglecting the fixed volumes of the connections, $V=\pi r^2L$ and the residence time is $\tau=V/Q_\text{reactor}=\pi r^2L/Q_\text{reactor}$.

This approach offers a robust alternative to complex sliding injectors. We employ standard PFA tubing (1/4-inch or 1/8-inch OD), whose chemical inertness and low surface reactivity make it particularly well suited to flow-tube experiments \cite{demingMeasurementsDelaysGasphase2019a,kennedyAircraftBasedThree2011}; over the residence times investigated here, wall losses can reasonably be considered negligible for many species \cite{liAssessmentLongTubing2023}, including some reactive radicals. This is supported by our previous study, which showed negligible interactions of organic peroxy radicals (RO$_2$) with PFA surfaces, whereas strong uptake was observed on metals \cite{durifStrongUptakeGasPhase2024a}. The interchangeable reaction tube is connected at both ends using commercially available Ultra-Torr-type vacuum fittings, allowing rapid replacement and preventing contamination carry-over between experiments.

Unlike conventional large-diameter flow tubes where achieving rapid radial mixing is challenging, our design relies on the efficiency of a simple tee junction combined with narrow tubing. Since the characteristic timescale for radial diffusion scales with the square of the diameter ($\tau_{\mathrm{diff}} \propto d^2/D_\mathrm{m}$), reducing the diameter strongly accelerates cross-sectional homogenization.

For the \SI{3.96}{\milli\meter} inner-diameter tubing used here, the characteristic molecular diffusion time across the tube radius is approximately \SI{0.4}{\second} under atmospheric conditions (see Appendix~\ref{app:rtd_characterization}). Increasing the tube diameter to \SI{3}{\centi\meter}, a value commonly used in conventional flow tubes, would increase this timescale by a factor of approximately 60, to about \SI{20}{\second}. This estimate is, however, only an order-of-magnitude indicator of radial transport, not an exact experimental mixing time.

Radial homogenization downstream of the tee junction depends on the injection geometry and on the ratio of the available residence time to the radial diffusion time. Because the radial diffusion time is independent of the bulk velocity, the axial distance required for a given degree of homogenization increases with the mean velocity. Molecular radial diffusion is expected to be rapid relative to the residence time for $\tau\gtrsim\SI{2}{\second}$. At the shortest residence times ($<\SI{2}{\second}$), however, entrance-mixing effects cannot be excluded \textit{a priori}.

Accordingly, varying the tube length from a few centimeters to several meters provides access to residence times ranging from approximately one second to several minutes while retaining the same narrow-tube geometry and minimizing wall-loss issues compared with traditional reactors made of glass or metal.

Moreover, this flow tube method offers low gas consumption and can in principle operate over a broad pressure range, subject to the pressure ratings of the tubing and fittings and to the detector requirements.

An exhaust line can be added downstream of the mixing point (Figure~\ref{fig:flowtube}a) to decouple the reactor pressure from the inlet flow rates, allowing the pressure to be controlled independently of the detector inlet pressure.

A key feature is the downstream capillary restrictor, which provides the flow resistance required to maintain the reactor at high pressure (e.g., atmospheric) while transferring a sampling flow $Q_0$ to the detector under vacuum. For the representative capillary restrictor used here ($L_c=\SI{5}{\centi\meter}$ and $r_c=\SI{65}{\micro\meter}$), the measured flow was $Q_0\approx\SI{50}{\sccm}$ for a pressure drop of approximately \SI{1}{\bar}. As discussed in Appendix~\ref{app:capillary_pressure_drop}, a simple incompressible laminar-flow model predicts $\Delta P\approx\SI{1.09}{\bar}$ at this flow, close to the measured pressure drop.

This method provides a reliable, easy-to-build setup for performing kinetic measurements. Further practical details are discussed in Appendix~\ref{app:flow_tube_discussion}.

\subsection{Variable-flow configuration}

In the variable-flow configuration (Figure~\ref{fig:flowtube}b), the reaction-tube length remains fixed and the residence time is varied by changing $Q_\text{reactor}$. The downstream mass flow controller sets the pumped flow $Q_\text{pump}$, while a separate sampling flow $Q_0$ is directed to the detector. The flow through the reaction tube is therefore $Q_\text{reactor}=Q_\text{pump}+Q_0$. Because the inlet flows $Q_A$ and $Q_B$ are held fixed, the excess gas leaves through the exhaust branch, with $Q_\text{exhaust}=Q_A+Q_B-Q_\text{reactor}$. Increasing $Q_\text{pump}$ therefore increases $Q_\text{reactor}$, shortens the residence time, and decreases $Q_\text{exhaust}$.

Two conditions define the accessible flow range. First, $Q_\text{exhaust}$ must remain positive to prevent back-diffusion through the exhaust line. Second, the pressure immediately upstream of the capillary restrictor must remain stable so that $Q_0$, and therefore the detector response, does not change.

In the variable-flow demonstration, with $Q_0=\SI{100}{\sccm}$, varying $Q_\text{pump}$ between \SIrange{0}{300}{\sccm} changed the residence time by a factor of four without measurably affecting $Q_0$. At higher $Q_\text{pump}$, the increasing pressure drop between the mixing junction and the detector inlet lowered the pressure upstream of the capillary restrictor and caused $Q_0$ to decrease. This pressure dependence limits the accessible variable-flow range. Varying the tube length is therefore preferred for routine kinetic measurements because it allows the flow, reactor pressure, and sampling conditions to remain fixed over a broader accessible residence-time range.

\section{Flow tube characterization}

The flow tube was characterized in two stages. First, the reactor's residence-time response was measured by monitoring acetone and acetonitrile as non-reactive tracers using a proton-transfer-reaction time-of-flight mass spectrometer (PTR-TOF-MS). Second, reactor performance was evaluated through the ozonolysis of 2,3-dimethyl-2-butene.

\subsection{Residence time characterization}

The residence-time response was characterized by tracking brief pulses of acetone and acetonitrile, which are readily detected by PTR-TOF-MS \cite{holzingerValidityLimitationsSimple2019}. Measurements were performed for five tube lengths ($L=\SIrange{0}{700}{\centi\meter}$) and reactor flow rates ranging from \SIrange{62}{917}{\sccm}, thereby sampling both length and flow dependence. Each length--flow condition included five to seven tracer pulses lasting \SIrange{0.5}{2}{\second} (Table~\ref{tab:RTD_symmetric}). Manual initiation of each pulse introduced an estimated uncertainty of approximately $\pm\SI{0.5}{\second}$ in its start time relative to the mass-spectral acquisition. During the tracer-pulse measurements, the PTR-TOF-MS signal was recorded at \SI{5}{\hertz}, corresponding to one mass spectrum every \SI{0.2}{\second}.

\begin{figure}[ht]
	\centering
	\includegraphics[width=\textwidth]{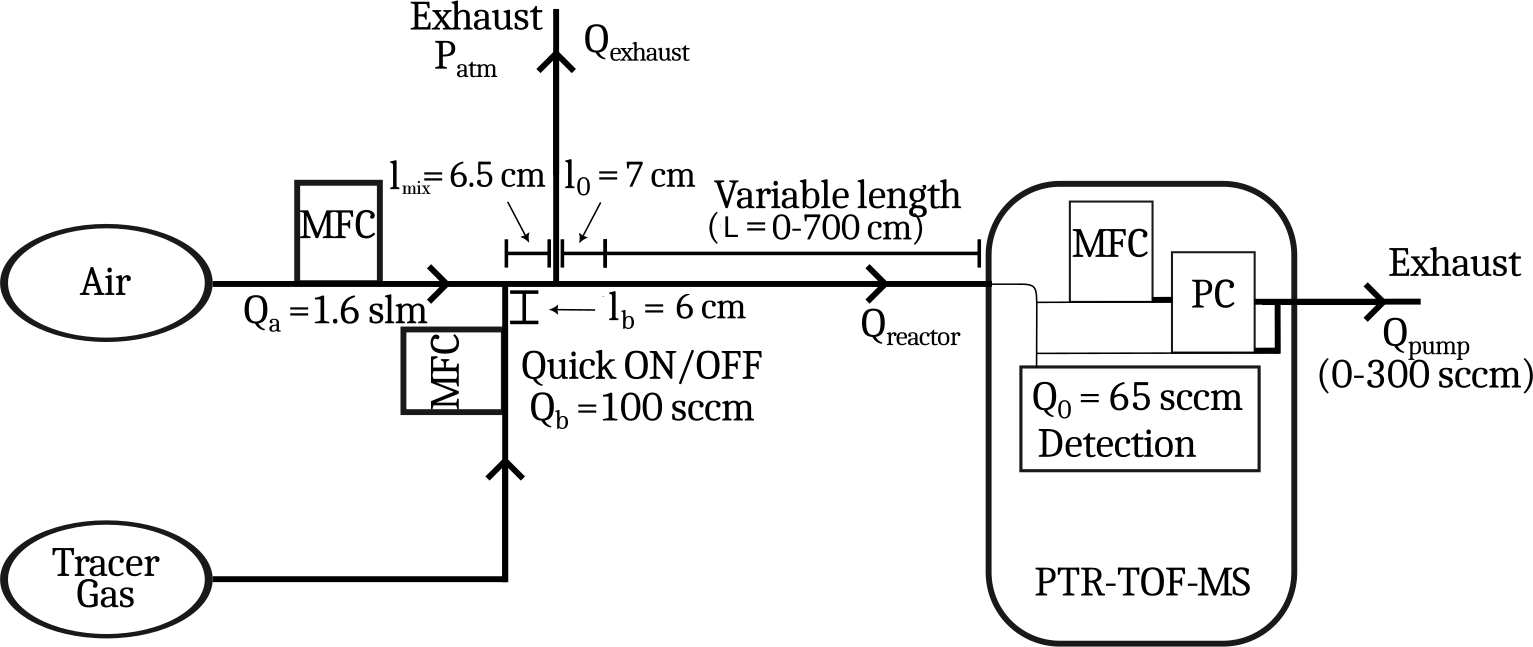}
    \caption{\textbf{Schematic of the setup used to characterize the residence time distribution.}
    The flow tube consists of PFA tubing with an internal diameter of \SI{3.96}{\milli\meter}. A tracer gas is injected as a rapid ON/OFF pulse through a dedicated inlet ($Q_B$, line length $\ell_B$) and mixed with a \SI{1.6}{\slm} flow of synthetic air ($Q_A$). The short section from the mixing tee to the exhaust is denoted $\ell_\text{mix}$. The total flow through the post-exhaust section, $Q_\text{reactor}$, is imposed downstream by a mass flow controller (MFC) located inside the PTR-TOF-MS. A pressure controller (PC), set to \SI{980}{\milli\bar}, ensures stable pressure when the flow through the MFC is varied. $Q_A+Q_B = Q_\text{exhaust} + Q_\text{reactor}$ and $Q_\text{reactor}=Q_\text{pump}+Q_0$.}
	\label{fig:setup_flowtube}
\end{figure}

Because the timing reference is the opening of the tracer-flow controller upstream of the mixing tee, the calculated tracer residence time $\tau$ includes transport through the tracer inlet, the fixed mixing section, and the post-exhaust flow tube (see Figure~\ref{fig:setup_flowtube}):
\begin{equation}
    \tau = \frac{\pi r^2\ell_B}{Q_B}
    + \frac{\pi r^2\ell_\text{mix}}{Q_A+Q_B}
    + \frac{\pi r^2(\ell_0+L)}{Q_\text{reactor}},
    \label{eq:arrival_time}
\end{equation}
where $r=\SI{1.98}{\milli\meter}$, $\ell_B=\SI{6.0}{\centi\meter}$, $Q_B=\SI{100}{\sccm}$, $\ell_\text{mix}=\SI{6.5}{\centi\meter}$, $Q_A=\SI{1.6}{\slm}$, $\ell_0=\SI{7}{\centi\meter}$, and $L$ ranges from \SIrange{0}{700}{\centi\meter}. The second term, representing transit through the fixed mixing section, is approximately \SI{0.03}{\second} and is negligible under the conditions studied, but it is retained for completeness. The transit time through branch B is approximately \SI{0.44}{\second}; it is small compared with most arrival times but is retained for accurate calibration, particularly for the shortest conditions. This branch-B contribution is specific to the tracer residence-time characterization; in kinetic measurements, the reaction time starts when reactant streams A and B merge at the mixing tee. Measurements at $L=0$ retain the tracer inlet line $\ell_B$, the fixed mixing section $\ell_\text{mix}$, and the fixed reactor section $\ell_0$.

The measured response to each tracer pulse was modeled with a Gaussian:
\begin{equation}
    S(t) = \frac{\alpha}{\sigma \, \sqrt{2 \, \pi}} \times \exp{\left( - \frac{(t - \mu)^2}{ 2 \, \sigma^2} \right)} + \epsilon, \label{eq:RTD}
\end{equation}
where $S(t)$ is the measured signal, $\alpha$ is the integrated Gaussian peak area, $\epsilon$ is the baseline signal, $\mu$ is the fitted mean arrival time, and $\sigma$ is the standard deviation of the measured response. For each pulse, the four parameters $\alpha$, $\mu$, $\sigma$, and $\epsilon$ were fitted independently by nonlinear least squares. The fitted width includes broadening from the finite tracer pulse and dispersion in the flow tube. The manual start-time uncertainty translates an entire response along the time axis and therefore changes its fitted $\mu$, but it does not appreciably affect the fitted width $\sigma$ of that individual response. Because Eq.~\eqref{eq:arrival_time} predicts the mean tracer arrival time directly from the relevant volumes and flow rates, the primary quantity of interest is not the fitted mean $\mu$ but the response width $\sigma$, which quantifies the temporal dispersion of the measured reactor response.

Across all calculated residence times, $\sigma$ remained below 13\% of the mean arrival time $\mu$. The typical temporal resolving power was $\mu/\Delta t\approx10$, with $\Delta t=\mathrm{FWHM}=2.355\sigma$ (Figure~\ref{fig:RTD}). An asymmetric Gaussian model was also evaluated to describe possible tailing (Appendix~\ref{app:rtd_characterization}), but the symmetric model was sufficient for the main analysis because the choice of line-shape model contributes an estimated model-form uncertainty of approximately 10\% to the response width and does not affect the conclusion that the response was narrow. As a secondary consistency check, comparison of the fitted centers with the calculated residence times gives $\mu=1.01\,\tau$, confirming that Eq.~\eqref{eq:arrival_time} accurately predicts the mean tracer arrival time. This comparison can also be inverted, for known volumes, to verify the calibration of the flow controllers. Conditions with $\tau\leq\SI{5}{\second}$ are excluded from this mean-time comparison because the approximately $\pm\SI{0.5}{\second}$ uncertainty in the manual injection start represents at least 10\% of the measured time; the fitted $\sigma$ values for these short-time responses are nevertheless retained in the dataset and summarized in Table~\ref{tab:RTD_symmetric}.

Further experimental details and supporting analyses of the residence-time characterization are provided in Appendix~\ref{app:rtd_characterization}.

\subsection{Kinetic demonstration}

\begin{figure}[ht]
	\centering
	\includegraphics[width=\textwidth]{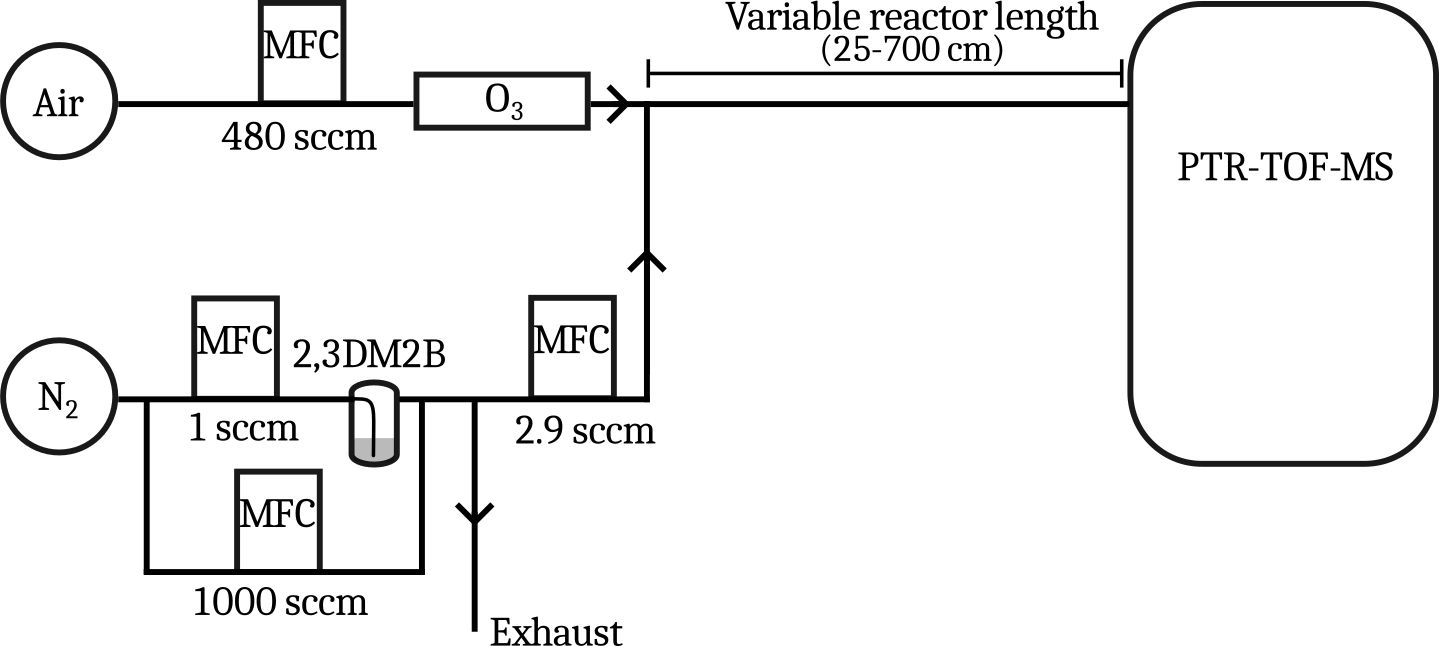}
	\caption{\textbf{Schematic of the setup used to measure the ozonolysis kinetics of 2,3-dimethyl-2-butene.}
		Ozone was produced by passing synthetic air through a cell equipped with lamps at \SI{185}{\nano\meter}. 2,3-Dimethyl-2-butene was introduced from a bubbler containing the pure liquid and then diluted with \ce{N2}. A small fraction of this mixture was combined with the ozone and synthetic-air flow. The reactor was maintained at atmospheric pressure with a pressure controller in the PTR-TOF-MS sampling inlet. Flow rates are controller set points; the calibrated total flow through the reactor, used to calculate the reaction times, was \SI{459}{\sccm}.}
	\label{fig:setup_ozonolysis}
\end{figure}

To demonstrate the reactor under kinetic conditions, we followed the disappearance of 2,3-dimethyl-2-butene (TME) during its ozonolysis:
\begin{equation}
    \ce{C6H12 + O3 ->[k] P} \label{eq:chemical_system}
\end{equation}
where $k$ is the rate coefficient for the TME + \ce{O3} reaction and P denotes the product ensemble. The setup is shown in Figure~\ref{fig:setup_ozonolysis}. Kinetic measurements were performed in the variable-length configuration at $Q_\text{reactor}=\SI{459}{\sccm}$, $T=\SI{293\pm1}{\kelvin}$, and atmospheric pressure, using nine tube lengths ranging from \SIrange{25}{700}{\centi\meter}. The reaction time $\tau_\mathrm{r}$ was calculated from the reaction-tube volume and $Q_\text{reactor}$, with time zero at the reactant-mixing junction.

\subsubsection{TME decay} \label{sec:tme_decay}

Experiments were performed under pseudo-first-order conditions ($[\text{O}_3] \gg [\text{TME}]$). The PTR-MS signal of TME was modeled as:
\begin{equation}
    S_\text{TME}(\tau_\mathrm{r}) = A_\text{TME}\exp(-k^\prime\tau_\mathrm{r}), \label{eq:TME}
\end{equation}
where $k^\prime$ is the observed first-order TME-loss coefficient and $A_\text{TME}$ is the fitted initial signal. We define the apparent bimolecular rate coefficient for TME loss as $k_\mathrm{app}=k^\prime/\ce{[O3]}$. 

\begin{samepage}
Nonlinear least-squares fitting of the full TME decay yielded $k^\prime=\SI{0.373 \pm 0.014}{\per\second}$ (Figure~\ref{fig:Kinetics}). The measured $\ce{[O3]}=\SI{7.1}{\ppm}$ (\SI{1.85e14}{\per\cubic\centi\meter}) gives \mbox{$k_\mathrm{app}=\SI{2.0(5)e-15}{\centi\meter\cubed\per\molecule\per\second}$}. The fit standard error (3.8\%), a conservative 5\% uncertainty assigned to the measured ozone concentration, and the 10.5\% sensitivity to fit weighting and background treatment give a combined relative uncertainty of 12.3\%, expanded to approximately 25\% with a coverage factor of 2. Appendix~\ref{app:ozonolysis} details this uncertainty budget and the sensitivity of $k^\prime$ to excluding short-residence-time data.\par
\end{samepage}

Because no OH scavenger was added, OH formed during ozonolysis can also consume TME. IUPAC reports an OH yield of $\phi_\mathrm{OH}=0.93\pm0.14$ from TME ozonolysis near atmospheric pressure. OH is produced by decomposition of the Criegee intermediate formed in the TME + \ce{O3} reaction \cite{iupacOxVOC41}. The exponential fit (Eq.~\eqref{eq:TME}) describes overall TME loss and does not separate this secondary consumption from the TME + \ce{O3} reaction.

The IUPAC recommendation combines three studies with different experimental observables and treatments of secondary chemistry \cite{iupacOxVOC41}. Huie and Herron \cite{Huie1975} followed ozone decay with TME in excess in a stopped-flow reactor coupled to mass spectrometry. Greene and Atkinson \cite{greeneRateConstantsGasPhase1992} measured TME loss relative to 2-methyl-2-butene with $n$-octane added to scavenge OH, then expressed the rate coefficient relative to propene. Witter et al.\ \cite{witterGasphaseOzonolysisRate2002} used comparable ozone and TME concentrations in a stopped-flow reactor at \SI{500}{\milli\bar}, without an OH scavenger. They fitted the TME + \ce{O3} rate coefficient and the OH yield to measured ozone decay profiles using three chemical reactions and a separately measured ozone wall-loss term.

\begin{samepage}
We therefore fitted the same TME decay data with a reaction network that takes into account OH production, competition for OH, radical recycling, and acetyl-peroxy addition to TME (Appendix~\ref{app:secondary_chemistry}). In the 44-channel model, OH accounts for 42\% of the TME consumed up to the last measurement. The secondary-reaction parameters were fixed from the literature or from explicitly stated assumptions; only $k$ and the signal amplitude were fitted. The IUPAC value of $k$ was used for comparison, not as a constraint on the fit.\par
\end{samepage}

\begingroup\clubpenalty=10000
At \SI{293}{\kelvin}, the reaction-network fit yields $k=\SI{1.18(19)e-15}{\centi\meter\cubed\per\molecule\per\second}$ (approximate expanded uncertainty, coverage factor 2; Appendix~\ref{app:secondary_chemistry}). This value is consistent with the IUPAC recommendation of \SI{1.08e-15}{\centi\meter\cubed\per\molecule\per\second} at the same temperature. The difference of 9\% is smaller than the estimated uncertainty of the recommendation near room temperature ($\Delta\log_{10} k=\pm0.08$ at \SI{298}{\kelvin}, approximately $-17\%/+20\%$) \cite{iupacOxVOC41}. A three-reaction model with no estimated secondary rate coefficients gives $k=\SI{1.09e-15}{\centi\meter\cubed\per\molecule\per\second}$ when fitted to the TME signal. Fitting its logarithm changes $k$ by 7.1\% in the three-reaction model and by 0.5\% in the 44-channel network (Appendix~\ref{app:secondary_chemistry}).\par\endgroup

The reaction-network fit reproduces the TME decay (Figure~\ref{fig:Kinetics}) with residuals comparable to the exponential fit. On a logarithmic scale, it also reproduces the slower decay at the longest reaction times, which a single exponential does not describe: as TME is consumed, a decreasing fraction of OH reacts with TME (Figure~\ref{fig:chemical_fits}). Across the weighting and normalization choices tested, the largest decreases from the nominal fits are 4\% for $k$ and 13\% for $k^\prime$ (Appendix~\ref{app:secondary_chemistry}). The exponential fit gives the apparent coefficient for total TME loss; the reaction network gives the coefficient for TME + \ce{O3}.

\subsubsection{Product formation} \label{sec:product_formation}

Acetone is a well-established major carbonyl product of TME ozonolysis \cite{copelandAlkeneOzoneReactions2011}. The acetone signal was first described by an exponential growth model:
\begin{equation}
    S_\text{acetone}(\tau_\mathrm{r}) = S_{\text{acetone},0}
    + A_\text{acetone}\left[1-\exp(-k^\prime\tau_\mathrm{r})\right]. \label{eq:Acetone}
\end{equation}
The acetone data were fitted with $k^\prime$ fixed to the value obtained from the TME decay.

\begin{figure}[ht] 
	\centering
	\includegraphics[width=\textwidth]{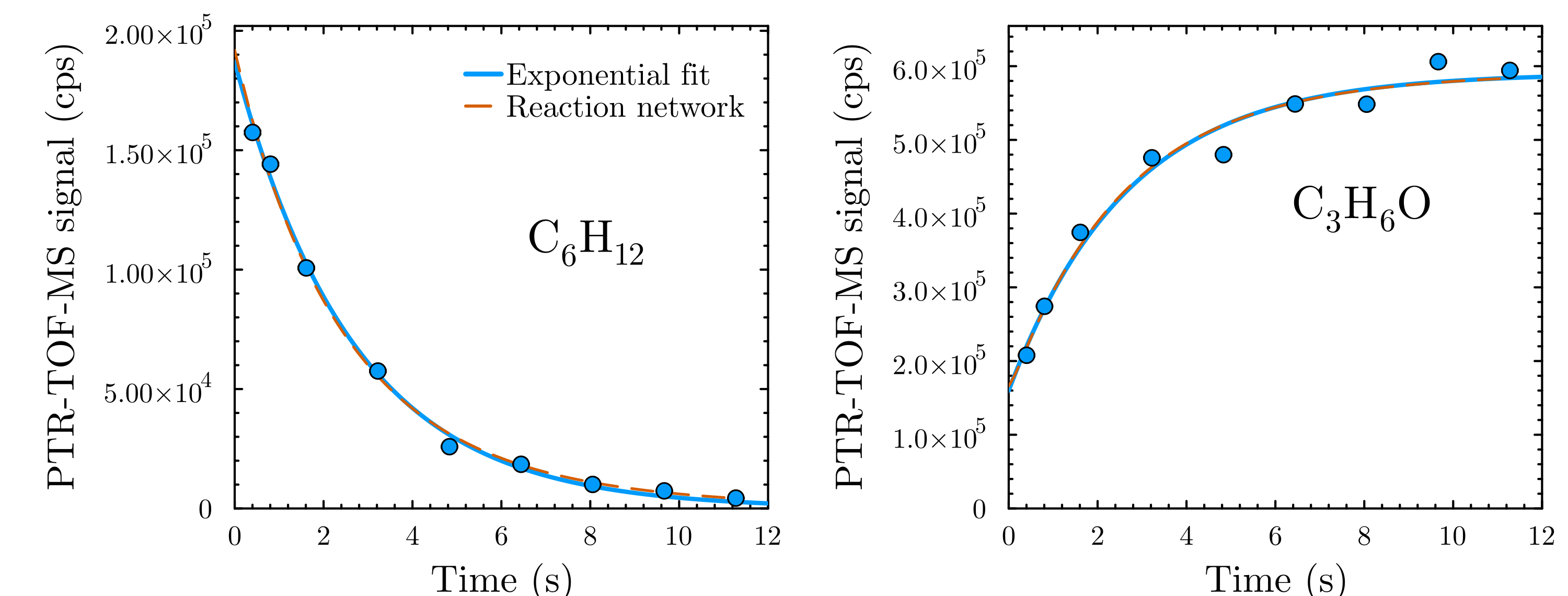}

	\caption{\textbf{Kinetics of the ozonolysis of 2,3-dimethyl-2-butene (TME).} Circles show measured PTR-TOF-MS count rates (cps), without normalization. \textbf{Left panel:} The solid line shows the exponential fit to the TME decay (Eq.~\eqref{eq:TME}), yielding $k^\prime=\SI{0.373 \pm 0.014}{\per\second}$. The dashed line shows the reaction-network fit including OH and peroxy-radical chemistry (Appendix~\ref{app:secondary_chemistry}). \textbf{Right panel:} The solid line shows the exponential fit to the \ce{C3H6O} signal assigned to acetone (Eq.~\eqref{eq:Acetone}), with $k^\prime$ fixed from TME and the signal offset and formation amplitude fitted. The dashed line shows the reaction-network prediction with $k$ and all chemical parameters fixed at the values used for the TME fit; only the signal offset and response factor are adjusted.}
	\label{fig:Kinetics}
\end{figure}

We also compared the acetone signal with the same reaction network, keeping $k$ and the chemical parameters fixed at the values used for the TME fit. Only the signal offset and response factor were adjusted. The calculated growth profile gives residuals comparable to the exponential fit (Figure~\ref{fig:Kinetics}b). Although this agreement supports temporal consistency between reactant loss and product formation, it does not determine the absolute acetone yield or the respective contributions of primary and secondary formation pathways.

We report 35 signals with assigned ion formulas: 29 increased with reaction time, five decreased, and one rose and then decayed (Appendix~\ref{app:ozonolysis}). The transient signal was assigned to protonated \ce{CH3O2}.

\begin{samepage}
Further experimental details and supporting analyses of the kinetic demonstration are provided in Appendix~\ref{app:ozonolysis}.\par
\end{samepage}

\clearpage
\section{Conclusions}

We have introduced the dual-arm flow tube reactor, a versatile flow tube design for gas-phase kinetics at ambient pressure. Assembled entirely from standard components, this setup reduces the complexity and cost of conventional movable-injector flow tubes while providing well-defined reaction times and low wall reactivity.

Residence-time characterization gives a typical temporal resolving power $\mu/\Delta t \approx 10$, while the ozonolysis demonstration yields an apparent rate coefficient for TME loss $k_\mathrm{app} = \num{2.0 \pm 0.5}\times\SI{e-15}{\centi\meter\cubed\per\molecule\per\second}$. A reaction-network fit including secondary OH and peroxy-radical chemistry reproduces the same TME signals, including the slower decay at the longest reaction times, with a freely fitted $k$ that is consistent with the IUPAC recommendation and robust to the weighting and normalization of the data. The reported $k_\mathrm{app}$ describes TME loss under our experimental conditions, rather than a reference determination of the rate coefficient for the TME + \ce{O3} reaction. Coupling with time-of-flight mass spectrometry enables simultaneous monitoring of numerous chemical species in a single steady-state experiment, revealing chemical complexity that can influence the observed kinetics and would be less apparent when only a limited set of targets is monitored.

Combining operational simplicity, low maintenance requirements, and broad adaptability, this reactor design provides a practical tool for atmospheric and physical chemistry, with possible extensions to multiphase studies by filling the reaction tube with a solid or liquid material.

\bibliographystyle{unsrtnat}
\bibliography{main}

@article{copelandAlkeneOzoneReactions2011,
  title = {A Study of the Alkene--Ozone Reactions, 2,3-Dimethyl-2-Butene + O$_3$ and 2-Methylpropene + O$_3$, with Photoelectron Spectroscopy: Measurement of Product Branching Ratios and Atmospheric Implications},
  author = {Copeland, Grant and Ghosh, Mariana V. and Shallcross, Dudley E. and Percival, Carl J. and Dyke, John M.},
  year = {2011},
  journal = {Physical Chemistry Chemical Physics},
  volume = {13},
  number = {39},
  pages = {17461--17472},
  doi = {10.1039/C1CP21922G},
}

@article{wilhelmyUeberGesetzNach1850,
  title = {Ueber Das {{Gesetz}}, Nach Welchem Die {{Einwirkung}} Der {{S{\"a}uren}} Auf Den {{Rohrzucker}} Stattfindet (The Law By Which the Action of Acids on Cane Sugar Occurs)},
  author = {Wilhelmy, Ludwig},
  year = {1850},
  journal = {Annalen der Physik},
  volume = {157},
  number = {12},
  pages = {499--526},
  issn = {1521-3889},
  doi = {10.1002/andp.18501571203},
  langid = {english},
}

@incollection{Ptacek18,
    author = {Petr Ptáček and Tomáš Opravil and František Šoukal},
    title = {A Brief Introduction to the History of Chemical Kinetics},
    booktitle = {Introducing the Effective Mass of Activated Complex and the Discussion on the Wave Function of this Instanton},
    publisher = {IntechOpen},
    address = {Rijeka},
    year = {2018},
    editor = {Petr Ptáček and Tomáš Opravil and František Šoukal},
    chapter = {1},
    doi = {10.5772/intechopen.78704},
    url = {https://doi.org/10.5772/intechopen.78704}
}

@article{osbornMultiplexedChemicalKinetic2008a,
  title = {The Multiplexed Chemical Kinetic Photoionization Mass Spectrometer: A New Approach to Isomer-Resolved Chemical Kinetics},
  shorttitle = {The Multiplexed Chemical Kinetic Photoionization Mass Spectrometer},
  author = {Osborn, David L. and Zou, Peng and Johnsen, Howard and Hayden, Carl C. and Taatjes, Craig A. and Knyazev, Vadim D. and North, Simon W. and Peterka, Darcy S. and Ahmed, Musahid and Leone, Stephen R.},
  year = {2008},
  month = oct,
  journal = {The Review of Scientific Instruments},
  volume = {79},
  number = {10},
  pages = {104103},
  issn = {1089-7623},
  doi = {10.1063/1.3000004},
  langid = {english},
  pmid = {19044733},
}

@article{howardKineticMeasurementsUsing1979b,
  title = {Kinetic Measurements Using Flow Tubes},
  author = {Howard, Carleton J.},
  year = {1979},
  month = jan,
  journal = {The Journal of Physical Chemistry},
  volume = {83},
  number = {1},
  pages = {3--9},
  issn = {0022-3654, 1541-5740},
  doi = {10.1021/j100464a001},
  langid = {english},
}

@article{luoSimultaneousDeterminationTransient2020,
  title = {Simultaneous Determination of Transient Free Radicals and Reaction Kinetics by High-Resolution Time-Resolved Dual-Comb Spectroscopy},
  author = {Luo, Pei-Ling and Horng, Er-Chien},
  year = {2020},
  month = jul,
  journal = {Communications Chemistry},
  volume = {3},
  number = {1},
  pages = {1--8},
  publisher = {Nature Publishing Group},
  issn = {2399-3669},
  doi = {10.1038/s42004-020-00353-6},
  langid = {english},
}

@article{glowackiDesignInitialResults2007a,
  title = {Design of and Initial Results from a {{Highly Instrumented Reactor}} for {{Atmospheric Chemistry}} ({{HIRAC}})},
  author = {Glowacki, D. R. and Goddard, A. and Hemavibool, K. and Malkin, T. L. and Commane, R. and Anderson, F. and Bloss, W. J. and Heard, D. E. and Ingham, T. and Pilling, M. J. and Seakins, P. W.},
  year = {2007},
  month = oct,
  journal = {Atmospheric Chemistry and Physics},
  volume = {7},
  number = {20},
  pages = {5371--5390},
  publisher = {Copernicus GmbH},
  issn = {1680-7316},
  doi = {10.5194/acp-7-5371-2007},
  langid = {english},
}

@article{osseiranDevelopmentValidationThermally2020,
  title = {Development and Validation of a Thermally Regulated Atmospheric Simulation Chamber ({{THALAMOS}}): {{A}} Versatile Tool to Simulate Atmospheric Processes},
  shorttitle = {Development and Validation of a Thermally Regulated Atmospheric Simulation Chamber ({{THALAMOS}})},
  author = {Osseiran, Noureddin and Romanias, Manolis N. and Gaudion, Vincent and Angelaki, Maria E. and Papadimitriou, Vassileios C. and Tomas, Alexandre and Coddeville, Patrice and Thevenet, Frederic},
  year = {2020},
  month = sep,
  journal = {Journal of Environmental Sciences},
  series = {Climate {{Friendly Air Pollution Control}}: {{Sources}}, {{Processes}}, {{Impacts}}, and {{Regulation}}},
  volume = {95},
  pages = {141--154},
  issn = {1001-0742},
  doi = {10.1016/j.jes.2020.03.036},
}

@article{geryContinuousStirredTank1985,
  title = {A Continuous Stirred Tank Reactor Investigation of the Gas-Phase Reaction of Hydroxyl Radicals and Toluene},
  author = {Gery, Michael W. and Fox, Donald L. and Jeffries, Harvey E. and Stockburger, Leonard and Weathers, Walter S.},
  year = {1985},
  month = sep,
  journal = {International Journal of Chemical Kinetics},
  volume = {17},
  number = {9},
  pages = {931--955},
  issn = {0538-8066},
  doi = {10.1002/kin.550170903},
  langid = {english},
}

@article{kleyBlitzlichtPhotolyseanordnungFurSchumannUV1963,
  title = {Blitzlicht-{{Photolyseanordnung}} F{\"u}r Das {{Schumann-UV}} Mit Hoher {{Zeitaufl{\"o}sung}} (Flash photolysis set-up for the Schumann UV with high time resolution)},
  author = {Kley, D. and Stuhl, F. and Welge, K. H.},
  year = {1963},
  month = sep,
  journal = {Zeitschrift f{\"u}r Naturforschung A},
  volume = {18},
  number = {8-9},
  pages = {906--913},
  publisher = {De Gruyter},
  issn = {1865-7109},
  doi = {10.1515/zna-1963-8-905},
  langid = {english},
}

@article{tollefsonReactionAtomicHydrogen1948,
  title = {The {{Reaction}} of {{Atomic Hydrogen}} with {{Acetylene}}},
  author = {Tollefson, E. L. and Le Roy, D. J.},
  year = {1948},
  month = nov,
  journal = {The Journal of Chemical Physics},
  volume = {16},
  number = {11},
  pages = {1057--1062},
  issn = {0021-9606, 1089-7690},
  doi = {10.1063/1.1746724},
  langid = {english},
}

@article{braunFlashPhotolysisMethane1966,
  title = {Flash {{Photolysis}} of {{Methane}} in the {{Vacuum Ultraviolet}}. {{I}}. {{End-Product Analysis}}},
  author = {Braun, W. and Welge, K. H. and McNesby, J. R.},
  year = {1966},
  month = oct,
  journal = {The Journal of Chemical Physics},
  volume = {45},
  number = {7},
  pages = {2650--2656},
  issn = {0021-9606, 1089-7690},
  doi = {10.1063/1.1727985},
  langid = {english},
}

@article{noziereUptakeMethylVinyl2006,
  title = {The {{Uptake}} of {{Methyl Vinyl Ketone}}, {{Methacrolein}}, and 2-{{Methyl-3-butene-2-ol}} onto {{Sulfuric Acid Solutions}}},
  author = {Nozi{\`e}re, Barbara and Voisin, Didier and Longfellow, Cheryl A. and Friedli, Hans and Henry, Bruce E. and Hanson, David R.},
  year = {2006},
  month = feb,
  journal = {The Journal of Physical Chemistry A},
  volume = {110},
  number = {7},
  pages = {2387--2395},
  publisher = {American Chemical Society},
  issn = {1089-5639},
  doi = {10.1021/jp0555899},
}

@article{hansonNH3MassAccommodation2003,
  title = {The {{NH}}{\textsubscript{3}} {{Mass Accommodation Coefficient}} for {{Uptake}} onto {{Sulfuric Acid Solutions}}},
  author = {Hanson, D. and Kosciuch, E.},
  year = {2003},
  month = apr,
  journal = {The Journal of Physical Chemistry A},
  volume = {107},
  number = {13},
  pages = {2199--2208},
  issn = {1089-5639, 1520-5215},
  doi = {10.1021/jp021570j},
}

@article{glowackiDesignInitialResults2007b,
  title = {Design of and Initial Results from a {{Highly Instrumented Reactor}} for {{Atmospheric Chemistry}} ({{HIRAC}})},
  author = {Glowacki, D R and Goddard, A and Hemavibool, K and Malkin, T L and Commane, R and Anderson, F and Bloss, W J and Heard, D E and Ingham, T and Pilling, M J and Seakins, P W},
  year = {2007},
  journal = {Atmos. Chem. Phys.},
  langid = {english},
}

@article{dehaanHeterogeneousChemistryTroposphere1999,
  title = {Heterogeneous Chemistry in the Troposphere: {{Experimental}} Approaches and Applications to the Chemistry of Sea Salt Particles},
  shorttitle = {Heterogeneous Chemistry in the Troposphere},
  author = {De Haan, David O. and Brauers, Theo and Oum, Kawon and Stutz, Jochen and Nordmeyer, Trent and {Finlayson-Pitts}, Barbara J.},
  year = {1999},
  month = jul,
  journal = {International Reviews in Physical Chemistry},
  volume = {18},
  number = {3},
  pages = {343--385},
  publisher = {Taylor \& Francis},
  issn = {0144-235X},
  doi = {10.1080/014423599229910},
}

@article{durifDesignLavalNozzles2022,
  title = {Design of de {{Laval}} Nozzles for Gas-Phase Molecular Studies in Uniform Supersonic Flow},
  author = {Durif, O.},
  year = {2022},
  month = jan,
  journal = {Physics of Fluids},
  volume = {34},
  number = {1},
  pages = {013605},
  issn = {1070-6631},
  doi = {10.1063/5.0060362},
}

@article{dupeyratDesignTestingAxisymmetric1985,
  title = {Design and Testing of Axisymmetric Nozzles for Ion-molecule Reaction Studies between 20\,{$^\circ$}{{K}} and 160\,{$^\circ$}{{K}}},
  author = {Dupeyrat, G. and Marquette, J. B. and Rowe, B. R.},
  year = {1985},
  month = may,
  journal = {The Physics of Fluids},
  volume = {28},
  number = {5},
  pages = {1273--1279},
  issn = {0031-9171},
  doi = {10.1063/1.865010},
}

@article{smithRateMeasurementsReactions1973,
  title = {Rate Measurements of Reactions of {{OH}} by Resonance Absorption. {{Part}} 2.---{{Reactions}} of {{OH}} with {{CO}}, {{C2H4}} and {{C2H2}}},
  author = {Smith, Ian W. M. and Zellner, Reinhard},
  year = {1973},
  month = jan,
  journal = {Journal of the Chemical Society, Faraday Transactions 2: Molecular and Chemical Physics},
  volume = {69},
  number = {0},
  pages = {1617--1627},
  publisher = {The Royal Society of Chemistry},
  issn = {0300-9238},
  doi = {10.1039/F29736901617},
}

@article{roweTechniquesStudyReaction1995,
  title = {Techniques for the Study of Reaction Kinetics at Low Temperatures: Application to the Atmospheric Chemistry of {{Titan}}},
  shorttitle = {Techniques for the Study of Reaction Kinetics at Low Temperatures},
  author = {Rowe, B. R. and Parent, D. C.},
  year = {1995},
  month = jan,
  journal = {Planetary and Space Science},
  series = {Exobiology},
  volume = {43},
  number = {1},
  pages = {105--114},
  issn = {0032-0633},
  doi = {10.1016/0032-0633(94)00094-8},
}

@article{trainorGasPhaseRecombination1973,
  title = {Gas Phase Recombination of Hydrogen and Deuterium Atoms},
  author = {Trainor, Daniel W. and Ham, David O. and Kaufman, Frederick},
  year = {1973},
  month = may,
  journal = {The Journal of Chemical Physics},
  volume = {58},
  number = {10},
  pages = {4599--4609},
  issn = {0021-9606},
  doi = {10.1063/1.1679024},
}

@article{durifStrongUptakeGasPhase2024a,
  title = {Strong {{Uptake}} of {{Gas-Phase Organic Peroxy Radicals}} ({{ROO}}{$\bullet$}) by {{Solid Surfaces Driven}} by {{Redox Reactions}}},
  author = {Durif, Olivier and Piel, Felix and Wisthaler, Armin and Nozi{\`e}re, Barbara},
  year = {2024},
  month = may,
  journal = {JACS Au},
  volume = {4},
  number = {5},
  pages = {1875--1882},
  publisher = {American Chemical Society},
  doi = {10.1021/jacsau.4c00060},
}

@article{demingMeasurementsDelaysGasphase2019a,
  title = {Measurements of Delays of Gas-Phase Compounds in a Wide Variety of Tubing Materials Due to Gas--Wall Interactions},
  author = {Deming, Benjamin L. and Pagonis, Demetrios and Liu, Xiaoxi and Day, Douglas A. and Talukdar, Ranajit and Krechmer, Jordan E. and {de Gouw}, Joost A. and Jimenez, Jose L. and Ziemann, Paul J.},
  year = {2019},
  month = jun,
  journal = {Atmospheric Measurement Techniques},
  volume = {12},
  number = {6},
  pages = {3453--3461},
  publisher = {Copernicus GmbH},
  issn = {1867-1381},
  doi = {10.5194/amt-12-3453-2019},
  langid = {english},
}

@article{kennedyAircraftBasedThree2011,
  title = {An Aircraft-Based Three-Channel Broadband Cavity-Enhanced Absorption Spectrometer for Simultaneous Measurements of {{NO}}$_3$, {{N}}$_2${{O}}$_5$, and {{NO}}$_2$},
  author = {Kennedy, O. J. and Ouyang, B. and Langridge, J. M. and Daniels, M. J. S. and Bauguitte, S. and Freshwater, R. and McLeod, M. W. and Ironmonger, C. and Sendall, J. and Norris, O. and Nightingale, R. and Ball, S. M. and Jones, R. L.},
  year = {2011},
  month = sep,
  journal = {Atmospheric Measurement Techniques},
  volume = {4},
  number = {9},
  pages = {1759--1776},
  issn = {1867-8548},
  doi = {10.5194/amt-4-1759-2011},
}

@article{liAssessmentLongTubing2023,
  title = {Assessment of Long Tubing in Measuring Atmospheric Trace Gases: Applications on Tall Towers},
  author = {Li, Xiao-Bing and Zhang, Chunsheng and Liu, Aiming and Yuan, Bin and Yang, Honglong and Liu, Chanfang and Wang, Sihang and Huangfu, Yibo and Qi, Jipeng and Liu, Zhijie and He, Xianjun and Song, Xin and Chen, Yubin and Peng, Yuwen and Zhang, Xiaoxiao and Zheng, E. and Yang, Lei and Yang, Qing and Qin, Guangzhi and Zhou, Jun and Shao, Min},
  year = {2023},
  journal = {Environmental Science: Atmospheres},
  volume = {3},
  number = {3},
  pages = {506--520},
  issn = {2634-3606},
  doi = {10.1039/D2EA00110A},
}

@article{ayassMixingstructureRelationshipJetstirred2016,
  title = {Mixing-Structure Relationship in Jet-Stirred Reactors},
  author = {Ayass, Wassim W. and Nasir, Ehson F. and Farooq, Aamir and Sarathy, S. Mani},
  year = {2016},
  month = jul,
  journal = {Chemical Engineering Research and Design},
  volume = {111},
  pages = {461--464},
  issn = {0263-8762},
  doi = {10.1016/j.cherd.2016.05.016},
}

@article{bartokMixingJetstirredReactor1960,
  title = {Mixing in a Jet-stirred Reactor},
  author = {Bartok, W. and Heath, C.E. and Weiss, M.A.},
  year = {1960},
  journal = {AIChE Journal},
  volume = {6},
  number = {4},
  pages = {685--687},
  doi = {10.1002/aic.690060433},
}

@article{longwellHighTemperatureReaction1955,
  title = {High {{Temperature Reaction Rates}} in {{Hydrocarbon Combustion}}},
  author = {Longwell, John P. and Weiss, Malcolm A.},
  year = {1955},
  month = aug,
  journal = {Industrial \& Engineering Chemistry},
  volume = {47},
  number = {8},
  pages = {1634--1643},
  publisher = {American Chemical Society},
  issn = {0019-7866},
  doi = {10.1021/ie50548a049},
}

@article{davidsonRecentAdvancesShock2009,
  title = {Recent Advances in Shock Tube/Laser Diagnostic Methods for Improved Chemical Kinetics Measurements},
  author = {Davidson, David F. and Hanson, R. K.},
  year = {2009},
  month = aug,
  journal = {Shock Waves},
  volume = {19},
  number = {4},
  pages = {271--283},
  issn = {1432-2153},
  doi = {10.1007/s00193-009-0203-0},
}

@article{tranterDesignHighpressureSingle2001,
  title = {Design of a High-Pressure Single Pulse Shock Tube for Chemical Kinetic Investigations},
  author = {Tranter, R. S. and Brezinsky, K. and Fulle, D.},
  year = {2001},
  month = jul,
  journal = {Review of Scientific Instruments},
  volume = {72},
  number = {7},
  pages = {3046--3054},
  issn = {0034-6748},
  doi = {10.1063/1.1379963},
}

@article{reslerInstrumentStudyRelaxation1955,
  title = {Instrument to {{Study Relaxation Rates}} behind {{Shock Waves}}},
  author = {Resler, Jr., E. L. and Scheibe, M.},
  year = {1955},
  month = sep,
  journal = {The Journal of the Acoustical Society of America},
  volume = {27},
  number = {5},
  pages = {932--938},
  issn = {0001-4966},
  doi = {10.1121/1.1908082},
}

@article{krechmerAlwaysLostNever2020,
  title = {Always {{Lost}} but {{Never Forgotten}}: {{Gas-Phase Wall Losses Are Important}} in {{All Teflon Environmental Chambers}}},
  shorttitle = {Always {{Lost}} but {{Never Forgotten}}},
  author = {Krechmer, Jordan E. and Day, Douglas A. and Jimenez, Jose L.},
  year = {2020},
  month = oct,
  journal = {Environmental Science \& Technology},
  volume = {54},
  number = {20},
  pages = {12890--12897},
  publisher = {American Chemical Society},
  issn = {0013-936X},
  doi = {10.1021/acs.est.0c03381},
}

@article{danckwertsContinuousFlowSystems1953,
  title = {Continuous Flow Systems: {{Distribution}} of Residence Times},
  shorttitle = {Continuous Flow Systems},
  author = {Danckwerts, P. V.},
  year = {1953},
  month = feb,
  journal = {Chemical Engineering Science},
  volume = {2},
  number = {1},
  pages = {1--13},
  issn = {0009-2509},
  doi = {10.1016/0009-2509(53)80001-1},
}

@article{yablonskyNewApproachDiagnostics2009,
  title = {A New Approach to Diagnostics of Ideal and Non-Ideal Flow Patterns: {{I}}. {{The}} Concept of Reactive-Mixing Index ({{REMI}}) Analysis},
  shorttitle = {A New Approach to Diagnostics of Ideal and Non-Ideal Flow Patterns},
  author = {Yablonsky, G. S. and Constales, D. and Marin, G. B.},
  year = {2009},
  month = dec,
  journal = {Chemical Engineering Science},
  volume = {64},
  number = {23},
  pages = {4875--4883},
  issn = {0009-2509},
  doi = {10.1016/j.ces.2009.07.026},
}

@article{danckwertsEffectIncompleteMixing1958,
  title = {The Effect of Incomplete Mixing on Homogeneous Reactions},
  author = {Danckwerts, P. V.},
  year = {1958},
  month = jan,
  journal = {Chemical Engineering Science},
  volume = {8},
  number = {1},
  pages = {93--102},
  issn = {0009-2509},
  doi = {10.1016/0009-2509(58)80040-8},
}

@article{shepsUVAbsorptionProbing2014,
  title = {{{UV}} Absorption Probing of the Conformer-Dependent Reactivity of a {{Criegee}} Intermediate {{CH3CHOO}}},
  author = {Sheps, Leonid and Scully, Ashley M. and Au, Kendrew},
  year = {2014},
  month = nov,
  journal = {Physical Chemistry Chemical Physics},
  volume = {16},
  number = {48},
  pages = {26701--26706},
  publisher = {The Royal Society of Chemistry},
  issn = {1463-9084},
  doi = {10.1039/C4CP04408H},
}

@article{lewisNovelMultiplexAbsorption2018,
  title = {A Novel Multiplex Absorption Spectrometer for Time-Resolved Studies},
  author = {Lewis, Thomas and Heard, Dwayne E. and Blitz, Mark A.},
  year = {2018},
  month = feb,
  journal = {Review of Scientific Instruments},
  volume = {89},
  number = {2},
  pages = {024101},
  issn = {0034-6748},
  doi = {10.1063/1.5006539},
}

@article{nilssonPhotochemicalReactorStudies2009,
  title = {A Photochemical Reactor for Studies of Atmospheric Chemistry},
  author = {Nilsson, E. J. K. and Eskebjerg, C. and Johnson, M. S.},
  year = {2009},
  month = jun,
  journal = {Atmospheric Environment},
  volume = {43},
  number = {18},
  pages = {3029--3033},
  issn = {1352-2310},
  doi = {10.1016/j.atmosenv.2009.02.034},
}

@article{fernholzKineticsReactionOH2024,
  title = {Kinetics of the Reaction of {{OH}} with Methyl Nitrate (223--343 {{K}})},
  author = {Fernholz, Christin and Baumann, Fabienne and Lelieveld, Jos and Crowley, John N.},
  year = {2024},
  month = feb,
  journal = {Physical Chemistry Chemical Physics},
  volume = {26},
  number = {8},
  pages = {6646--6654},
  publisher = {The Royal Society of Chemistry},
  issn = {1463-9084},
  doi = {10.1039/D4CP00054D},
}

@article{romanInvestigationsGasphasePhotolysis2022,
  title = {Investigations into the Gas-Phase Photolysis and {{OH}} Radical Kinetics of Nitrocatechols: Implications of Intramolecular Interactions on Their Atmospheric Behaviour},
  shorttitle = {Investigations into the Gas-Phase Photolysis and {{OH}} Radical Kinetics of Nitrocatechols},
  author = {Roman, Claudiu and Arsene, Cecilia and Bejan, Iustinian Gabriel and Olariu, Romeo Iulian},
  year = {2022},
  month = feb,
  journal = {Atmospheric Chemistry and Physics},
  volume = {22},
  number = {4},
  pages = {2203--2219},
  publisher = {Copernicus GmbH},
  issn = {1680-7316},
  doi = {10.5194/acp-22-2203-2022},
}

@article{debnathInvestigationKineticsMechanistic2023,
  title = {Investigation of Kinetics and Mechanistic Insights of the Reaction of Criegee Intermediate ({{CH2OO}}) with Methyl-Ethyl Ketone ({{MEK}}) under Tropospherically Relevant Conditions},
  author = {Debnath, Amit and Rajakumar, Balla},
  year = {2023},
  month = jan,
  journal = {Chemosphere},
  volume = {312},
  pages = {137217},
  issn = {0045-6535},
  doi = {10.1016/j.chemosphere.2022.137217},
}

@article{xinRateCoefficientsReactions2024,
  title = {Rate Coefficients for the Reactions of {{OH}} Radicals with {{C}}{\textsubscript{3}}--{{C}}{\textsubscript{11}} Alkanes Determined by the Relative-Rate Technique},
  author = {Xin, Yanyan and Liu, Chengtang and Lun, Xiaoxiu and Xie, Shuyang and Liu, Junfeng and Mu, Yujing},
  year = {2024},
  month = oct,
  journal = {Atmospheric Chemistry and Physics},
  volume = {24},
  number = {19},
  pages = {11409--11429},
  publisher = {Copernicus GmbH},
  issn = {1680-7316},
  doi = {10.5194/acp-24-11409-2024},
}

@article{atkinsonKineticsMechanismsGasphase1986,
  title = {Kinetics and Mechanisms of the Gas-Phase Reactions of the Hydroxyl Radical with Organic Compounds under Atmospheric Conditions},
  author = {Atkinson, Roger},
  year = {1986},
  month = feb,
  journal = {Chemical Reviews},
  volume = {86},
  number = {1},
  pages = {69--201},
  publisher = {American Chemical Society},
  issn = {0009-2665},
  doi = {10.1021/cr00071a004},
}

@article{horneRateHabstractionOH1967,
  title = {Rate of {{H-abstraction}} by {{OH}} from {{Hydrocarbons}}},
  author = {Horne, D. G. and Norrish, R. G. W.},
  year = {1967},
  month = sep,
  journal = {Nature},
  volume = {215},
  number = {5108},
  pages = {1373--1374},
  publisher = {Nature Publishing Group},
  issn = {1476-4687},
  doi = {10.1038/2151373a0},
}

@article{greinerHydroxylRadicalKineticsKinetic1967,
  title = {Hydroxyl-{{Radical Kinetics}} by {{Kinetic Spectroscopy}}. {{I}}. {{Reactions}} with {{H2}}, {{CO}}, and {{CH4}} at 300{$^\circ$}{{K}}},
  author = {Greiner, N. R.},
  year = {1967},
  month = apr,
  journal = {The Journal of Chemical Physics},
  volume = {46},
  number = {7},
  pages = {2795--2799},
  issn = {0021-9606},
  doi = {10.1063/1.1841115},
}

@article{gorsePhotochemistryGaseousHydrogen1972,
  title = {Photochemistry of the Gaseous Hydrogen Peroxide-Carbon Monoxide System: {{Rate}} Constants for Hydroxyl Radical Reactions with Hydrogen Peroxide and Isobutane by Competitive Kinetics},
  shorttitle = {Photochemistry of the Gaseous Hydrogen Peroxide-Carbon Monoxide System},
  author = {Gorse, R. A. and Volman, D. H.},
  year = {1972},
  month = jan,
  journal = {Journal of Photochemistry},
  volume = {1},
  number = {1},
  pages = {1--10},
  issn = {0047-2670},
  doi = {10.1016/0047-2670(72)80001-7},
}

@article{simonaitisReactionO1DH2O1973,
  title = {The Reaction of {{O}}({{1D}}) with {{H2O}} and the Reaction of {{OH}} with {{C3H6}}},
  author = {Simonaitis, R. and Heicklen, Julian},
  year = {1973},
  journal = {International Journal of Chemical Kinetics},
  volume = {5},
  number = {2},
  pages = {231--241},
  issn = {1097-4601},
  doi = {10.1002/kin.550050206},
}

@article{doyleGasPhaseKinetic1975,
  title = {Gas Phase Kinetic Study of Relative Rates of Reaction of Selected Aromatic Compounds with Hydroxyl Radicals in an Environmental Chamber},
  author = {Doyle, George J. and Lloyd, Alan C. and Darnall, K. R. and Winer, Arthur M. and Pitts, James N.},
  year = {1975},
  month = mar,
  journal = {Environmental Science \& Technology},
  volume = {9},
  number = {3},
  pages = {237--241},
  issn = {0013-936X, 1520-5851},
  doi = {10.1021/es60101a002},
}

@incollection{raviChapter2Flow2017,
  title = {Chapter 2 - {{Flow Characteristics}} of {{Reactors}}---{{Flow Modeling}}},
  booktitle = {Coulson and {{Richardson}}'s {{Chemical Engineering}} ({{Fourth Edition}})},
  author = {Ravi, Ramamurthy},
  editor = {Ravi, R. and Vinu, R. and Gummadi, S. N.},
  year = {2017},
  month = jan,
  pages = {103--160},
  publisher = {Butterworth-Heinemann},
  doi = {10.1016/B978-0-08-101096-9.00002-9},
}

@article{howardLaserMagneticResonance1974a,
  title = {Laser Magnetic Resonance Study of the Gas Phase Reactions of {{OH}} with {{CO}}, {{NO}}, and {{NO2}}},
  author = {Howard, Carleton J. and Evenson, K. M.},
  year = {1974},
  month = sep,
  journal = {The Journal of Chemical Physics},
  volume = {61},
  number = {5},
  pages = {1943--1952},
  issn = {0021-9606, 1089-7690},
  doi = {10.1063/1.1682195},
}

@article{westenbergAtomMoleculeKinetics1967,
  title = {Atom---{{Molecule Kinetics}} at {{High Temperature Using ESR Detection}}. {{Technique}} and {{Results}} for {{O}} +{{H2}}, {{O}} +{{CH4}}, and {{O}} +{{C2H6}}},
  author = {Westenberg, A. A. and {de Haas}, N.},
  year = {1967},
  month = jan,
  journal = {The Journal of Chemical Physics},
  volume = {46},
  number = {2},
  pages = {490--501},
  issn = {0021-9606},
  doi = {10.1063/1.1840694},
}

@article{andersonGasPhaseRecombination1974,
  title = {Gas Phase Recombination of {{OH}} with {{NO}} and {{NO2}}},
  author = {Anderson, J. G. and Margitan, J. J. and Kaufman, F.},
  year = {1974},
  month = apr,
  journal = {The Journal of Chemical Physics},
  volume = {60},
  number = {8},
  pages = {3310--3317},
  issn = {0021-9606},
  doi = {10.1063/1.1681522},
}

@article{atkinsonEvaluatedKineticPhotochemical2004,
  title = {Evaluated Kinetic and Photochemical Data for Atmospheric Chemistry: {{Volume I}} - Gas Phase Reactions of {{O}}{\textsubscript{x}}, {{HO}}{\textsubscript{x}}, {{NO}}{\textsubscript{x}} and {{SO}}{\textsubscript{x}} Species},
  shorttitle = {Evaluated Kinetic and Photochemical Data for Atmospheric Chemistry},
  author = {Atkinson, R. and Baulch, D. L. and Cox, R. A. and Crowley, J. N. and Hampson, R. F. and Hynes, R. G. and Jenkin, M. E. and Rossi, M. J. and Troe, J.},
  year = {2004},
  month = sep,
  journal = {Atmospheric Chemistry and Physics},
  volume = {4},
  number = {6},
  pages = {1461--1738},
  publisher = {Copernicus GmbH},
  issn = {1680-7316},
  doi = {10.5194/acp-4-1461-2004},
}

@article{atkinsonEvaluatedKineticPhotochemical2006c,
  title = {Evaluated Kinetic and Photochemical Data for Atmospheric Chemistry: {{Volume II}} \&ndash; Gas Phase Reactions of Organic Species},
  shorttitle = {Evaluated Kinetic and Photochemical Data for Atmospheric Chemistry},
  author = {Atkinson, R. and Baulch, D. L. and Cox, R. A. and Crowley, J. N. and Hampson, R. F. and Hynes, R. G. and Jenkin, M. E. and Rossi, M. J. and Troe, J. and IUPAC Subcommittee},
  year = {2006},
  month = sep,
  journal = {Atmospheric Chemistry and Physics},
  volume = {6},
  number = {11},
  pages = {3625--4055},
  publisher = {Copernicus GmbH},
  issn = {1680-7316},
  doi = {10.5194/acp-6-3625-2006},
}

@article{onelIntercomparisonHO2Measurements2017,
  title = {An Intercomparison of {{HO}}{\textsubscript{2}} Measurements by Fluorescence Assay by Gas Expansion and Cavity Ring-down Spectroscopy within {{HIRAC}} ({{Highly Instrumented Reactor}} for {{Atmospheric Chemistry}})},
  author = {Onel, Lavinia and Brennan, Alexander and Gianella, Michele and Ronnie, Grace and Lawry Aguila, Ana and Hancock, Gus and Whalley, Lisa and Seakins, Paul W. and Ritchie, Grant A. D. and Heard, Dwayne E.},
  year = {2017},
  month = dec,
  journal = {Atmospheric Measurement Techniques},
  volume = {10},
  number = {12},
  pages = {4877--4894},
  issn = {1867-8548},
  doi = {10.5194/amt-10-4877-2017},
}

@article{wangDesignCharacterizationSmog2014,
  title = {Design and Characterization of a Smog Chamber for Studying Gas-Phase Chemical Mechanisms and Aerosol Formation},
  author = {Wang, X. and Liu, T. and Bernard, F. and Ding, X. and Wen, S. and Zhang, Y. and Zhang, Z. and He, Q. and L{\"u}, S. and Chen, J. and Saunders, S. and Yu, J.},
  year = {2014},
  month = jan,
  journal = {Atmospheric Measurement Techniques},
  volume = {7},
  number = {1},
  pages = {301--313},
  publisher = {Copernicus GmbH},
  issn = {1867-1381},
  doi = {10.5194/amt-7-301-2014},
}

@article{friedVaporPressuresDensities1971,
  title = {Vapor Pressures and Densities of 2,3-Dimethyl-2-Butene and 3,3-Dimethyl-1-Butene},
  author = {Fried, {\relax Vojtech}. and Baghdoyan, {\relax Armen}. and Malik, {\relax Jana}.},
  year = {1971},
  month = jan,
  journal = {Journal of Chemical \& Engineering Data},
  volume = {16},
  number = {1},
  pages = {96--97},
  publisher = {American Chemical Society},
  issn = {0021-9568},
  doi = {10.1021/je60048a009},
}

@article{CatalystPLOSCompBio2023,
 doi = {10.1371/journal.pcbi.1011530},
 author = {Loman, Torkel E. AND Ma, Yingbo AND Ilin, Vasily AND Gowda, Shashi AND Korsbo, Niklas AND Yewale, Nikhil AND Rackauckas, Chris AND Isaacson, Samuel A.},
 journal = {PLOS Computational Biology},
 publisher = {Public Library of Science},
 title = {Catalyst: Fast and flexible modeling of reaction networks},
 year = {2023},
 month = {10},
 volume = {19},
 url = {https://doi.org/10.1371/journal.pcbi.1011530},
 pages = {1-19},
 number = {10},
}

@article{Huie1975,
  author = {Huie, R. E. and Herron, J. T.},
  year = {1975},
  title = {Temperature dependence of the rate constants for reactions of ozone with some olefins},
  journal = {International Journal of Chemical Kinetics},
  volume = {7},
  number = {S1},
  pages = {165--181},
  url = {https://www.osti.gov/biblio/7291300}
}

@article{greeneRateConstantsGasPhase1992,
  author = {Greene, Chipper R. and Atkinson, Roger},
  title = {Rate constants for the gas-phase reactions of {O$_3$} with a series of alkenes at {296 $\pm$ 2 K}},
  journal = {International Journal of Chemical Kinetics},
  year = {1992},
  volume = {24},
  number = {9},
  pages = {803--811},
  doi = {10.1002/kin.550240905}
}

@article{witterGasphaseOzonolysisRate2002,
  title = {Gas-Phase Ozonolysis: {{Rate}} Coefficients for a Series of Terpenes and Rate Coefficients and {{OH}} Yields for 2-Methyl-2-Butene and 2,3-Dimethyl-2-Butene},
  shorttitle = {Gas-Phase Ozonolysis},
  author = {Witter, M. and Berndt, T. and B{\"o}ge, O. and Stratmann, F. and Heintzenberg, J.},
  year = {2002},
  journal = {International Journal of Chemical Kinetics},
  volume = {34},
  number = {6},
  pages = {394--403},
  issn = {1097-4601},
  doi = {10.1002/kin.10063},
}

@article{hagenUeberBewegungWassers1839,
  title = {Ueber die Bewegung des Wassers in engen cylindrischen R{\"o}hren},
  author = {Hagen, G. H. L.},
  year = {1839},
  journal = {Annalen der Physik},
  volume = {122},
  number = {3},
  pages = {423--442},
  doi = {10.1002/andp.18391220304},
}

@book{weisbachExperimentalHydraulik1855,
  title = {Die Experimental-Hydraulik: Eine Anleitung zur Ausf{\"u}hrung Hydraulischer Versuche im Kleinen},
  author = {Weisbach, Julius},
  year = {1855},
  publisher = {Engelhardt},
  address = {Freiberg},
}

@article{holzingerValidityLimitationsSimple2019,
  title = {Validity and limitations of simple reaction kinetics to calculate concentrations of organic compounds from ion counts in {{PTR-MS}}},
  author = {Holzinger, R. and Acton, W. J. F. and Bloss, W. J. and Breitenlechner, M. and Crilley, L. R. and Dusanter, S. and Gonin, M. and Gros, V. and Keutsch, F. N. and Kiendler-Scharr, A. and Kramer, L. J. and Krechmer, J. E. and Languille, B. and Locoge, N. and Lopez-Hilfiker, F. and Materi\'c, D. and Moreno, S. and Nemitz, E. and Qu\'el\'ever, L. L. J. and Sarda Esteve, R. and Sauvage, S. and Schallhart, S. and Sommariva, R. and Tillmann, R. and Wedel, S. and Worton, D. R. and Xu, K. and Zaytsev, A.},
  year = {2019},
  journal = {Atmospheric Measurement Techniques},
  volume = {12},
  number = {11},
  pages = {6193--6208},
  doi = {10.5194/amt-12-6193-2019},
  url = {https://amt.copernicus.org/articles/12/6193/2019/},
}

@misc{durifFlowTube2026,
  title = {{Flow Tube}: Precursor Dilution and {UV} Photolysis Calculator},
  author = {Durif, O.},
  year = {2026},
  note = {Version 1.0.0},
  url = {https://gitlab.com/odurif/flowtube},
}

@misc{durifMassSpecJuliaPackage2025,
  title = {MassSpec.jl: A Julia Package for Mass Spectrometry Data Processing},
  author = {Durif, O.},
  year = {2025},
  url = {https://gitlab.com/odurif/MassSpec.jl},
}

@article{orkinOHButenes2023,
  title = {Experimental kinetic study of the reactions between {OH} radicals and three 2-butenes over the temperature range 220--370 {K} and pressure range 0.67--40 {kPa} (5--300 {Torr})},
  author = {Orkin, Vladimir L. and Khamaganov, Victor G. and Kurylo, Michael J.},
  journal = {International Journal of Chemical Kinetics},
  year = {2023},
  volume = {55},
  number = {5},
  pages = {221--237},
  doi = {10.1002/kin.21630},
}

@article{nozierePeroxyAlkenes2021,
  title = {Reactions of organic peroxy radicals, {RO$_2$}, with substituted and biogenic alkenes at room temperature: unsuspected sinks for some {RO$_2$} in the atmosphere?},
  author = {Nozi{\`e}re, Barbara and Fache, Fabienne},
  journal = {Chemical Science},
  year = {2021},
  volume = {12},
  number = {35},
  pages = {11676--11683},
  doi = {10.1039/D1SC02263F},
}

@article{frandsenTmeOzonolysis2025,
  title = {Detailed Investigation of 2,3-Dimethyl-2-butene Ozonolysis-Derived Hydroxyl, Peroxy, and Alkoxy Radical Chemistry},
  author = {Frandsen, Benjamin N. and Franzon, Lauri and Meder, Melissa and Pasik, Dominika and Ahongshangbam, Emelda and Vinkvist, Netta and Myllys, Nanna and Iyer, Siddharth and Rissanen, Matti P. and Ehn, Mikael and Kurt{\'e}n, Theo C.},
  journal = {ACS Earth and Space Chemistry},
  year = {2025},
  volume = {9},
  number = {6},
  pages = {1322--1337},
  doi = {10.1021/acsearthspacechem.4c00355},
}

@article{jenkinPeroxy2019,
  title = {Estimation of rate coefficients and branching ratios for reactions of organic peroxy radicals for use in automated mechanism construction},
  author = {Jenkin, Michael E. and Valorso, Richard and Aumont, Bernard and Rickard, Andrew R.},
  journal = {Atmospheric Chemistry and Physics},
  year = {2019},
  volume = {19},
  number = {11},
  pages = {7691--7717},
  doi = {10.5194/acp-19-7691-2019},
}

@misc{iupacSecondaryKinetics,
  title = {Evaluated gas-phase kinetic data for atmospheric chemistry},
  author = {{IUPAC Task Group on Atmospheric Chemical Kinetic Data Evaluation}},
  year = {n.d.},
  url = {https://iupac-aeris.ipsl.fr/},
  note = {Individual data sheets identified in the accompanying code archive; accessed 24 September 2026},
}

@misc{iupacOxVOC41,
  title = {{Data sheet Ox\_VOC41: O$_3$ + 2,3-dimethylbut-2-ene}},
  author = {{IUPAC Task Group on Atmospheric Chemical Kinetic Data Evaluation}},
  year = {2020},
  url = {https://iupac-aeris.ipsl.fr/datasheets/pdf/Ox_VOC41.pdf},
}

@article{taylorDispersionSolubleSolute1953,
  title = {Dispersion of Soluble Matter in Solvent Flowing Slowly Through a Tube},
  author = {Taylor, G.},
  journal = {Proceedings of the Royal Society A},
  volume = {219},
  pages = {186--203},
  year = {1953},
  doi = {10.1098/rspa.1953.0139},
}

@article{arisDispersionSoluteFluid1956,
  title = {On the Dispersion of a Solute in a Fluid Flowing Through a Tube},
  author = {Aris, R.},
  journal = {Proceedings of the Royal Society A},
  volume = {235},
  pages = {67--77},
  year = {1956},
  doi = {10.1098/rspa.1956.0065},
}

@article{nagyUncertaintyArrheniusParameters2011,
  title = {Uncertainty of {{Arrhenius}} parameters},
  author = {Nagy, T. and Tur{\'a}nyi, T.},
  journal = {International Journal of Chemical Kinetics},
  year = {2011},
  volume = {43},
  number = {7},
  pages = {359--378},
  doi = {10.1002/kin.20551}
}

@article{hopleyResidenceTimeDistribution2019,
  title = {Residence time distribution in coil and plate micro-reactors},
  author = {Hopley, A. and Doyle, B. J. and Roberge, D. M. and Macchi, A.},
  journal = {Chemical Engineering Science},
  year = {2019},
  volume = {207},
  pages = {181--193},
  doi = {10.1016/j.ces.2019.06.016}
}

\clearpage



\clearpage

\appendix
\renewcommand{\thefigure}{\thesection\arabic{figure}}
\renewcommand{\thetable}{\thesection\arabic{table}}
\renewcommand{\theequation}{\thesection\arabic{equation}}
\renewcommand{\theHfigure}{\thesection\arabic{figure}}
\renewcommand{\theHtable}{\thesection\arabic{table}}
\renewcommand{\theHequation}{\thesection\arabic{equation}}

\section{Flow tube discussion} \label{app:flow_tube_discussion}
\setcounter{figure}{0}
\setcounter{table}{0}
\setcounter{equation}{0}

In our setup, the reactor operates near atmospheric pressure (approximately \SI{1}{\bar}) while the detector is under vacuum, so a large pressure drop is imposed at the outlet. The capillary restrictor is about \SI{5}{\centi\meter} long with an internal radius of about \SI{65}{\micro\meter}; for $\Delta P \approx \SI{1}{\bar}$, the sampled flow is typically on the order of \SIrange{50}{100}{\sccm}. Under these conditions, frictional losses in the restrictor dominate over singular losses.

The sampled flow through the restrictor can be measured directly with a flow meter. When no flow meter is available, it can instead be inferred from the flow-controller readings by closing the control valve (see Figure~\ref{fig:flowtube}a) and adjusting the total inlet flow until the reactor pressure reaches atmospheric pressure.

When an exhaust is used, the flow through the restrictor depends only on its geometry and on the imposed pressure differential ($\Delta P$), not on the upstream inlet flow. This decouples the operating pressure from the inlet flows of gases A and B, which is a practical advantage. The exhaust can be placed anywhere between the reactant mixing point and the detector, but its position changes the flow rate in the reaction section and therefore the residence time.

For atmospheric-pressure studies, connecting the exhaust to a ventilation line at room pressure is often convenient. This provides flexibility in both the inlet and sampled flow rates, provided that $Q_A + Q_B \geq Q_\text{reactor}$; otherwise, backflow would occur through the exhaust line.

An exhaust placed close to the detector favors shorter residence times, whereas an exhaust located closer to the mixing point reduces the flow through the reaction section and therefore favors longer residence times.

To ensure stable operating conditions and verify that no significant pressure drop develops along the tube, the reactor pressure can be monitored with a standard pressure gauge, as in other flow tube experiments. If dead volume associated with the gauge connection is a concern, the gauge can be placed upstream of the flow tube or on the exhaust arm.

Whenever possible, we recommend adjusting the reaction time by changing the reactor length rather than the reactor flow. In routine operation, this can be done without interrupting the gas flows or modifying the restrictor. This strategy provides a wider accessible time range while maintaining very stable pressure and flow conditions. In specific situations, however, such as experiments involving toxic gases or reactor pressures below room pressure, the user may prefer not to disconnect the reactor, thereby avoiding exposure or backflow. In such cases, the alternative variable-flow configuration may be preferable.

\section{Pseudo-first-order method} \label{app:pseudo_first_order}
\setcounter{figure}{0}
\setcounter{table}{0}
\setcounter{equation}{0}

Consider the system:
\begin{align}
    \ce{A + B & ->[k] P}, \\
    \ce{B & ->[k_{loss}] \text{loss}}.
\end{align}

The differential equation governing the temporal evolution of B is:
\begin{equation}
    \frac{d \ce{[B]}}{dt} = - k \ce{[A]} \ce{[B]} - k_\text{loss} \ce{[B]}.
\end{equation}

When \ce{[A](t)} = \ce{[A]_0} is constant, defining $k_0 = k \ce{[A]_0}$ gives
\begin{equation}
    \frac{d \ce{[B]}}{dt} = - (k_0 + k_\text{loss}) \ce{[B]} \Rightarrow \ce{[B](t)\vert_{[A]_0}} = \ce{[B]_0} e^{- (k_0 + k_\text{loss}) \, t}.
\end{equation}

Similarly, when \ce{[A](t)} = \ce{[A]_1} is constant, defining $k_1 = k \ce{[A]_1}$ gives
\begin{equation}
    \frac{d \ce{[B]}}{dt} = - (k_1 + k_\text{loss}) \ce{[B]} \Rightarrow \ce{[B](t)\vert_{[A]_1}} = \ce{[B]_0} e^{- (k_1 + k_\text{loss}) \, t}.
\end{equation}

Experimentally, the measured quantities are $k_0^\prime = k_0 + k_\text{loss}$ and $k_1^\prime = k_1 + k_\text{loss}$.

Two otherwise identical measurements are required, differing only in the concentration of A: one at \ce{[A]_0} (which may be zero in a reference experiment) and one at \ce{[A]_1}. Provided that $k_\text{loss}$, including gas renewal, wall losses, and any other background first-order losses, remains unchanged between the paired measurements, subtracting the two measured pseudo-first-order rate constants and solving for $k$ gives
\begin{equation}
    k_1^\prime-k_0^\prime=\left(k\ce{[A]_1}+k_\text{loss}\right)-\left(k\ce{[A]_0}+k_\text{loss}\right)=k\left(\ce{[A]_1}-\ce{[A]_0}\right) \quad \Rightarrow \quad k=\frac{k_1^\prime-k_0^\prime}{\ce{[A]_1}-\ce{[A]_0}}.
\end{equation}

This procedure also avoids the need to know the absolute concentration of B because only the fitted decay rates are used. The assumption that \ce{[A]_0} and \ce{[A]_1} remain constant must be well satisfied.

\section{Capillary restrictor pressure drop} \label{app:capillary_pressure_drop}
\setcounter{figure}{0}
\setcounter{table}{0}
\setcounter{equation}{0}

To compare the measured capillary performance with a simple hydraulic model, the pressure drop is calculated using an incompressible laminar-flow approximation. The total pressure drop is written as the sum of frictional and local contraction terms:
\begin{equation}
    \Delta P = \Delta P_\text{friction} + \Delta P_\text{local}. \label{eq:Plosses}
\end{equation}

The frictional term is given by the Darcy--Weisbach equation:
\begin{equation}
    \Delta P_\text{friction} = f_D \frac{L_c}{d_c} \frac{\rho v_c^2}{2}, \label{eq:Pregular}
\end{equation}
where $f_D$ is the Darcy friction factor, $L_c$ is the restrictor length, $d_c=2r_c$ is its internal diameter, and $v_c$ is the gas velocity within the restrictor.
For a circular pipe in laminar flow \cite{hagenUeberBewegungWassers1839},
\begin{equation}
    f_D=\frac{64}{\mathrm{Re}}, \label{eq:f}
\end{equation}
where $\mathrm{Re}$ is the Reynolds number.
This value is given by
\begin{equation}
    \mathrm{Re}=\frac{\rho v_c 2 r_c}{\mu}, \label{eq:Reynolds}
\end{equation}
where $\rho$ is the density and $\mu$ is the dynamic viscosity of the gas.

The velocity $v_c$ can be expressed in terms of the sampling flow $Q_0$ and the restrictor radius $r_c$ as
\begin{equation}
    v_c = \frac{Q_0}{\pi r_c^2}, \label{eq:v0}
\end{equation}

Substituting Eq.~\eqref{eq:v0} into Eq.~\eqref{eq:Reynolds} and then into Eq.~\eqref{eq:f} gives
\begin{equation}
    f_D = \frac{32 \pi r_c \mu}{\rho Q_0}. \label{eq:f2}
\end{equation}

Substituting Eqs.~\eqref{eq:v0} and \eqref{eq:f2} into Eq.~\eqref{eq:Pregular} gives
\begin{equation}
    \Delta P_\text{friction} = \frac{8 \mu L_c Q_0}{\pi r_c^4}. \label{eq:regular}
\end{equation}

The local contraction term is given by the loss coefficient method \cite{weisbachExperimentalHydraulik1855}:
\begin{equation}
    \Delta P_\text{local} = \kappa \frac{\rho v_c^2}{2},
\end{equation}
where $\kappa$ is the local loss coefficient; a representative value $\kappa=0.5$ is used for the sudden contraction.
This term can be rewritten as a function of $Q_0$:
\begin{equation}
    \Delta P_\text{local} = \frac{\kappa \rho}{2} \frac{Q_0^2}{\pi^2 r_c^4}. \label{eq:singular}
\end{equation}

Finally, combining Eqs.~\eqref{eq:regular} and \eqref{eq:singular} in Eq.~\eqref{eq:Plosses} gives
\begin{equation}
    \Delta P = \frac{8 \mu L_c Q_0}{\pi r_c^4}  + \frac{\kappa \rho Q_0^2}{2 \pi^2 r_c^4}.
\end{equation}

For a representative configuration, a restrictor with $r_c=\SI{65}{\micro\meter}$ and $L_c=\SI{5}{\centi\meter}$ gave a measured sampling flow $Q_{0,\text{meas}}\approx\SI{50}{\sccm}$ at $\Delta P_\text{meas}\approx\SI{1}{\bar}$. Using $\mu=\SI{1.81e-5}{\pascal\second}$ and $\rho=\SI{1.20}{\kilo\gram\per\cubic\meter}$ for air at \SI{293}{\kelvin}, together with $\kappa=0.5$, the incompressible model predicts $\Delta P_\text{friction}\approx\SI{1.08}{\bar}$ and $\Delta P_\text{local}\approx\SI{0.012}{\bar}$. The resulting total pressure drop, $\Delta P_\text{model}\approx\SI{1.09}{\bar}$, is close to the measured value. The local contraction term accounts for only approximately \SI{1.1}{\percent} of the modeled pressure drop, and the corresponding Reynolds number, $\mathrm{Re}\approx541$, is consistent with laminar flow. Conversely, setting $\Delta P_\text{model}$ to \SI{1}{\bar} gives $Q_{0,\text{model}}\approx\SI{46}{\sccm}$, approximately \SI{8}{\percent} below the measured flow. This agreement should not be overinterpreted: the model neglects gas expansion along the restrictor, which is significant for a pressure drop to vacuum, and $Q_0$ scales as $r_c^4$, so a deviation of a few micrometers from the nominal radius changes $Q_0$ by tens of percent. $Q_0$ should therefore be measured directly with a flow meter whenever possible.


\section{Residence time distribution characterization} \label{app:rtd_characterization}
\setcounter{figure}{0}
\setcounter{table}{0}
\setcounter{equation}{0}

\subsection{Experimental setup}

The calculated tracer residence time $\tau$ is obtained from Eq.~\eqref{eq:arrival_time}. The timing reference is the opening of the tracer-flow controller, so the calculation includes the tracer inlet line ($\ell_B=\SI{6.0}{\centi\meter}$ at $Q_B=\SI{100}{\sccm}$), the fixed section from the mixing tee to the exhaust ($\ell_\text{mix}=\SI{6.5}{\centi\meter}$ at $Q_A+Q_B=\SI{1.7}{\slm}$), and the post-exhaust flow tube ($\ell_0+L$, with $\ell_0=\SI{7}{\centi\meter}$ and $L=\SIrange{0}{700}{\centi\meter}$) at $Q_\text{reactor}$. The internal tube radius is $r=\SI{1.98}{\milli\meter}$ and the sampling flow is $Q_0=\SI{65}{\sccm}$. The branch-B contribution is approximately \SI{0.44}{\second}, while the tee-to-exhaust contribution is approximately \SI{0.03}{\second}.

Measurements at $L=0$ retain the tracer line $\ell_B$, the mixing section $\ell_\text{mix}$, and the fixed post-exhaust section $\ell_0$.

\subsection{Finite tracer pulse and fixed-length response}

\begin{figure}[ht]
	\centering
	\includegraphics[width=\textwidth]{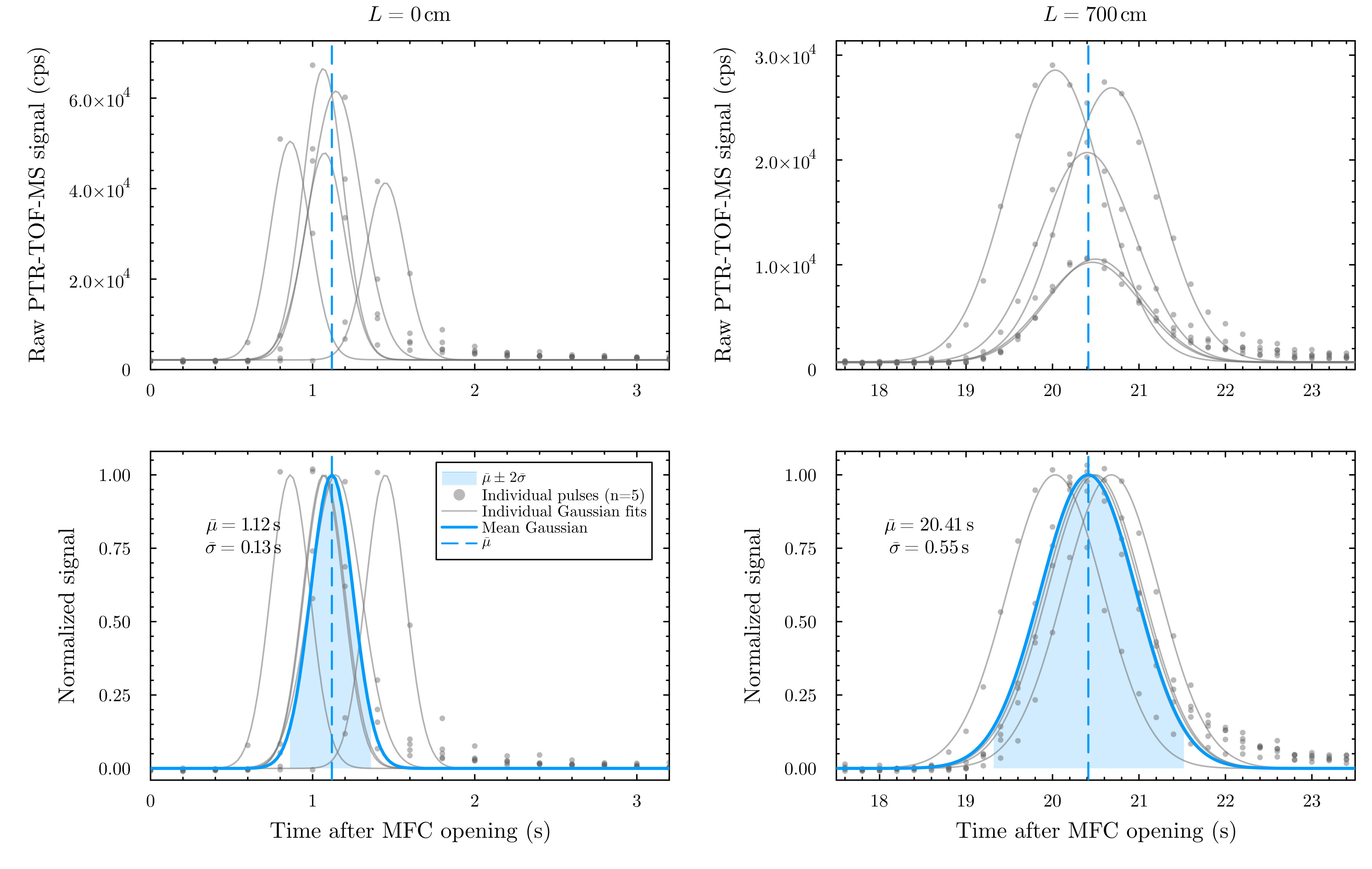}
	\caption{\textbf{Raw and normalized comparison of repeated tracer responses.} Five representative acetonitrile pulses measured at $Q_\text{reactor}=\SI{252}{\sccm}$ with \textbf{(left column)} $L=0$ and \textbf{(right column)} $L=\SI{700}{\centi\meter}$. \textbf{Top row:} Raw PTR-TOF-MS responses (circles) and their individual Gaussian fits including the fitted baselines (gray lines), showing pulse-to-pulse changes in detected signal amplitude. \textbf{Bottom row:} The same responses after subtraction of their fitted baselines and normalization to their fitted peak amplitudes. This normalization removes amplitude differences and allows direct comparison of fitted centers and widths. The finite tracer pulse and transport dispersion both contribute to each fitted width and are not separated. Manual start-time variability shifts the fitted center $\mu$ without broadening an individual recorded profile. Each condition is represented by the mean and standard deviation of five to seven independently fitted pulses, so random pulse-to-pulse variations are averaged and retained in the reported scatter. Solid blue lines are Gaussians constructed from the mean $\mu$ and $\sigma$; dashed lines indicate $\bar{\mu}$, and shaded regions show $\bar{\mu}\pm2\bar{\sigma}$. The time resolution is \SI{0.2}{\second}. All 139 acquired pulse traces are available on Zenodo.}
	\label{fig:RTD_length_comparison}
\end{figure}

The tracer pulses lasted \SIrange{0.5}{2}{\second}. Their finite duration contributes to the width of the measured response together with dispersion in the flow tube. Consequently, the fitted $\sigma$ is treated as an effective response width rather than as a deconvolved intrinsic RTD width. Because the tracer flow rate is set by a mass flow controller, varying the pulse duration changes the total amount of injected tracer---and thus the signal amplitude---more than it changes the response width. At fixed $L$ and $Q_\text{reactor}$, the pulse-to-pulse shifts in $\mu$ arise from the approximately $\pm\SI{0.5}{\second}$ uncertainty in the manually initiated injection. This uncertainty shifts each complete profile without changing its individual width; $\sigma$ is therefore obtained by fitting each pulse separately rather than by averaging unaligned pulses. If greater precision were required for this secondary calibration, automatic synchronization of the tracer-flow-controller trigger with the start of mass-spectral acquisition would improve the reproducibility of the fitted centers to the precision allowed by the acquisition clock. This possible refinement would be specific to the auxiliary mean-time comparison and would not affect the widths obtained from the independently fitted pulses or the conclusions of the dispersion analysis. Figure~\ref{fig:RTD_length_comparison} compares representative $L=0$ and $L=700$~cm measurements at the same reactor flow rate. The $L=0$ measurements show the response with only the fixed sections of the setup. The broader individual responses with the 7~m tube demonstrate the additional contribution of flow-tube dispersion.

\subsection{Symmetric-Gaussian characterization}

Fits were constrained to yield finite values for all four parameters, with the fitted peak time $\mu$ located within the time interval selected for each pulse and the fitted width $\sigma$ positive and no greater than the duration of that interval.

\begin{figure}[ht]
	\centering
	\includegraphics[width=\textwidth]{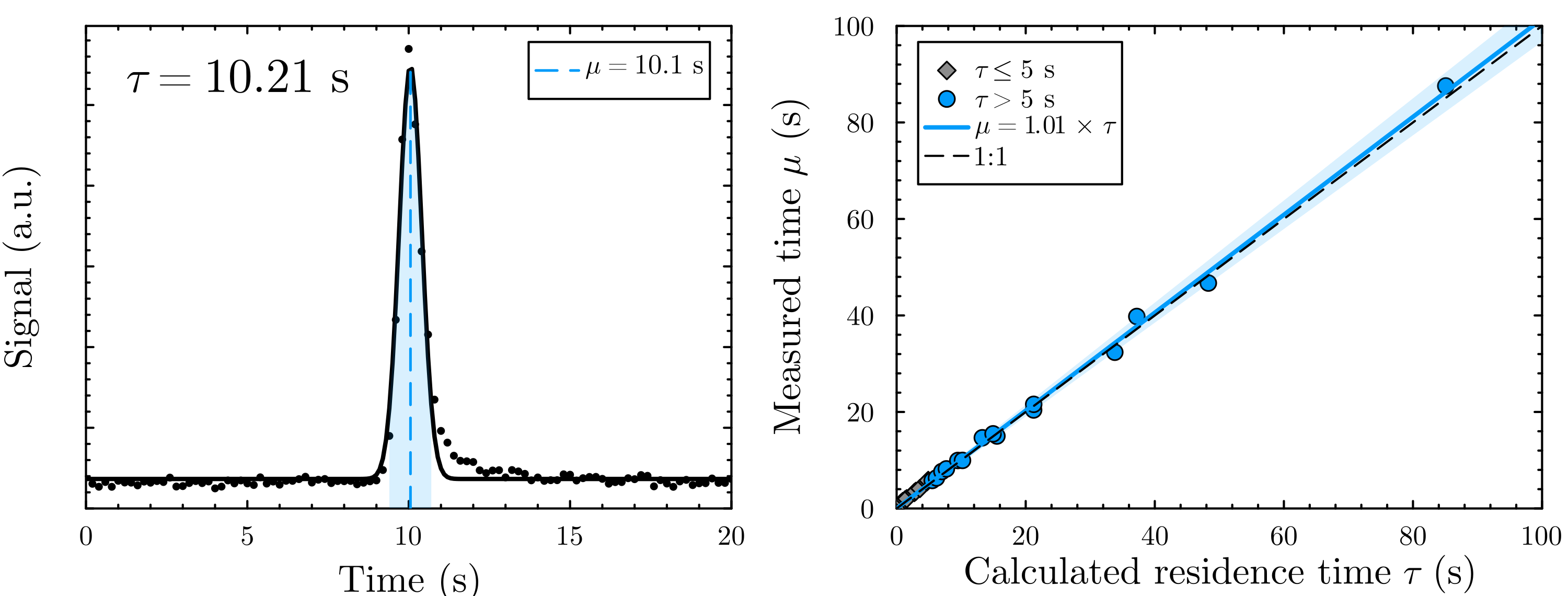}
	\caption{\textbf{Characterization of the measured tracer response using a symmetric Gaussian.} \textbf{Left panel:} Representative single acetonitrile pulse measured with a 7~m tube at approximately \SI{500}{\sccm}. The solid line is the fit from Eq.~\eqref{eq:RTD}, the dashed line indicates $\mu$, and the shaded region shows $\mu\pm2\sigma$. \textbf{Right panel:} Measured mean arrival time $\mu$ versus calculated residence time $\tau$ for different tube lengths and flow rates. Gray diamonds denote $\tau\leq\SI{5}{\second}$ and blue circles denote the conditions included in the proportional fit. The blue line shows $\mu=1.01\,\tau$ for $\tau>\SI{5}{\second}$; including all points also gives $\mu=1.01\,\tau$ (unrounded proportionality coefficients 1.014 and 1.015, respectively). The shaded region represents the fitted $\pm2\sigma$ response width, and the black dashed line is the 1:1 relationship.}
	\label{fig:RTD}
\end{figure}

Figure~\ref{fig:RTD} summarizes two distinct aspects of the symmetric-Gaussian analysis. The primary quantity of interest is the temporal dispersion, quantified by the width $\sigma$ of each independently fitted pulse. Across all tabulated conditions, $\sigma/\mu$ ranges from 2.3 to 12.9\%, corresponding to $\mu/\Delta t=3.3$--18.9. The short-time conditions with $\tau\leq\SI{5}{\second}$ span $\mu/\Delta t=3.3$--8.4, whereas for $\tau>\SI{5}{\second}$, $\sigma/\mu$ remains below 7\% and $\mu/\Delta t$ ranges from 6.3 to 18.9. The comparison of $\mu$ with $\tau$ is a secondary consistency check because the nominal mean time follows directly from the relevant $V/Q$ terms in Eq.~\eqref{eq:arrival_time}. A proportional fit over this range gives $\mu=1.01\,\tau$, corresponding to an average scale difference of approximately 1\%. A linear fit of measured against calculated arrival times over all conditions gives a slope of $1.007\pm0.010$, consistent with unity, and an intercept of \SI{0.31 \pm 0.25}{\second}, comparable to the uncertainty in manual injection timing. Conversely, if the volumes are known, the same relation can be inverted to verify or calibrate the delivered flow rates $Q$. Measurements with $\tau\leq\SI{5}{\second}$ are shown but excluded from this proportional fit because the approximately $\pm\SI{0.5}{\second}$ uncertainty in manual injection timing becomes at least 10\% in relative terms. This exclusion applies only to the comparison of mean times, not to the characterization of the individual response widths; the short-time widths are included in the ranges above, although their normalization by $\mu$ inherits the greater relative timing uncertainty.

\subsection{Asymmetric Gaussian analysis}

In addition to the symmetric Gaussian model used in the main text (Eq.~\eqref{eq:RTD}), the RTD data were also fitted with an asymmetric Gaussian:
\begin{equation}
    {RTD}_\text{asym} = \frac{\alpha}{\sigma \, \sqrt{2 \, \pi}} \times \exp{\left( - \frac{(t - \mu_0)^2}{ 2 \, \sigma^2} \right)} \times \left[ 1 + \erf{ \left( \frac{\beta \, (t - \mu_0)}{\sigma \sqrt{2}} \right) } \right] + \epsilon, \label{eq:RTD_asym}
\end{equation}
where $\beta$ is a skewness parameter and $\mu_0$ is the position parameter, distinct from the symmetric-Gaussian mean $\mu$. Ignoring the constant offset $\epsilon$, the mean of the normalized asymmetric profile is:
\begin{equation}
    \text{mean} = \mu_0 \, + \, \sigma \,  \beta \, \sqrt{\frac{2}{\pi}} \, \dfrac{1}{\sqrt{1 + \beta^2}}, \label{eq:mean}
\end{equation}
where $\sigma$ is the scale parameter and $\beta$ the asymmetry parameter. For comparison with the symmetric-Gaussian width, the standard deviation of the normalized asymmetric profile is $\sigma_\text{asym}=\sigma\sqrt{1-(2/\pi)\beta^2/(1+\beta^2)}$. For $\tau>\SI{5}{\second}$, this equivalent width differed from the symmetric-Gaussian width by about 8\% on average for both tracers. We therefore estimate a typical model-form uncertainty of approximately 10\% in the reported response width. The fitted quantities are reported in Table~\ref{tab:RTD_asym}.

\subsection{Discussion on RTD asymmetry}

The asymmetric model was used to assess possible tailing in the measured response. According to Taylor--Aris dispersion theory \cite{taylorDispersionSolubleSolute1953,arisDispersionSoluteFluid1956}, a tracer response approaches a symmetric Gaussian when the transit time is much longer than the radial diffusion time $\tau_\text{diff}=r^2/D_\mathrm{m}$. With $r=\SI{1.98}{\milli\meter}$ and a representative $D_\mathrm{m}\approx\SI{e-5}{\meter\squared\per\second}$, $\tau_\text{diff}\approx\SI{0.4}{\second}$. Most measurements with $\tau>\SI{5}{\second}$ are therefore in the regime where a nearly symmetric response is expected. At the shortest times, radial diffusion, entrance mixing, and the finite pulse duration occur on comparable timescales.

Typical long-time responses show moderate positive asymmetry. Several contributions may be involved, including the finite tracer pulse, flow-controller response, entrance effects, and flow-tube dispersion. Because these contributions cannot be measured separately, the asymmetry is not assigned to a single physical source. The asymmetric analysis is therefore used as a sensitivity check rather than as the primary response-width characterization.

For the shortest $L=0$ responses, the asymmetric fits approach the high-skewness limit of the model, where the profile becomes only weakly sensitive to further increases in $\beta$. The corresponding large $\beta$ values in Table~\ref{tab:RTD_asym} therefore indicate that the fitted profiles approach this limiting shape. The narrowest symmetric fit has a FWHM of approximately \SI{0.31}{\second}, with a sampling interval of \SI{0.2}{\second}.

\begin{figure}[ht]
	\centering
	\includegraphics[width=\textwidth]{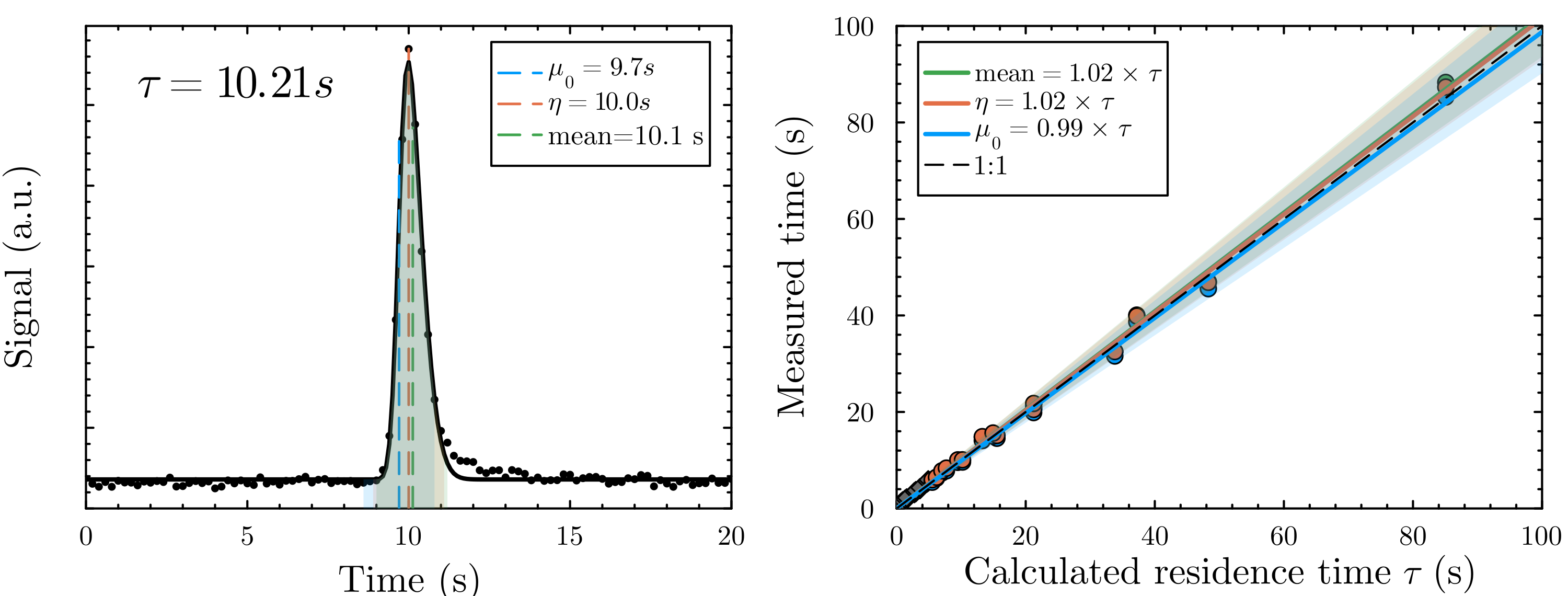}
	\caption{\textbf{Asymmetric-Gaussian analysis of the measured tracer response.} \textbf{Left panel:} Representative response fitted with Eq.~\eqref{eq:RTD_asym}, showing the position parameter $\mu_0$, the peak position $\eta$, and the distribution mean. Shaded regions indicate $\pm2\sigma$ around each characteristic time for visual comparison. \textbf{Right panel:} Values of $\mu_0$, $\eta$, and the asymmetric-distribution mean plotted against $\tau$. Short-time conditions are shown as gray diamonds and are excluded from the proportional fits. The black dashed line indicates the 1:1 relationship.}
	\label{fig:RTD_SI}
\end{figure}

The dataset combines five tube lengths with multiple flow rates at each length. No systematic separation between length-dependent and flow-dependent conditions is visible in the $\mu$--$\tau$ relation within the operating range where the sampling pressure and $Q_0$ remained stable.

\subsection{Validity of the plug flow approximation}

The plug-flow model assumes a narrow distribution around the nominal reaction time. Under the present conditions (tube diameter $d=\SI{3.96}{\milli\meter}$, air at \SI{293}{\kelvin}, and approximately atmospheric pressure), the Reynolds number $\mathrm{Re}=\rho v d/\mu_\mathrm{g}$ ranges from approximately 22 at \SI{62}{\sccm} to 330 at \SI{917}{\sccm}, where $\mu_\mathrm{g}$ is the gas dynamic viscosity. All measurements are therefore well within the laminar regime.

A diffusion-free parabolic velocity profile would produce a broad, non-Gaussian response. In the present millimeter-scale tube, however, radial molecular diffusion acts on a timescale of approximately \SI{0.4}{\second}; at longer arrival times, repeated radial exchange leads toward the Taylor--Aris regime and a nearly Gaussian response. Across the full dataset, $\sigma$ remained below 13\% of $\mu$. The tighter result, $\sigma/\mu<7\%$ for $\tau>\SI{5}{\second}$, supports the use of a plug-flow approximation over this range. At shorter times, the effective response is more strongly influenced by pulse duration and entrance effects, and the kinetic sensitivity analysis in Appendix~\ref{app:ozonolysis} examines how excluding these points changes the fitted rate coefficient.

\begin{table}[ht] 
	\centering
		\caption{\textbf{RTD parameters from symmetric Gaussian fit (Eq.~\eqref{eq:RTD}).}
			The first column, $\tau$, gives the calculated tracer residence time from Eq.~\eqref{eq:arrival_time}. The second column, $L$, gives the variable tube length. The third column, $Q_\text{reactor}$, gives the post-exhaust flow rate. Parameters $\mu$ and $\sigma$ are obtained by fitting Eq.~\eqref{eq:RTD} to the measured tracer response. Each condition includes five to seven valid repeated tracer pulses with durations ranging from \SIrange{0.5}{2}{\second}. Each value is reported as the mean, with the standard deviation in parentheses.}
	\label{tab:RTD_symmetric}
	\tiny
	\begin{tabular}{|c|cc|cccc|}
    \hline
		\multicolumn{1}{|c|}{} & \multicolumn{2}{c|}{} & \multicolumn{2}{c|}{\textbf{Acetone}} & \multicolumn{2}{c|}{\textbf{Acetonitrile}} \\
    \hline
$\tau$ (s)  & $L$ (cm) & $Q_\text{reactor}$ (sccm) & $\mu$ (s) & $\sigma$ (s) & $\mu$ (s) & $\sigma$ (s) \\
		\hline
0.68	& 0	& 252	& 1.12(0.21)	& 0.13(0.01)	& 1.12(0.21)	& 0.13(0.02)\\
0.95	& 0	& 109	& 1.43(0.23)	& 0.17(0.01)	& 1.44(0.23)	& 0.17(0.01)\\
1.31	& 0	& 62	& 1.78(0.42)	& 0.23(0.04)	& 1.79(0.42)	& 0.23(0.04)\\
1.56	& 30	& 252	& 2.01(0.13)	& 0.20(0.02)	& 2.01(0.13)	& 0.19(0.02)\\
2.75	& 100	& 347	& 3.31(0.19)	& 0.25(0.02)	& 3.30(0.19)	& 0.24(0.02)\\
2.97	& 30	& 109	& 3.54(0.14)	& 0.27(0.02)	& 3.55(0.13)	& 0.27(0.02)\\
3.61	& 100	& 252	& 4.14(0.29)	& 0.28(0.02)	& 4.14(0.28)	& 0.28(0.02)\\
4.70	& 300	& 537	& 5.53(0.06)	& 0.29(0.01)	& 5.52(0.06)	& 0.28(0.01)\\
4.90	& 30	& 62	& 5.79(0.14)	& 0.36(0.01)	& 5.81(0.14)	& 0.36(0.02)\\
5.52	& 100	& 157	& 5.83(1.02)	& 0.39(0.04)	& 5.80(1.01)	& 0.34(0.03)\\
6.17	& 700	& 917	& 6.34(0.15)	& 0.27(0.02)	& 6.33(0.15)	& 0.26(0.02)\\
7.01	& 300	& 347	& 7.66(0.06)	& 0.33(0.01)	& 7.66(0.07)	& 0.31(0.02)\\
7.71	& 100	& 109	& 8.19(0.25)	& 0.42(0.06)	& 8.20(0.23)	& 0.40(0.05)\\
9.48	& 300	& 252	& 9.96(0.10)	& 0.43(0.06)	& 9.95(0.11)	& 0.39(0.03)\\
10.21	& 700	& 537	& 9.97(0.15)	& 0.35(0.00)	& 9.96(0.16)	& 0.34(0.01)\\
13.28	& 100	& 62	& 14.63(0.30)	& 0.60(0.05)	& 14.65(0.30)	& 0.58(0.04)\\
14.94	& 300	& 157	& 15.49(0.21)	& 0.56(0.05)	& 15.47(0.21)	& 0.51(0.05)\\
15.54	& 700	& 347	& 15.03(0.21)	& 0.50(0.03)	& 15.02(0.21)	& 0.47(0.02)\\
21.22	& 700	& 252	& 20.45(0.24)	& 0.61(0.02)	& 20.41(0.24)	& 0.55(0.00)\\
21.24	& 300	& 109	& 21.63(0.08)	& 0.69(0.02)	& 21.58(0.09)	& 0.61(0.01)\\
33.80	& 700	& 157	& 32.46(0.16)	& 0.87(0.04)	& 32.38(0.15)	& 0.77(0.02)\\
37.21	& 300	& 62	& 39.91(0.14)	& 1.19(0.08)	& 39.79(0.15)	& 1.07(0.06)\\
48.29	& 700	& 109	& 46.86(0.30)	& 1.27(0.15)	& 46.71(0.30)	& 1.11(0.10)\\
85.08	& 700	& 62	& 88.08(0.29)	& 2.52(0.43)	& 87.47(0.37)	& 1.98(0.14)\\
		\hline
	\end{tabular}
\end{table}

\begin{table}[ht] 
	\centering
		\caption{\textbf{RTD parameters from asymmetric Gaussian fit (Eq.~\eqref{eq:RTD_asym}).}
			The first column, $\tau$, gives the calculated tracer residence time from Eq.~\eqref{eq:arrival_time}. The second column, $L$, gives the variable tube length. The third column, $Q_\text{reactor}$, gives the post-exhaust flow rate. Parameters $\mu_0$, $\sigma$, and $\beta$ are obtained by fitting Eq.~\eqref{eq:RTD_asym} to the measured response. Parameter $\eta$ is the peak position, and the distribution mean is obtained from Eq.~\eqref{eq:mean}. Each condition includes five to seven valid repeated tracer pulses with durations ranging from \SIrange{0.5}{2}{\second}. Each value is reported as the mean, with the standard deviation in parentheses.}
	\label{tab:RTD_asym}
	\tiny
	\resizebox{\textwidth}{!}{%
	\begin{tabular}{|c|cc|ccccc|ccccc|} 
    \hline
		\multicolumn{1}{|c|}{} & \multicolumn{2}{c|}{} & \multicolumn{5}{c|}{\textbf{Acetone}} & \multicolumn{5}{c|}{\textbf{Acetonitrile}} \\ 
		\hline
$\tau$ (s)  & $L$ (cm) & $Q_\text{reactor}$ (sccm) & $\mu_0$ (s) & $\sigma$ (s)  & $\beta$  & $\eta$ (s) & mean (s) & $\mu_0$ (s) & $\sigma$ (s) & $\beta$ & $\eta$ (s) & mean (s) \\
		\hline
0.68	& 0	& 252	& 0.95(0.18)	& 0.26(0.02)	& 249.14(549.93)	& 1.28(0.23)	& 1.15	& 0.94(0.19)	& 0.27(0.03)	& 211.22(463.16)	& 1.28(0.23)	& 1.15\\
0.95	& 0	& 109	& 1.25(0.23)	& 0.29(0.02)	& 39.51(96.68)	& 1.60(0.23)	& 1.48	& 1.24(0.24)	& 0.30(0.04)	& 222.77(375.65)	& 1.60(0.23)	& 1.48\\
1.31	& 0	& 62	& 1.55(0.41)	& 0.37(0.06)	& 2.90(1.33)	& 1.91(0.45)	& 1.83	& 1.56(0.40)	& 0.38(0.06)	& 3.14(1.42)	& 1.94(0.44)	& 1.85\\
1.56	& 30	& 252	& 1.81(0.13)	& 0.32(0.05)	& 2.59(0.72)	& 2.20(0.13)	& 2.05	& 1.81(0.14)	& 0.33(0.07)	& 29.75(65.42)	& 2.20(0.13)	& 2.07\\
2.75	& 100	& 347	& 3.07(0.22)	& 0.39(0.05)	& 2.44(0.60)	& 3.52(0.23)	& 3.35	& 3.07(0.22)	& 0.38(0.06)	& 2.53(0.59)	& 3.48(0.18)	& 3.35\\
2.97	& 30	& 109	& 3.26(0.13)	& 0.45(0.05)	& 2.69(0.42)	& 3.71(0.11)	& 3.60	& 3.27(0.12)	& 0.45(0.05)	& 2.94(0.54)	& 3.71(0.11)	& 3.61\\
3.61	& 100	& 252	& 3.88(0.30)	& 0.44(0.04)	& 2.29(0.67)	& 4.28(0.30)	& 4.20	& 3.88(0.30)	& 0.43(0.05)	& 2.35(0.86)	& 4.32(0.30)	& 4.20\\
4.70	& 300	& 537	& 5.24(0.07)	& 0.46(0.03)	& 2.54(0.42)	& 5.64(0.09)	& 5.59	& 5.24(0.07)	& 0.45(0.03)	& 2.64(0.34)	& 5.64(0.09)	& 5.58\\
4.90	& 30	& 62	& 5.45(0.14)	& 0.56(0.04)	& 2.23(0.41)	& 5.94(0.15)	& 5.85	& 5.44(0.14)	& 0.59(0.06)	& 2.67(0.46)	& 5.94(0.15)	& 5.89\\
5.52	& 100	& 157	& 5.44(1.00)	& 0.62(0.06)	& 2.59(0.23)	& 6.00(1.04)	& 5.90	& 5.44(0.98)	& 0.60(0.08)	& 3.34(0.66)	& 5.96(1.01)	& 5.90\\
6.17	& 700	& 917	& 6.27(0.26)	& 0.31(0.12)	& 0.69(1.44)	& 6.47(0.16)	& 6.41	& 6.26(0.26)	& 0.30(0.13)	& 0.75(1.58)	& 6.47(0.16)	& 6.41\\
7.01	& 300	& 347	& 7.37(0.07)	& 0.47(0.01)	& 1.74(0.12)	& 7.84(0.09)	& 7.70	& 7.38(0.08)	& 0.45(0.02)	& 1.78(0.12)	& 7.80(0.00)	& 7.69\\
7.71	& 100	& 109	& 7.79(0.26)	& 0.65(0.14)	& 2.16(0.65)	& 8.32(0.30)	& 8.26	& 7.81(0.27)	& 0.64(0.14)	& 2.41(0.72)	& 8.36(0.22)	& 8.28\\
9.48	& 300	& 252	& 9.55(0.15)	& 0.66(0.13)	& 2.13(0.45)	& 10.13(0.16)	& 10.03	& 9.57(0.14)	& 0.60(0.09)	& 2.11(0.37)	& 10.07(0.10)	& 10.01\\
10.21	& 700	& 537	& 9.63(0.16)	& 0.54(0.02)	& 2.22(0.20)	& 10.12(0.18)	& 10.03	& 9.63(0.16)	& 0.54(0.02)	& 2.36(0.23)	& 10.08(0.11)	& 10.02\\
13.28	& 100	& 62	& 14.03(0.33)	& 0.95(0.13)	& 2.49(0.64)	& 14.80(0.32)	& 14.74	& 14.07(0.33)	& 0.93(0.09)	& 2.47(0.24)	& 14.88(0.23)	& 14.77\\
14.94	& 300	& 157	& 14.98(0.23)	& 0.83(0.11)	& 1.91(0.29)	& 15.67(0.21)	& 15.57	& 15.00(0.22)	& 0.76(0.11)	& 1.98(0.50)	& 15.63(0.23)	& 15.54\\
15.54	& 700	& 347	& 14.54(0.19)	& 0.78(0.07)	& 2.29(0.27)	& 15.20(0.32)	& 15.11	& 14.56(0.19)	& 0.74(0.06)	& 2.30(0.27)	& 15.12(0.23)	& 15.10\\
21.22	& 700	& 252	& 19.86(0.24)	& 0.95(0.05)	& 2.27(0.26)	& 20.56(0.22)	& 20.56	& 19.88(0.23)	& 0.86(0.02)	& 2.21(0.16)	& 20.52(0.23)	& 20.50\\
21.24	& 300	& 109	& 20.95(0.10)	& 1.09(0.06)	& 2.30(0.16)	& 21.80(0.14)	& 21.75	& 21.00(0.09)	& 0.94(0.04)	& 2.18(0.25)	& 21.76(0.17)	& 21.68\\
33.80	& 700	& 157	& 31.58(0.14)	& 1.42(0.08)	& 2.68(0.28)	& 32.52(0.30)	& 32.64	& 31.60(0.14)	& 1.25(0.06)	& 2.57(0.27)	& 32.52(0.23)	& 32.54\\
37.21	& 300	& 62	& 38.64(0.22)	& 2.11(0.25)	& 3.65(1.42)	& 39.56(0.43)	& 40.26	& 38.67(0.19)	& 1.84(0.15)	& 3.06(0.26)	& 39.88(0.11)	& 40.06\\
48.29	& 700	& 109	& 45.53(0.38)	& 2.20(0.34)	& 3.22(0.55)	& 46.77(0.41)	& 47.20	& 45.55(0.34)	& 1.90(0.27)	& 2.96(0.45)	& 46.87(0.35)	& 46.99\\
85.08	& 700	& 62	& 85.30(0.50)	& 4.76(0.87)	& 4.87(0.74)	& 87.20(0.77)	& 89.02	& 85.29(0.45)	& 3.71(0.41)	& 4.39(1.04)	& 87.28(0.41)	& 88.18\\
		\hline
	\end{tabular}
	}
\end{table}

\clearpage
\section{2,3-Dimethyl-2-butene ozonolysis} \label{app:ozonolysis}
\setcounter{figure}{0}
\setcounter{table}{0}
\setcounter{equation}{0}

\subsection{Chemical kinetic setup}

The setup used for measuring the 2,3-dimethyl-2-butene ozonolysis kinetics is shown in Figure~\ref{fig:setup_ozonolysis}.

Kinetics were measured by varying the reactor length between \SIrange{25}{700}{\centi\meter}. The reactor had an inner diameter of \SI{3.96}{\milli\meter} and the calibrated total flow rate was \SI{459}{\sccm}, corresponding to reaction times ranging from \SI{0.40}{\second} to \SI{11.3}{\second}.

\paragraph{Acquisition and error bars.}
During the ozonolysis measurements, the PTR-TOF-MS signal was recorded at \SI{1}{\hertz}. At each reaction time, the signal plotted in Figure~\ref{fig:Kinetics} is the mean of 51 consecutive spectra collected during one steady-state lamp-on acquisition window, and the error bar is $\pm1$ sample standard deviation across those spectra. The corresponding relative standard deviations are 0.43--2.26\% for TME and 0.20--0.48\% for \ce{C3H6O}. These bars quantify short-term signal variability rather than reproducibility across independent experiments.

\paragraph{Sensitivity to short reaction times.}
The \SI{5}{\second} threshold used in the residence-time characterization applies only to the tracer-pulse comparison of measured mean arrival times and calculated residence times. It is not a lower limit for the continuous kinetic experiment, where reaction time starts at the reactant-mixing tee and no pulse timing is involved. Nevertheless, to test sensitivity to possible short-time mixing effects, we repeated the fit after excluding short-time points. Removing the first point gives $k^\prime=\SI{0.388 \pm 0.015}{\per\second}$ (change $+4.0\%$), and removing the first two points, both below \SI{1}{\second}, gives $k^\prime=\SI{0.369 \pm 0.018}{\per\second}$ (change $-1.1\%$). Excluding data with $\tau_\mathrm{r}<\SI{2}{\second}$ (6 of 9 points remaining) yields $k^\prime=\SI{0.372 \pm 0.037}{\per\second}$, essentially identical to the full-dataset value ($\SI{0.373}{\per\second}$, change below 0.5\%). Excluding points with $\tau_\mathrm{r}<\SI{5}{\second}$ leaves only four low-signal points and gives $k^\prime=\SI{0.307 \pm 0.031}{\per\second}$; the 95\% confidence interval, \SIrange{0.176}{0.439}{\per\second}, includes the full-dataset value. The quoted uncertainties are fit standard errors. For an exponential fit with a free amplitude, a common time shift $\delta$ changes only the amplitude, leaving $k^\prime$ unchanged, since $A\exp[-k^\prime(t+\delta)]=(A\exp[-k^\prime\delta])\exp(-k^\prime t)$.

\paragraph{Assessment of the non-zero \ce{C3H6O} offset.}
The non-zero fitted \ce{C3H6O} offset was substantially larger than the lamp-off background (see Figure~\ref{fig:Kinetics}). After paired lamp-off subtraction and PTR reagent-ion normalization, \ce{C3H6O} formation remained closely correlated with TME loss ($R^2=0.990$), indicating an ozone-dependent prompt contribution rather than a simple instrumental baseline. Including the TME-derived effective delay of \SI{0.54}{\second} reduces the fitted acetone offset from 166 to 66~kcps. The delay is comparable to the tracer residual in Table~\ref{tab:RTD_symmetric}. This effective time is not a measured dead volume. The origin of the remaining offset cannot be resolved from the present data and may involve rapid secondary chemistry or an ozone-induced inlet, surface, or PTR response. The reagent-ion-normalized lamp-off TME signal showed no decrease with reactor length. At each reactor length, the lamp-on TME signal was divided by the lamp-off signal measured at the same length after both signals were normalized to the summed PTR reagent-ion signal. Fitting the logarithm of this ratio against reaction time gave $k^\prime=\SI{0.334 \pm 0.009}{\per\second}$, 10.5\% below the unweighted fit of the raw lamp-on signal. Most of this difference arises from the weighting rather than from the signal treatment: a logarithmic fit of the raw lamp-on signal gives $k^\prime=\SI{0.335}{\per\second}$, whereas reagent-ion normalization alone changes the unweighted fit by $-3.4\%$. A logarithmic fit gives the late, low-signal points the same relative weight as the early points, and the decay slows at the longest reaction times (Appendix~\ref{app:secondary_chemistry}); a single exponential therefore yields a weighting-dependent average rate. This 10.5\% difference defines the observed sensitivity to fit weighting and background treatment and remains within the overall uncertainty; the other combinations of weighting and normalization tested in Appendix~\ref{app:secondary_chemistry} also lie within the expanded uncertainty. The product trace is therefore used to test the modeled temporal evolution, while the rate coefficient derived from TME decay is regarded as an apparent upper bound.

\paragraph{Uncertainty budget.}
The fit standard error of \SI{0.014}{\per\second} corresponds to 3.8\% of $k^\prime$. A conservative relative uncertainty of 5\% was assigned to the measured ozone concentration to account for analyzer calibration and instrumental performance. Combining these terms with the 10.5\% sensitivity to fit weighting and background treatment in quadrature gives a relative standard uncertainty of 12.3\%. Applying a coverage factor of 2 gives an expanded relative uncertainty of approximately 25\%, corresponding to \SI{0.5e-15}{\centi\meter\cubed\per\molecule\per\second}. A common scale error in the calculated reaction times would change $k^\prime$ proportionally; the tracer comparison in Appendix~\ref{app:rtd_characterization} shows no systematic time-scale error beyond the percent level, which is small compared with these components. This budget applies to the measured apparent coefficient. The reaction-network analysis below assesses the chemical interpretation separately: its fitted $k$ values depend on the assumed secondary chemistry, and their conditional fit errors are not a total uncertainty budget for the TME + \ce{O3} rate coefficient.

The initial concentration of 2,3-dimethyl-2-butene (98\%, Sigma-Aldrich) was calculated as $\ce{[C6H12]_0} = \SI{1.96e13}{\per\cubic\centi\meter}$ using a vapor pressure of $V_p = \SI{13240}{\pascal}$~\cite{friedVaporPressuresDensities1971} at $T = \SI{293.15}{\kelvin}$. The upstream \ce{N2} flow was corrected for saturation by $1/(1-V_p/P_\mathrm{b})$, with $P_\mathrm{b}=\SI{1.19}{\bar}$ absolute. These dilution calculations are also implemented in the Flow Tube calculator~\cite{durifFlowTube2026}. The liquid TME sample in the bubbler was kept at ambient laboratory temperature. The room temperature was monitored by the laboratory thermostat and remained at $\SI{293 \pm 1}{\kelvin}$. Ozone was produced with two lamps at \SI{185}{\nano\meter} (SOG-3, 97-0068-02, Analytik Jena), and its concentration was measured to be \SI{7.1}{\ppm} at the reactor inlet with an ozone analyzer (Serinus 10, Acoem), corresponding to \SI{1.85e14}{\per\cubic\centi\meter} at the reactor pressure of \SI{1055}{\hecto\pascal} and \SI{293}{\kelvin}. Ozone was also measured at the reactor outlet (7~m tube, maximum reaction time \SI{11.3}{\second}). No measurable loss of ozone was detected. To verify the validity of the pseudo-first-order approximation, a simple kinetic model corresponding to Eq.~\eqref{eq:chemical_system} was computed using \texttt{Catalyst.jl}~\cite{CatalystPLOSCompBio2023}. The numerical bimolecular solution closely matched the analytical pseudo-first-order solution over the measured time range. This single-reaction calculation checks the effect of ozone depletion, not the contribution of secondary chemistry; within this check, the derived coefficient is not strongly sensitive to the estimated initial TME concentration.

\Needspace{8\baselineskip}
\subsection{Secondary chemistry: TME decay} \label{app:secondary_chemistry}

\paragraph{Reaction network and fitting procedure.}
We fitted the TME decay with a reduced \texttt{Catalyst.jl} network \cite{CatalystPLOSCompBio2023} comprising 44 reaction channels (see Table~\ref{tab:reaction_network}). The TME balance follows from its three reactions with \ce{O3}, \ce{OH} and acetyl peroxy:
\begin{equation}
    -\frac{\mathrm{d}X}{\mathrm{d}t}=kZX+k_\mathrm{OH}HX+k_\mathrm{AcP}[\mathrm{AcP}]X,
    \label{eq:tme_secondary_balance}
\end{equation}
where $X=\ce{[TME]}$, $Z=\ce{[O3]}$ and $H=\ce{[OH]}$. AcP denotes acetyl peroxy (\ce{CH3C(O)O2}) and $k_\mathrm{AcP}$ its addition coefficient for TME. The other peroxy radicals are AP (acetonyl peroxy, \ce{CH3C(O)CH2O2}), $R_\mathrm{T}$ (\ce{C6H13O3}, formed by OH addition to TME and oxygen capture), $R_8$ (\ce{C8H15O5}, formed from AcP + TME and oxygen capture), and methyl peroxy (\ce{CH3O2}). Ozone and radical concentrations are integrated explicitly. The nominal calculation uses $T=\SI{293}{\kelvin}$, $Z_0=\SI{1.85e14}{\per\cubic\centi\meter}$, $X_0=\SI{1.96e13}{\per\cubic\centi\meter}$ and dry air at \SI{101325}{\pascal}, with zero radical and product concentrations at $t=0$ and no radical wall loss. The calculated signal is
\begin{equation}
    S_\mathrm{model}(t)=A\frac{X(t;k)}{X_0}.
\end{equation}
Unweighted least squares fits $k$ and $A$ to the same nine lamp-on mean signals as Eq.~\eqref{eq:TME}. $X_0$ and all other chemical parameters are fixed; $A$ converts the calculated TME fraction to signal. The search spans \numrange{0.05}{10}\,$\times$\,\SI{e-15}{\centi\meter\cubed\per\molecule\per\second}, without an IUPAC prior or penalty.

Ozonolysis produces prompt acetone and, with yield $\phi_\mathrm{OH}=0.93$, OH and AP; the remaining carbon is assigned to unreactive products. This source term groups the Criegee steps. OH and \ce{HO2} chemistry includes OH regeneration by \ce{HO2 + O3}, and organic peroxy radicals undergo self-, cross-, \ce{HO2} and OH reactions (Table~\ref{tab:reaction_network}). Rapid oxygen addition and alkoxy fragmentation are represented by net products: AP can form AcP and formaldehyde, AcP can form methyl peroxy and \ce{CO2}, and $R_\mathrm{T}$ can form two acetone molecules and \ce{HO2} \cite{frandsenTmeOzonolysis2025}. The minor OH-abstraction branch is omitted. Product reoxidation is limited to acetone, formaldehyde, methanol and hydrogen peroxide.

Secondary coefficients use IUPAC data where available \cite{iupacSecondaryKinetics}. The OH-addition coefficient for TME follows Orkin et al.\ \cite{orkinOHButenes2023}:
\begin{equation}
    k_\mathrm{OH}(T)=\SI{14.08e-11}{\centi\meter\cubed\per\molecule\per\second}
    \left(\frac{T}{\SI{298}{\kelvin}}\right)^{-1.714}
    \exp\!\left(-\frac{\SI{95.7}{\kelvin}}{T}\right).
\end{equation}
Peroxy-radical estimates follow Jenkin et al.\ \cite{jenkinPeroxy2019}. For $R_\mathrm{T}$ and $R_8$ reacting with non-acyl peroxy radicals, we use $2\sqrt{k_ik_j}$, with self-reaction coefficients $k_i$ and $k_j$, multiplied by the tertiary-peroxy factor $f_\mathrm{RO_2}=2$. Applying this peroxy-pool parameterization to explicit cross-reactions is an assumption tested below.

AcP + TME forms $R_8$ with $k_\mathrm{AcP}=\SI{1.2e-13}{\centi\meter\cubed\per\molecule\per\second}$, measured at \SI{298}{\kelvin} with a factor-five uncertainty \cite{nozierePeroxyAlkenes2021}. Its subsequent alkoxy formation can yield two acetone molecules, \ce{CO2} and methyl peroxy \cite{frandsenTmeOzonolysis2025}. The adduct is followed explicitly, with unmeasured fate coefficients estimated from those of $R_\mathrm{T}$ and the generic OH + peroxy expression for its OH reaction. The assumed alkoxy-product fraction is 0.8 for self- and cross-reactions of $R_\mathrm{T}$ and, independently, for $R_8$. OH reactions with the larger peroxy radicals nominally form persistent termination products.

For the published methyl-peroxy and acetonyl-peroxy + TME coefficients \cite{nozierePeroxyAlkenes2021,frandsenTmeOzonolysis2025}, integrated loss fluxes along the nominal trajectory account for less than 0.001\% of TME consumed for either reaction; these channels are omitted. We do not assign a coefficient to OH-derived TME peroxy + TME, whose possible contribution to TME loss remains unquantified. Including this channel could further lower the rate coefficient inferred for TME + \ce{O3}.

Table~\ref{tab:reaction_network} gives net products and nominal mean yields. $\mathrm{C}_\mathrm{p}$ counts carbon atoms in products not subsequently reacted; it is a balance variable, not a chemical species. Untracked non-carbon products and bath-gas oxygen are omitted. The formaldehyde fraction in reaction~15 is $f=[1+498\exp(-1160/T)]^{-1}=0.09522$ at \SI{293}{\kelvin}, with $T$ in kelvin.

\begingroup
\small
\setlength{\tabcolsep}{4pt}
\setlength{\LTleft}{0pt}
\setlength{\LTcapwidth}{\linewidth}
\setlength{\LTright}{\fill}
\renewcommand{\arraystretch}{1.1}
\begin{longtable}{@{}p{\dimexpr\linewidth-69pt-110pt-4\tabcolsep\relax} p{69pt} p{110pt}@{}}
\caption{\textbf{The 44 chemical channels of the nominal reaction network.} Coefficients $k_i$ are in \si{\centi\meter\cubed\per\molecule\per\second} at \SI{293}{\kelvin}; the source or estimation basis is given beside each value. $k_1=k$ is fitted to the TME decay. Product yields use the nominal branching fractions.}\label{tab:reaction_network}\\
\hline
\raggedright\hspace*{1.8em}Effective reaction & \raggedright $k_i(293\,\mathrm{K})$ & \raggedright Reference \tabularnewline[3pt]
\hline
\endfirsthead
\multicolumn{3}{@{}l}{Table~\thetable\ (continued).} \\[5pt]
\hline
\raggedright\hspace*{1.8em}Effective reaction & \raggedright $k_i(293\,\mathrm{K})$ & \raggedright Reference \tabularnewline[3pt]
\hline
\endhead
\hline
\multicolumn{3}{r@{}}{Continued on next page.} \\
\endfoot
\hline
\endlastfoot
\multicolumn{3}{@{}l}{\textit{Ozonolysis and OH/HO$_2$ chemistry}} \\*[3pt]
\raggedright\hangindent=1.8em\hangafter=1\makebox[1.4em][r]{1}\nobreak\hspace{0.4em}$\mathrm{TME} + \ce{O3} \rightarrow \mathrm{acetone} + 0.93\ce{OH} + 0.93\mathrm{AP} + 0.21\mathrm{C}_\mathrm{p}$ & \raggedright \num{1.175e-15} & \raggedright Fitted; this work \tabularnewline[3pt]
\raggedright\hangindent=1.8em\hangafter=1\makebox[1.4em][r]{2}\nobreak\hspace{0.4em}$\ce{OH} + \mathrm{TME} \rightarrow R_\mathrm{T}$ & \raggedright \num{1.046e-10} & \raggedright \cite{orkinOHButenes2023}, OH addition \tabularnewline[3pt]
\raggedright\hangindent=1.8em\hangafter=1\makebox[1.4em][r]{3}\nobreak\hspace{0.4em}$\ce{OH} + \ce{O3} \rightarrow \ce{HO2}$ & \raggedright \num{6.873e-14} & \raggedright \cite{iupacSecondaryKinetics}, \href{https://iupac-aeris.ipsl.fr/datasheets/pdf/HOx13.pdf}{HOx13} \tabularnewline[3pt]
\raggedright\hangindent=1.8em\hangafter=1\makebox[1.4em][r]{4}\nobreak\hspace{0.4em}$\ce{HO2} + \ce{O3} \rightarrow \ce{OH}$ & \raggedright \num{1.940e-15} & \raggedright \cite{iupacSecondaryKinetics}, \href{https://iupac-aeris.ipsl.fr/datasheets/pdf/HOx15.pdf}{HOx15} \tabularnewline[3pt]
\raggedright\hangindent=1.8em\hangafter=1\makebox[1.4em][r]{5}\nobreak\hspace{0.4em}$\ce{OH} + \ce{HO2} \rightarrow \mathrm{products}$ & \raggedright \num{1.033e-10} & \raggedright \cite{iupacSecondaryKinetics}, \href{https://iupac-aeris.ipsl.fr/datasheets/pdf/HOx11.pdf}{HOx11} \tabularnewline[3pt]
\raggedright\hangindent=1.8em\hangafter=1\makebox[1.4em][r]{6}\nobreak\hspace{0.4em}$2\ce{HO2} \rightarrow \ce{H2O2}$ & \raggedright \num{3.016e-12} & \raggedright \cite{iupacSecondaryKinetics}, \href{https://iupac-aeris.ipsl.fr/datasheets/pdf/HOx14.pdf}{HOx14} \tabularnewline[3pt]
\raggedright\hangindent=1.8em\hangafter=1\makebox[1.4em][r]{7}\nobreak\hspace{0.4em}$\ce{OH} + \ce{H2O2} \rightarrow \ce{HO2}$ & \raggedright \num{1.680e-12} & \raggedright \cite{iupacSecondaryKinetics}, \href{https://iupac-aeris.ipsl.fr/datasheets/pdf/HOx12.pdf}{HOx12} \tabularnewline[3pt]
\raggedright\hangindent=1.8em\hangafter=1\makebox[1.4em][r]{8}\nobreak\hspace{0.4em}$\ce{OH} + \mathrm{acetone} \rightarrow \mathrm{AP}$ & \raggedright \num{1.693e-13} & \raggedright \cite{iupacSecondaryKinetics}, \href{https://iupac-aeris.ipsl.fr/datasheets/pdf/HOx_VOC19.pdf}{HOx\_VOC19} \tabularnewline[3pt]
\raggedright\hangindent=1.8em\hangafter=1\makebox[1.4em][r]{9}\nobreak\hspace{0.4em}$\ce{OH} + \ce{HCHO} \rightarrow \ce{HO2} + \mathrm{C}_\mathrm{p}$ & \raggedright \num{8.560e-12} & \raggedright \cite{iupacSecondaryKinetics}, \href{https://iupac-aeris.ipsl.fr/datasheets/pdf/HOx_VOC11.pdf}{HOx\_VOC11} \tabularnewline[3pt]
\raggedright\hangindent=1.8em\hangafter=1\makebox[1.4em][r]{10}\nobreak\hspace{0.4em}$\ce{OH} + \ce{CH3OH} \rightarrow \ce{HCHO} + \ce{HO2}$ & \raggedright \num{8.780e-13} & \raggedright \cite{iupacSecondaryKinetics}, \href{https://iupac-aeris.ipsl.fr/datasheets/pdf/HOx_VOC23.pdf}{HOx\_VOC23} \tabularnewline[3pt]
\multicolumn{3}{@{}l}{\textit{Organic peroxy radicals + HO$_2$}} \\*[3pt]
\raggedright\hangindent=1.8em\hangafter=1\makebox[1.4em][r]{11}\nobreak\hspace{0.4em}$\mathrm{AP} + \ce{HO2} \rightarrow 0.15\ce{OH} + 0.15\mathrm{AcP} + 0.15\ce{HCHO} + 2.55\mathrm{C}_\mathrm{p}$ & \raggedright \num{9.000e-12} & \raggedright \cite{iupacSecondaryKinetics}, \href{https://iupac-aeris.ipsl.fr/datasheets/pdf/HOx_VOC57.pdf}{HOx\_VOC57} \tabularnewline[3pt]
\raggedright\hangindent=1.8em\hangafter=1\makebox[1.4em][r]{12}\nobreak\hspace{0.4em}$\mathrm{AcP} + \ce{HO2} \rightarrow 2\mathrm{C}_\mathrm{p}$ & \raggedright \num{7.719e-12} & \raggedright \cite{iupacSecondaryKinetics}, \href{https://iupac-aeris.ipsl.fr/datasheets/pdf/HOx_VOC54.pdf}{HOx\_VOC54} \tabularnewline[3pt]
\raggedright\hangindent=1.8em\hangafter=1\makebox[1.4em][r]{13}\nobreak\hspace{0.4em}$\mathrm{AcP} + \ce{HO2} \rightarrow 2\mathrm{C}_\mathrm{p} + \ce{O3}$ & \raggedright \num{2.982e-12} & \raggedright \cite{iupacSecondaryKinetics}, \href{https://iupac-aeris.ipsl.fr/datasheets/pdf/HOx_VOC54.pdf}{HOx\_VOC54} \tabularnewline[3pt]
\raggedright\hangindent=1.8em\hangafter=1\makebox[1.4em][r]{14}\nobreak\hspace{0.4em}$\mathrm{AcP} + \ce{HO2} \rightarrow \ce{OH} + \ce{CH3O2} + \mathrm{C}_\mathrm{p}$ & \raggedright \num{1.039e-11} & \raggedright \cite{iupacSecondaryKinetics}, \href{https://iupac-aeris.ipsl.fr/datasheets/pdf/HOx_VOC54.pdf}{HOx\_VOC54} \tabularnewline[3pt]
\raggedright\hangindent=1.8em\hangafter=1\makebox[1.4em][r]{15}\nobreak\hspace{0.4em}$\ce{CH3O2} + \ce{HO2} \rightarrow f\ce{HCHO} + (1-f)\mathrm{C}_\mathrm{p}$ & \raggedright \num{5.315e-12} & \raggedright \cite{iupacSecondaryKinetics}, \href{https://iupac-aeris.ipsl.fr/datasheets/pdf/HOx_VOC51.pdf}{HOx\_VOC51} \tabularnewline[3pt]
\raggedright\hangindent=1.8em\hangafter=1\makebox[1.4em][r]{16}\nobreak\hspace{0.4em}$R_\mathrm{T} + \ce{HO2} \rightarrow 6\mathrm{C}_\mathrm{p}$ & \raggedright \num{1.500e-11} & \raggedright \cite{jenkinPeroxy2019}, Table~7 \tabularnewline[3pt]
\multicolumn{3}{@{}l}{\textit{Organic peroxy self-reactions}} \\*[3pt]
\raggedright\hangindent=1.8em\hangafter=1\makebox[1.4em][r]{17}\nobreak\hspace{0.4em}$2\mathrm{AP} \rightarrow 1.26\mathrm{AcP} + 1.26\ce{HCHO} + 2.22\mathrm{C}_\mathrm{p}$ & \raggedright \num{8.000e-12} & \raggedright \cite{iupacSecondaryKinetics}, \href{https://iupac-aeris.ipsl.fr/datasheets/pdf/ROO_34.pdf}{ROO\_34} \tabularnewline[3pt]
\raggedright\hangindent=1.8em\hangafter=1\makebox[1.4em][r]{18}\nobreak\hspace{0.4em}$2\mathrm{AcP} \rightarrow 2\ce{CH3O2} + 2\mathrm{C}_\mathrm{p}$ & \raggedright \num{1.598e-11} & \raggedright \cite{iupacSecondaryKinetics}, \href{https://iupac-aeris.ipsl.fr/datasheets/pdf/ROO_29.pdf}{ROO\_29} \tabularnewline[3pt]
\raggedright\hangindent=1.8em\hangafter=1\makebox[1.4em][r]{19}\nobreak\hspace{0.4em}$2\ce{CH3O2} \rightarrow 2\ce{HCHO} + 2\ce{HO2}$ & \raggedright \num{1.154e-13} & \raggedright \cite{iupacSecondaryKinetics}, \href{https://iupac-aeris.ipsl.fr/datasheets/pdf/ROO_22.pdf}{ROO\_22} \tabularnewline[3pt]
\raggedright\hangindent=1.8em\hangafter=1\makebox[1.4em][r]{20}\nobreak\hspace{0.4em}$2\ce{CH3O2} \rightarrow \ce{HCHO} + \ce{CH3OH}$ & \raggedright \num{2.130e-13} & \raggedright \cite{iupacSecondaryKinetics}, \href{https://iupac-aeris.ipsl.fr/datasheets/pdf/ROO_22.pdf}{ROO\_22} \tabularnewline[3pt]
\raggedright\hangindent=1.8em\hangafter=1\makebox[1.4em][r]{21}\nobreak\hspace{0.4em}$2R_\mathrm{T} \rightarrow 3.2\mathrm{acetone} + 1.6\ce{HO2} + 2.4\mathrm{C}_\mathrm{p}$ & \raggedright \num{4.635e-15} & \raggedright \cite{jenkinPeroxy2019}, Table~10 \tabularnewline[3pt]
\multicolumn{3}{@{}l}{\textit{Organic peroxy cross-reactions}} \\*[3pt]
\raggedright\hangindent=1.8em\hangafter=1\makebox[1.4em][r]{22}\nobreak\hspace{0.4em}$\mathrm{AP} + \ce{CH3O2} \rightarrow 3\mathrm{C}_\mathrm{p} + \ce{CH3OH}$ & \raggedright \num{1.900e-12} & \raggedright \cite{iupacSecondaryKinetics}, \href{https://iupac-aeris.ipsl.fr/datasheets/pdf/ROO_24.pdf}{ROO\_24} \tabularnewline[3pt]
\raggedright\hangindent=1.8em\hangafter=1\makebox[1.4em][r]{23}\nobreak\hspace{0.4em}$\mathrm{AP} + \ce{CH3O2} \rightarrow 3\mathrm{C}_\mathrm{p} + \ce{HCHO}$ & \raggedright \num{7.600e-13} & \raggedright \cite{iupacSecondaryKinetics}, \href{https://iupac-aeris.ipsl.fr/datasheets/pdf/ROO_24.pdf}{ROO\_24} \tabularnewline[3pt]
\raggedright\hangindent=1.8em\hangafter=1\makebox[1.4em][r]{24}\nobreak\hspace{0.4em}$\mathrm{AP} + \ce{CH3O2} \rightarrow \mathrm{AcP} + 2\ce{HCHO} + \ce{HO2}$ & \raggedright \num{1.140e-12} & \raggedright \cite{iupacSecondaryKinetics}, \href{https://iupac-aeris.ipsl.fr/datasheets/pdf/ROO_24.pdf}{ROO\_24} \tabularnewline[3pt]
\raggedright\hangindent=1.8em\hangafter=1\makebox[1.4em][r]{25}\nobreak\hspace{0.4em}$\mathrm{AP} + \mathrm{AcP} \rightarrow \mathrm{AcP} + \ce{CH3O2} + \ce{HCHO} + \mathrm{C}_\mathrm{p}$ & \raggedright \num{2.500e-12} & \raggedright \cite{iupacSecondaryKinetics}, \href{https://iupac-aeris.ipsl.fr/datasheets/pdf/ROO_30.pdf}{ROO\_30} \tabularnewline[3pt]
\raggedright\hangindent=1.8em\hangafter=1\makebox[1.4em][r]{26}\nobreak\hspace{0.4em}$\mathrm{AP} + \mathrm{AcP} \rightarrow 5\mathrm{C}_\mathrm{p}$ & \raggedright \num{2.500e-12} & \raggedright \cite{iupacSecondaryKinetics}, \href{https://iupac-aeris.ipsl.fr/datasheets/pdf/ROO_30.pdf}{ROO\_30} \tabularnewline[3pt]
\raggedright\hangindent=1.8em\hangafter=1\makebox[1.4em][r]{27}\nobreak\hspace{0.4em}$\mathrm{AcP} + \ce{CH3O2} \rightarrow \ce{CH3O2} + \ce{HCHO} + \ce{HO2} + \mathrm{C}_\mathrm{p}$ & \raggedright \num{9.917e-12} & \raggedright \cite{iupacSecondaryKinetics}, \href{https://iupac-aeris.ipsl.fr/datasheets/pdf/ROO_23.pdf}{ROO\_23} \tabularnewline[3pt]
\raggedright\hangindent=1.8em\hangafter=1\makebox[1.4em][r]{28}\nobreak\hspace{0.4em}$\mathrm{AcP} + \ce{CH3O2} \rightarrow \ce{HCHO} + 2\mathrm{C}_\mathrm{p}$ & \raggedright \num{1.102e-12} & \raggedright \cite{iupacSecondaryKinetics}, \href{https://iupac-aeris.ipsl.fr/datasheets/pdf/ROO_23.pdf}{ROO\_23} \tabularnewline[3pt]
\raggedright\hangindent=1.8em\hangafter=1\makebox[1.4em][r]{29}\nobreak\hspace{0.4em}$R_\mathrm{T} + \mathrm{AP} \rightarrow 1.6\mathrm{acetone} + 0.8\ce{HO2} + 0.8\mathrm{AcP} + 0.8\ce{HCHO} + 1.8\mathrm{C}_\mathrm{p}$ & \raggedright \num{7.703e-13} & \raggedright \cite{jenkinPeroxy2019}, Estimate: $2f_\mathrm{RO_2}\sqrt{k_{21}k_{17}}$ \tabularnewline[3pt]
\raggedright\hangindent=1.8em\hangafter=1\makebox[1.4em][r]{30}\nobreak\hspace{0.4em}$R_\mathrm{T} + \mathrm{AcP} \rightarrow 1.6\mathrm{acetone} + 0.8\ce{HO2} + 0.8\ce{CH3O2} + 2.4\mathrm{C}_\mathrm{p}$ & \raggedright \num{1.102e-11} & \raggedright \cite{iupacSecondaryKinetics}, Estimate: $k_{27}+k_{28}$ \tabularnewline[3pt]
\raggedright\hangindent=1.8em\hangafter=1\makebox[1.4em][r]{31}\nobreak\hspace{0.4em}$R_\mathrm{T} + \ce{CH3O2} \rightarrow 1.6\mathrm{acetone} + 1.6\ce{HO2} + \ce{HCHO} + 1.2\mathrm{C}_\mathrm{p}$ & \raggedright \num{1.561e-13} & \raggedright \cite{jenkinPeroxy2019}, Estimate: $2f_\mathrm{RO_2}\sqrt{k_{21}(k_{19}+k_{20})}$ \tabularnewline[3pt]
\multicolumn{3}{@{}l}{\textit{OH + organic peroxy radicals}} \\*[3pt]
\raggedright\hangindent=1.8em\hangafter=1\makebox[1.4em][r]{32}\nobreak\hspace{0.4em}$\ce{OH} + \mathrm{AP} \rightarrow 3\mathrm{C}_\mathrm{p}$ & \raggedright \num{1.222e-10} & \raggedright \cite{jenkinPeroxy2019}, Generic estimate, Eq.~8 \tabularnewline[3pt]
\raggedright\hangindent=1.8em\hangafter=1\makebox[1.4em][r]{33}\nobreak\hspace{0.4em}$\ce{OH} + \mathrm{AcP} \rightarrow 2\mathrm{C}_\mathrm{p}$ & \raggedright \num{1.222e-10} & \raggedright \cite{jenkinPeroxy2019}, Generic estimate, Eq.~8 \tabularnewline[3pt]
\raggedright\hangindent=1.8em\hangafter=1\makebox[1.4em][r]{34}\nobreak\hspace{0.4em}$\ce{OH} + R_\mathrm{T} \rightarrow 6\mathrm{C}_\mathrm{p}$ & \raggedright \num{1.222e-10} & \raggedright \cite{jenkinPeroxy2019}, Generic estimate, Eq.~8 \tabularnewline[3pt]
\raggedright\hangindent=1.8em\hangafter=1\makebox[1.4em][r]{35}\nobreak\hspace{0.4em}$\ce{OH} + \ce{CH3O2} \rightarrow \ce{HCHO} + 2\ce{HO2}$ & \raggedright \num{1.136e-10} & \raggedright \cite{jenkinPeroxy2019}, Generic estimate; 0.93 branch \tabularnewline[3pt]
\raggedright\hangindent=1.8em\hangafter=1\makebox[1.4em][r]{36}\nobreak\hspace{0.4em}$\ce{OH} + \ce{CH3O2} \rightarrow \ce{CH3OH}$ & \raggedright \num{8.552e-12} & \raggedright \cite{jenkinPeroxy2019}, Generic estimate; 0.07 branch \tabularnewline[3pt]
\multicolumn{3}{@{}l}{\textit{Acetyl-peroxy addition to TME and C8-adduct reactions}} \\*[3pt]
\raggedright\hangindent=1.8em\hangafter=1\makebox[1.4em][r]{37}\nobreak\hspace{0.4em}$\mathrm{AcP} + \mathrm{TME} \rightarrow R_8$ & \raggedright \num{1.200e-13} & \raggedright \cite{nozierePeroxyAlkenes2021}, SI Table~S5 \tabularnewline[3pt]
\raggedright\hangindent=1.8em\hangafter=1\makebox[1.4em][r]{38}\nobreak\hspace{0.4em}$R_8 + \ce{HO2} \rightarrow 8\mathrm{C}_\mathrm{p}$ & \raggedright \num{1.500e-11} & \raggedright \cite{jenkinPeroxy2019}, Estimate: $k_{16}$ \tabularnewline[3pt]
\raggedright\hangindent=1.8em\hangafter=1\makebox[1.4em][r]{39}\nobreak\hspace{0.4em}$2R_8 \rightarrow 3.2\mathrm{acetone} + 1.6\ce{CH3O2} + 4.8\mathrm{C}_\mathrm{p}$ & \raggedright \num{4.635e-15} & \raggedright \cite{jenkinPeroxy2019}, Estimate: $k_{21}$ \tabularnewline[3pt]
\raggedright\hangindent=1.8em\hangafter=1\makebox[1.4em][r]{40}\nobreak\hspace{0.4em}$R_8 + \mathrm{AP} \rightarrow 1.6\mathrm{acetone} + 0.8\ce{CH3O2} + 0.8\mathrm{AcP} + 0.8\ce{HCHO} + 3\mathrm{C}_\mathrm{p}$ & \raggedright \num{7.703e-13} & \raggedright \cite{jenkinPeroxy2019}, Estimate: $k_{29}$ \tabularnewline[3pt]
\raggedright\hangindent=1.8em\hangafter=1\makebox[1.4em][r]{41}\nobreak\hspace{0.4em}$R_8 + \mathrm{AcP} \rightarrow 1.6\mathrm{acetone} + 1.6\ce{CH3O2} + 3.6\mathrm{C}_\mathrm{p}$ & \raggedright \num{1.102e-11} & \raggedright \cite{iupacSecondaryKinetics}, Estimate: $k_{30}$ \tabularnewline[3pt]
\raggedright\hangindent=1.8em\hangafter=1\makebox[1.4em][r]{42}\nobreak\hspace{0.4em}$R_8 + \ce{CH3O2} \rightarrow 1.6\mathrm{acetone} + 0.8\ce{CH3O2} + 0.8\ce{HO2} + \ce{HCHO} + 2.4\mathrm{C}_\mathrm{p}$ & \raggedright \num{1.561e-13} & \raggedright \cite{jenkinPeroxy2019}, Estimate: $k_{31}$ \tabularnewline[3pt]
\raggedright\hangindent=1.8em\hangafter=1\makebox[1.4em][r]{43}\nobreak\hspace{0.4em}$R_8 + R_\mathrm{T} \rightarrow 3.2\mathrm{acetone} + 0.8\ce{CH3O2} + 0.8\ce{HO2} + 3.6\mathrm{C}_\mathrm{p}$ & \raggedright \num{1.854e-14} & \raggedright \cite{jenkinPeroxy2019}, Estimate: $2f_\mathrm{RO_2}k_{21}$ \tabularnewline[3pt]
\raggedright\hangindent=1.8em\hangafter=1\makebox[1.4em][r]{44}\nobreak\hspace{0.4em}$R_8 + \ce{OH} \rightarrow 8\mathrm{C}_\mathrm{p}$ & \raggedright \num{1.222e-10} & \raggedright \cite{jenkinPeroxy2019}, Generic estimate: $k_{32}$ \tabularnewline[3pt]
\end{longtable}
\endgroup

Self-reactions $2R\rightarrow\mathrm{products}$ have event rate $k_i[R]^2$ and reactant-loss rate $2k_i[R]^2$. The \ce{HO2} self-reaction coefficient uses dry air at \SI{101325}{\pascal}. Seven additional first-order wall-loss channels act on OH, \ce{HO2}, AP, AcP, \ce{CH3O2}, $R_\mathrm{T}$ and $R_8$, with a common coefficient of zero nominally and 1 or \SI{10}{\per\second} in sensitivity tests. They are excluded from the 44 chemical channels.

\begin{figure}[p]
    \centering
    \includegraphics[width=\linewidth]{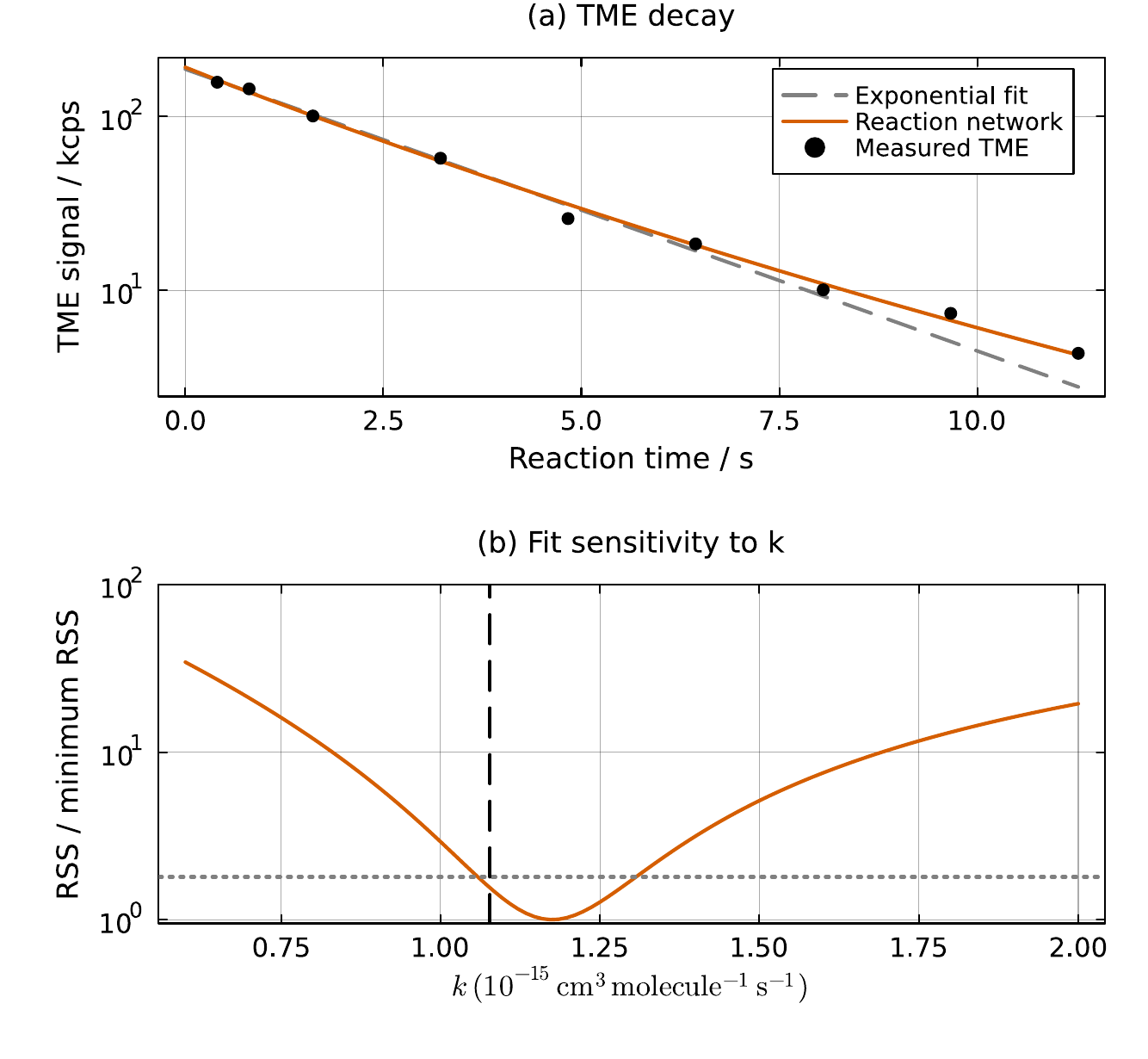}
    \caption{\textbf{TME decay interpreted with secondary chemistry.} (a) The same nine mean signals as Figure~\ref{fig:Kinetics} on a logarithmic scale, with the exponential fit (grey dashed) and the nominal reaction-network fit (orange). The exponential falls below the three longest-time points, whereas the network follows them. (b) Residual sum of squares (RSS) for trial $k$ values, with the amplitude reoptimized and RSS divided by its minimum. The black dashed line is the IUPAC reference, not a fitting constraint. The grey dotted line, $1+F_{1,7}(0.95)/7$, defines an approximate 95\% profile interval with the chemical parameters fixed. This interval does not include chemical uncertainty.}
    \label{fig:chemical_fits}
\end{figure}

\paragraph{TME fit and curvature.}
The fit gives $k=(1.175\pm0.053)\times\SI{e-15}{\centi\meter\cubed\per\molecule\per\second}$ (fit standard error), with a 95\% interval of $\numrange{1.056}{1.308}\times\SI{e-15}{\centi\meter\cubed\per\molecule\per\second}$, both conditional on the fixed chemical parameters and unweighted residual model. The RMSE is \num{3292} cps, compared with \num{3072} cps for the unchanged exponential fit (Figure~\ref{fig:chemical_fits}).

As TME is consumed, its loss by OH decreases relative to its loss by ozone. Along the nominal trajectory, this ratio falls from 0.85 to 0.34 across the measured time range, while the local decay constant falls from \SI{0.403}{\per\second} between the first two reaction times to \SI{0.284}{\per\second} between the last two. This curvature follows from the fixed secondary chemistry, with no additional fit parameter, and supports a contribution from secondary reactions to the observed TME loss.

\paragraph{Uncertainty.}
Joint variation of 39 inputs in 640 refits, using evaluated uncertainties where available and stated ranges for estimated rates and branching fractions, gives a chemical-parameter standard deviation of $(0.047\text{--}0.057)\times\SI{e-15}{\centi\meter\cubed\per\molecule\per\second}$, depending on the input distributions. Dependencies between reaction channels are retained.

Assuming independent chemical, fit and ozone contributions, their quadrature combination gives a standard uncertainty of approximately \SI{0.094e-15}{\centi\meter\cubed\per\molecule\per\second}. The ozone contribution is \SI{0.061e-15}{\centi\meter\cubed\per\molecule\per\second}, from the assigned 5\% concentration uncertainty. We therefore report $k=\SI{1.18(19)e-15}{\centi\meter\cubed\per\molecule\per\second}$ with a coverage factor of 2. This approximate budget is conditional on the network and input distributions; omitted pathways, wall losses and signal-treatment choices are excluded.

\paragraph{Sensitivity tests.}
The 43 one-at-a-time tests complement this uncertainty calculation. Chemical and concentration tests using all nine raw signals give $k=\numrange{1.117}{1.255}\times\SI{e-15}{\centi\meter\cubed\per\molecule\per\second}$, a range of separate tests, not a confidence interval or combined bound. They vary $\phi_\mathrm{OH}=0.79$--1, $X_0$ from half to twice nominal, ozone by $\pm5\%$, secondary coefficients, branching fractions, radical wall loss and water. Branching tests include alkoxy fractions, cross-reaction estimates and prompt-product fractions from OH + peroxy reactions. The yield test is capped at one primary OH per ozonolysis and does not propagate the full reported yield uncertainty. A separate check using the measured pressure of \SI{1055}{\hecto\pascal} changes $k$ by less than 0.01\%.

Nominally, AcP + TME consumes 1.7\% of the TME lost at the longest measured time. Its experimental factor-five coefficient limits give $k=\numrange{1.137}{1.186}\times\SI{e-15}{\centi\meter\cubed\per\molecule\per\second}$. Using the calculated addition coefficient of \SI{3.2e-15}{\centi\meter\cubed\per\molecule\per\second} at \SI{296}{\kelvin} \cite{frandsenTmeOzonolysis2025} gives $k=\SI{1.189e-15}{\centi\meter\cubed\per\molecule\per\second}$; omitting the addition increases $k$ by 1.2\%. Varying the C8 alkoxy fraction from zero to one, or scaling its estimated fate coefficients by 0.1 or 10, changes $k$ by less than 0.2\% separately. The calculated ozone loss is 6.0\% at the longest time; holding it constant changes $k$ by $-3.2\%$.

Removing the first point, the first two, or all points below \SI{2}{\second} gives $k=1.253$, 1.220 and $1.331\times\SI{e-15}{\centi\meter\cubed\per\molecule\per\second}$ ($+6.6\%$, $+3.8\%$ and $+13.3\%$). The last exclusion increases the conditional standard error to \SI{0.141e-15}{\centi\meter\cubed\per\molecule\per\second}; the four points above \SI{5}{\second} give a broad conditional 95\% interval of $\numrange{0.680}{2.046}\times\SI{e-15}{\centi\meter\cubed\per\molecule\per\second}$.

Reagent-ion normalization and logarithmic fitting of the paired normalized ON/OFF ratio give $k=1.133$ and $1.160\times\SI{e-15}{\centi\meter\cubed\per\molecule\per\second}$. In a separate weighting analysis across five combinations of weighting (unweighted or logarithmic) and signal treatment (raw, normalized or paired ratio), the exponential $k^\prime$ spans \SIrange{0.326}{0.373}{\per\second}, compared with $k=\numrange{1.128}{1.175}\times\SI{e-15}{\centi\meter\cubed\per\molecule\per\second}$ for the network. Relative to the nominal fits to the unnormalized signal, the largest decreases are 12.7\% and 4.0\%, respectively. Logarithmic fitting gives equal relative weight to each point. With two fitted parameters per model, the logarithmic RMS residuals for the network and exponential are 0.078 and 0.118 with raw signals, and 0.053 and 0.085 with the paired ratio.

A common $+\SI{0.4}{\second}$ time shift increases the network fit's $k$ by 1.8\%; exact shift invariance applies only to the exponential fit. The fit to the paired ON/OFF ratio gives a free amplitude of 0.80. Fixing it at unity and fitting an effective delay gives $\delta=\SI{0.54}{\second}$ and $k=\SI{1.188e-15}{\centi\meter\cubed\per\molecule\per\second}$, 2.4\% above the free-amplitude fit, with a comparable residual. This delay is comparable to the tracer residual of \SIrange{0.44}{0.49}{\second} (Table~\ref{tab:RTD_symmetric}). Imposing it on the raw-signal fit gives $k=\SI{1.204e-15}{\centi\meter\cubed\per\molecule\per\second}$, 2.4\% above the raw-signal fit without delay.

\FloatBarrier

\paragraph{Sensitivity to the choice of reaction model.}
We retained only channels 1--3 of Table~\ref{tab:reaction_network}: TME + \ce{O3}, OH + TME, and OH + \ce{O3}. The OH yield in channel 1 was fixed at 0.93 from the IUPAC evaluation \cite{iupacOxVOC41}. The two secondary coefficients at \SI{293}{\kelvin}, $1.046\times10^{-10}$ for OH + TME and $6.873\times10^{-14}$ for OH + \ce{O3}, in \si{\centi\meter\cubed\per\molecule\per\second}, were fixed to the same values as in the 44-channel network \cite{orkinOHButenes2023,iupacSecondaryKinetics}. Both models integrate ozone and OH explicitly. The three-reaction model omits \ce{HO2} recycling and the other radical reactions. Only $k$ and the TME signal amplitude were fitted in either model, using the same nine measurements. The nominal 44-channel model attributes 56.3\%, 42.0\%, and 1.7\% of the integrated TME loss up to the last measurement to ozone, OH, and acetyl peroxy, respectively.

\begin{table}[!ht]
    \centering
    \caption{\textbf{Comparison of the two reaction models.} The coefficients $k$ are reported in units of $10^{-15}$~\si{\centi\meter\cubed\per\molecule\per\second}. The last column gives the RMS of $\ln(S_\mathrm{meas}/S_\mathrm{fit})$ for each fit.}
    \label{tab:model_comparison}
    \begin{tabular}{llrrr}
        \hline
        Fit objective & Model & $k$ & RMSE (cps) & Log RMS \\
        \hline
        Signal & 44 channels & 1.175 & 3292 & 0.079 \\
        Log signal & 44 channels & 1.170 & 3517 & 0.078 \\
        Signal & 3 reactions & 1.088 & 3078 & 0.146 \\
        Log signal & 3 reactions & 1.011 & 5180 & 0.102 \\
        \hline
    \end{tabular}
\end{table}

\begin{figure}[!ht]
    \centering
    \includegraphics[width=\linewidth]{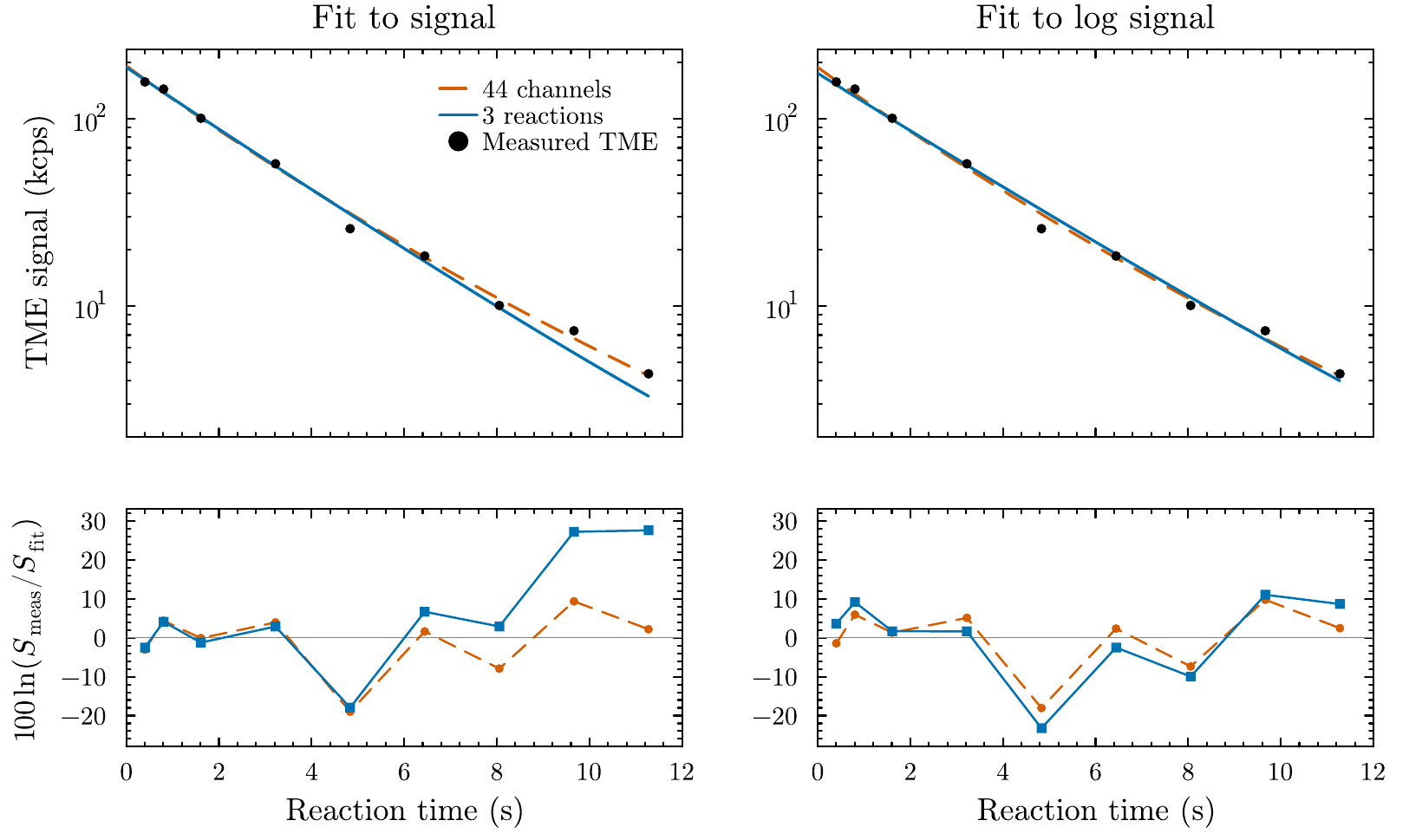}
    \caption{\textbf{TME decay and residuals for both reaction models.} The same nine measurements are fitted in each panel. Top: measured and fitted signals on a log scale. Bottom: log residuals for the respective fits.}
    \label{fig:model_comparison}
\end{figure}

The three-reaction model underpredicts the last two measurements by about 24\% when fitted to the signal (Figure~\ref{fig:model_comparison}). Compared with the 44-channel model, it gives a $k$ that is 7.4\% lower when fitting the signal and 13.6\% lower when fitting its logarithm. These differences are reported separately from the expanded uncertainty.

\FloatBarrier

\subsection{Mass-spectral signals associated with the reaction}

\begin{figure}[ht] 
	\centering
    \includegraphics[width=436.00296bp]{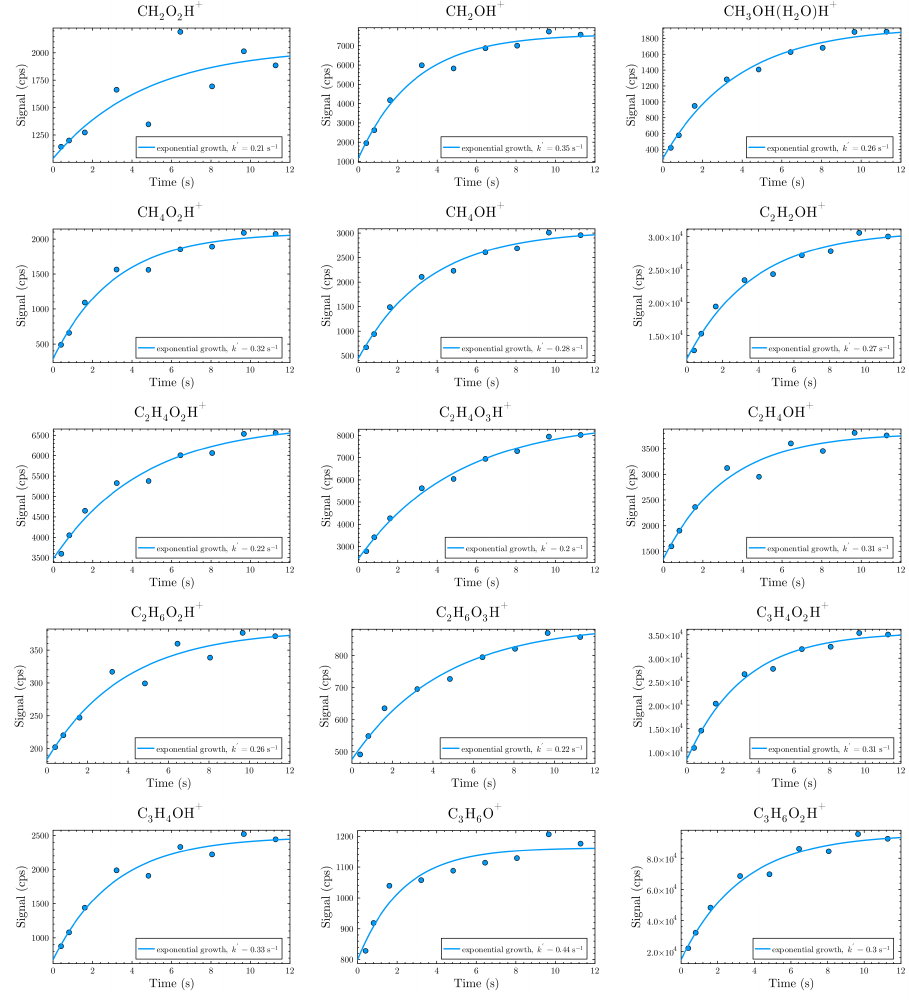}
	\caption{\textbf{Increasing mass-spectral signals in the \ce{C6H12 + O3} reaction system (Part 1).} Each panel shows the evolution of a signal assigned a molecular formula and classified as increasing. Experimental data (circles) are fitted with Eq.~\eqref{eq:products}. The fitted rate coefficients $k^{\prime}$ are reported in Table~\ref{tab:products}.}
	\label{fig:products1} 
\end{figure}

\begin{figure}[ht] 
	\centering
    \includegraphics[width=436.00296bp]{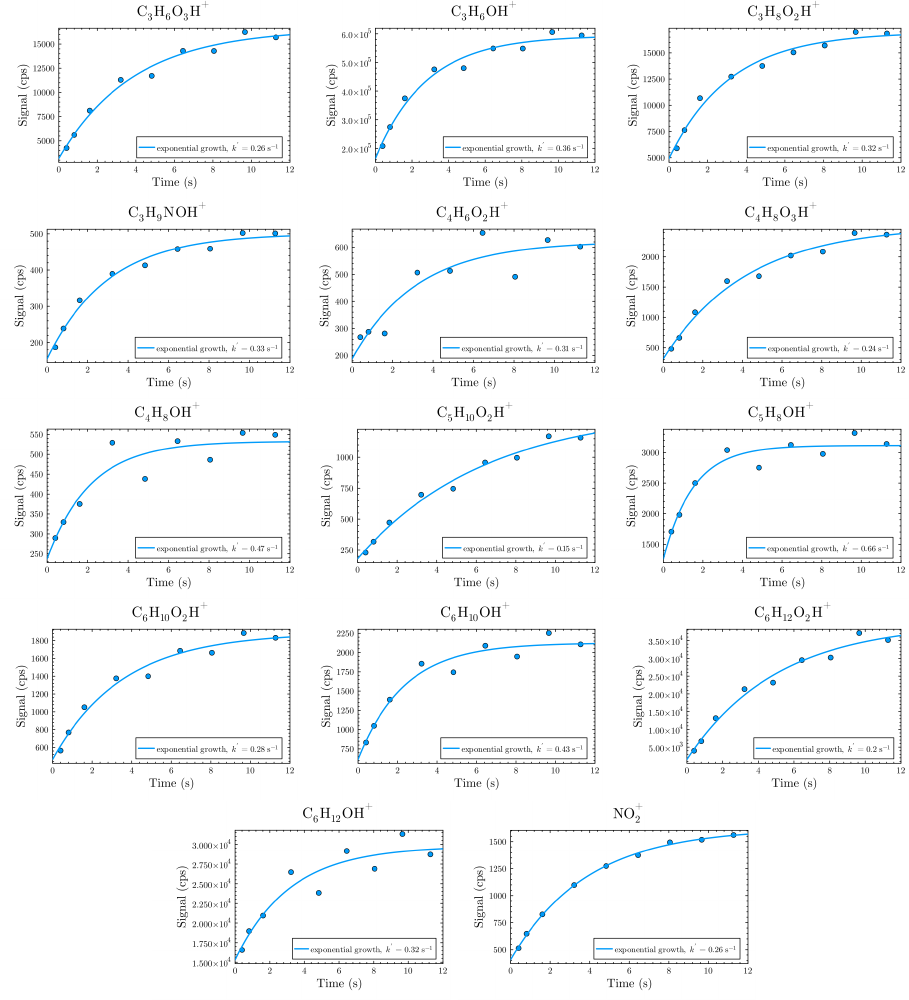}
	\caption{\textbf{Increasing mass-spectral signals in the \ce{C6H12 + O3} reaction system (Part 2).} Each panel shows the evolution of a signal assigned a molecular formula and classified as increasing. Experimental data (circles) are fitted with Eq.~\eqref{eq:products}. The fitted rate coefficients $k^{\prime}$ are reported in Table~\ref{tab:products}.}
	\label{fig:products2} 
\end{figure}

\begin{figure}[ht] 
	\centering
    \includegraphics[width=457.08903bp]{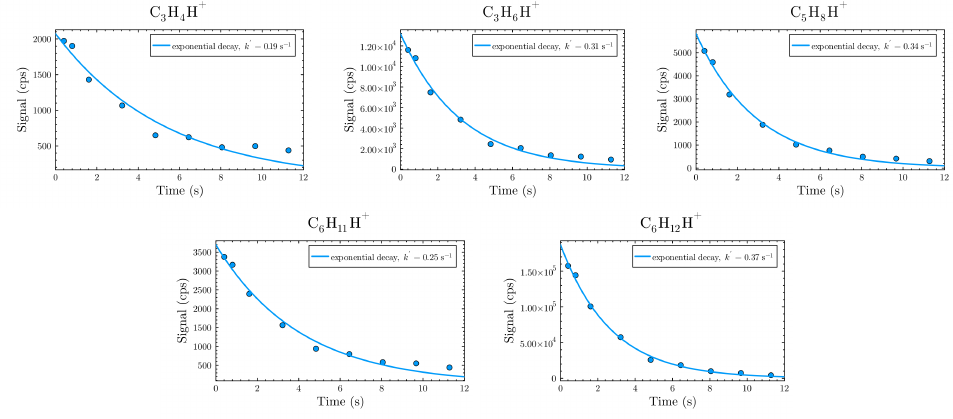}
	\caption{\textbf{Decreasing mass-spectral signals in the \ce{C6H12 + O3} reaction system.} Only the parent \ce{C6H12H+} signal is used for the kinetic determination; the molecular origins of the other decreasing signals cannot be established uniquely from these data. Experimental data (circles) are fitted with Eq.~\eqref{eq:reactants}. The fitted apparent signal-decay constants $k^{\prime}$ are reported in Table~\ref{tab:reactants}.}
	\label{fig:reactants} 
\end{figure}

\begin{figure}[ht] 
	\centering
    \includegraphics[width=0.32\textwidth]{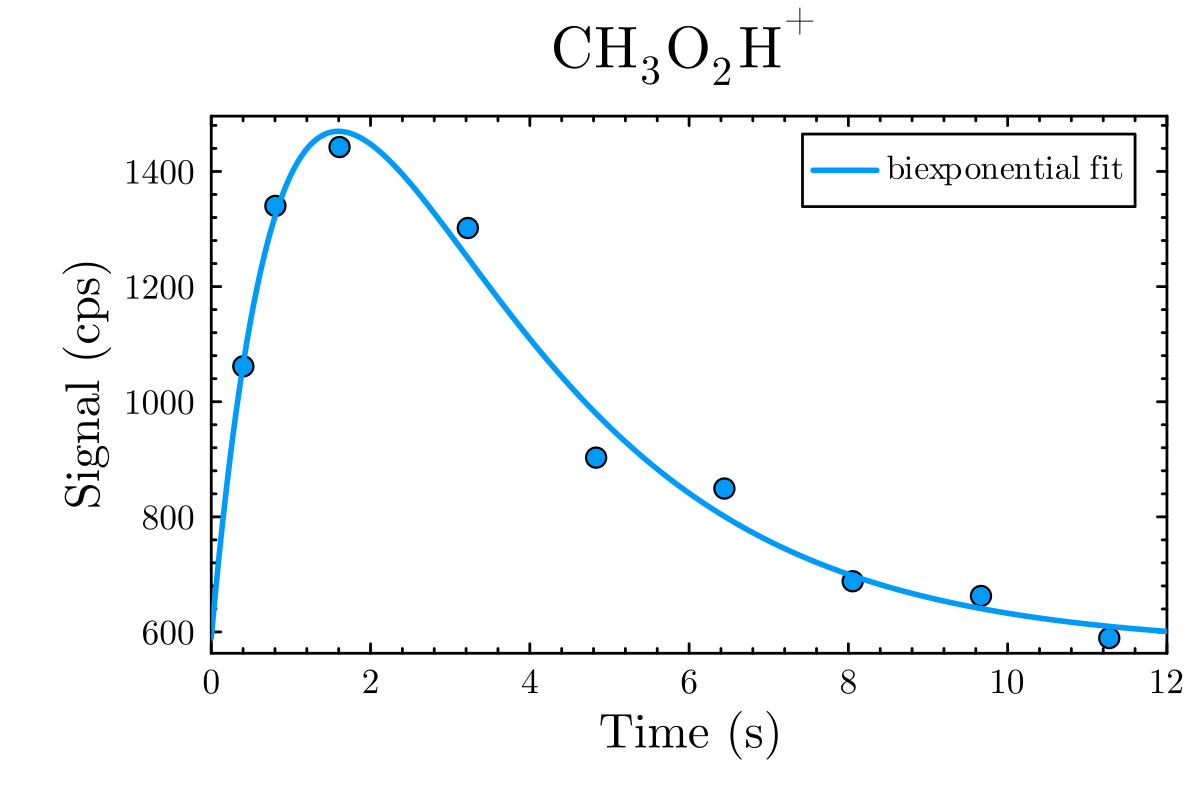} 
    \caption{\textbf{Transient signal assigned to protonated \ce{CH3O2} in the \ce{C6H12 + O3} reaction system.} The clear rise-and-decay behavior is characteristic of an intermediate that is both formed and consumed. Experimental data (circles) are fitted with Eq.~\eqref{eq:intermediate}; the fitted rate coefficients $k^{\prime}_\text{grow}$ and $k^{\prime}_\text{decay}$ are reported in Table~\ref{tab:intermediate}.}
	\label{fig:intermediate} 
\end{figure}

The 35 signals selected for reporting comprise five decreasing signals (Figure~\ref{fig:reactants}), one transient signal (Figure~\ref{fig:intermediate}), and 29 increasing signals (Figures~\ref{fig:products1} and~\ref{fig:products2}). The nearly flat \ce{NO+} trace was excluded. Data processing was performed with the in-house Julia package \texttt{MassSpec.jl} \cite{durifMassSpecJuliaPackage2025} and is fully reproducible from the raw spectra and analysis scripts available in the Zenodo archive.

Mass calibration uses the reflectron model
\begin{equation}
    t(m)=a\sqrt{m}+b+\frac{c}{\sqrt{m}},
\end{equation}
where $m$ denotes $m/z$ and $t$ is the time-of-flight coordinate in bins. The model is inverted on its increasing branch to assign a mass-to-charge ratio to each bin.

These three parameters are obtained by a robust fit over the reference ions of each acquisition: 51 anchor masses spanning $m/z$~18 to~133, each assigned from a curated catalogue and checked against its exact mass. The fit follows the drift of the time axis through the acquisition, in blocks of cycles whose length is chosen by cross-validation. The residual scatter of the anchors with respect to their exact masses, $\tau=\SI{16}{\ppm}$, is reported with the calibration.

Detected peaks were matched to catalogue ion formulas within a mass-dependent tolerance; ambiguous matches were left unassigned. The resulting assignments were reviewed manually.

Once the list of peaks is established, the signal of each species is integrated over the window of its peak in each spectrum. The peak table is built from a consensus over the nine steady-state lamp-on acquisition windows rather than from a single spectrum, so that a weak ion detected in one window only is kept, and each peak centre is taken from the summed signal. The resulting values are averaged over the 51 spectra of a window to obtain one signal value per kinetic time point; the procedure is then repeated for all kinetic time points.

The temporal evolution of the signal assigned to each molecular formula is fitted with an increasing, decreasing, or biexponential function. The fitted results are subsequently filtered using the criterion $k^\prime \geq \SI{0.02}{\per\second}$, corresponding to approximately \SI{5}{\percent} of the rate observed for \ce{C6H12} ($k^\prime=\SI{0.37}{\per\second}$).

Increasing signals were fitted using
\begin{equation}
    G(t) = S_0 + A\left(1-\exp(-k^\prime t)\right), \label{eq:products}
\end{equation}
Decreasing signals were fitted using
\begin{equation}
    D(t) = A\exp(-k^\prime t), \label{eq:reactants}
\end{equation}
Transient signals were fitted using
\begin{equation}
    I(t) = S_\infty-A_\text{grow}\exp(-k^\prime_\text{grow}t)
    +A_\text{decay}\exp(-k^\prime_\text{decay}t). \label{eq:intermediate}
\end{equation}
These parameterizations avoid introducing a time-offset parameter that would be correlated with the fitted amplitudes.

The species identified by this methodology are reported in Tables~\ref{tab:products},~\ref{tab:reactants}, and~\ref{tab:intermediate} and are displayed in Figures~\ref{fig:products1},~\ref{fig:products2},~\ref{fig:reactants}, and~\ref{fig:intermediate}.


\begin{table}[ht]
	\centering
	\caption{\textbf{Detected ions associated with increasing mass-spectral signals in the \ce{C6H12 + O3} reaction system.} The first column reports the detected ion formula; most signals were detected in protonated form. $k^\prime$ was obtained by fitting Eq.~\eqref{eq:products}; the ions are listed in order of increasing $k^\prime/k^\prime_\mathrm{TME}$.}
	\label{tab:products} 
    \footnotesize
	\begin{tabular}{lcc} 
		\\
		\hline
        \bf{Detected ion} & \bm{$k^\prime$} (s$^{-1}$) & \bf{ratio} \\
        \hline
        \ce{C5H10O2H+} & 0.15 & 0.41 \\
        \ce{C6H12O2H+} & 0.20 & 0.53 \\
        \ce{C2H4O3H+} & 0.20 & 0.54 \\
        \ce{CH2O2H+} & 0.21 & 0.57 \\
        \ce{C2H4O2H+} & 0.22 & 0.59 \\
        \ce{C2H6O3H+} & 0.22 & 0.60 \\
        \ce{C4H8O3H+} & 0.24 & 0.63 \\
        \ce{C3H6O3H+} & 0.26 & 0.69 \\
        \ce{NO2+} & 0.26 & 0.70 \\
        \ce{C2H6O2H+} & 0.26 & 0.70 \\
        \ce{CH3OH(H2O)H+} & 0.26 & 0.71 \\
        \ce{C2H2OH+} & 0.27 & 0.72 \\
        \ce{CH4OH+} & 0.28 & 0.75 \\
        \ce{C6H10O2H+} & 0.28 & 0.76 \\
        \ce{C3H6O2H+} & 0.30 & 0.81 \\
        \ce{C4H6O2H+} & 0.31 & 0.84 \\
        \ce{C2H4OH+} & 0.31 & 0.84 \\
        \ce{C3H4O2H+} & 0.31 & 0.84 \\
        \ce{C3H8O2H+} & 0.32 & 0.85 \\
        \ce{C6H12OH+} & 0.32 & 0.85 \\
        \ce{CH4O2H+} & 0.32 & 0.87 \\
        \ce{C3H4OH+} & 0.33 & 0.87 \\
        \ce{C3H9NOH+} & 0.33 & 0.89 \\
        \ce{CH2OH+} & 0.35 & 0.95 \\
        \ce{C3H6OH+} & 0.36 & 0.96 \\
        \ce{C6H10OH+} & 0.43 & 1.15 \\
        \ce{C3H6O+} & 0.44 & 1.18 \\
        \ce{C4H8OH+} & 0.47 & 1.25 \\
        \ce{C5H8OH+} & 0.66 & 1.76 \\
			\hline
		\end{tabular}
\end{table}

\begin{table}[ht]
	\centering
	\caption{\textbf{Detected ions associated with decreasing mass-spectral signals in the \ce{C6H12 + O3} reaction system.}
		\ce{C6H12} (detected as \ce{C6H12H+}, $m/z$~85) is the only primary reactant with ozone. The other decreasing signals have formulas compatible with possible fragments of protonated TME, but their lower fitted constants show that they are not pure proxies for TME decay. The cause of these differences cannot be established from the present data. Their fitted $k^\prime$ values are therefore reported as apparent signal-decay constants and are not interpreted as chemical rate coefficients. The kinetic analysis is based only on the parent ion at $m/z$~85. $k^\prime$ was obtained by fitting the data with Eq.~\eqref{eq:reactants}.}
	\label{tab:reactants} 
    \footnotesize
	\begin{tabular}{lcc} 
		\\
		\hline
        \bf{Detected ion} & \bm{$k^\prime$} (s$^{-1}$) & \bf{ratio} \\
        \hline
        \ce{C3H4H+} & 0.19 & 0.50 \\
        \ce{C6H11H+} & 0.25 & 0.66 \\
        \ce{C3H6H+} & 0.31 & 0.83 \\
        \ce{C5H8H+} & 0.34 & 0.91 \\
        \ce{C6H12H+} & 0.37 & 1.00 \\
		\hline
	\end{tabular}
\end{table}

\begin{table}[ht]
	\centering
	\caption{\textbf{Detected ion associated with the transient signal in the \ce{C6H12 + O3} reaction system.}
		Only the \ce{CH3O2H+} signal, assigned to protonated \ce{CH3O2}, was reliably detected. A biexponential fit was performed, and $k^\prime_\text{grow}$ and $k^\prime_\text{decay}$ were obtained by fitting the \ce{CH3O2H+} kinetic trace with Eq.~\eqref{eq:intermediate}.}
	\label{tab:intermediate} 
    \footnotesize
	\begin{tabular}{lcc} 
		\\
		\hline
        \bf{Detected ion} & \bm{$k^\prime_\text{grow}$} (s$^{-1}$) & \bm{$k^\prime_\text{decay}$} (s$^{-1}$) \\
        \hline
        \ce{CH3O2H+} & 0.94 & 0.39 \\
		\hline
	\end{tabular}
\end{table}

\clearpage




\clearpage
\section*{Code and Data Availability}
The public Julia package \texttt{MassSpec.jl} used for PTR-TOF-MS data processing is available at \url{https://gitlab.com/odurif/MassSpec.jl}. The exact analysis scripts used to generate the figures and tables in this study, together with all raw PTR-TOF-MS data, processed RTD data, and derived tables supporting this study, are archived on Zenodo at \url{https://doi.org/10.5281/zenodo.24399593}.

\section*{Author Contributions}
Olivier Durif led the study, including reactor design and assembly, experiments, data processing, interpretation, and manuscript preparation. Barbara Nozi\`ere contributed through scientific discussion, critical review, and funding acquisition.

\section*{Competing Interests}
The authors declare that they have no conflict of interest.

\section*{Funding}
This work is part of the ERC Advanced Grant Project EPHEMERAL (grant no. 884532) and has received funding from the European Research Council (ERC) under the European Union's Horizon 2020 research and innovation programme.

\end{document}